\documentstyle[11pt,epsf,rotate]{article}
\input{psfig}

\newcommand{\gsim}{~{}_{\textstyle\sim}^{\textstyle >}~}

\def\OEE{\Omega_{\eta+\eta'}}

\newcommand{\RE}{{\rm Re}}
\newcommand{\IM}{{\rm Im}}
\newcommand{\vcb}{|V_{cb}|}
\newcommand{\vtd}{|V_{td}|}
\newcommand{\vub}{|V_{ub}/V_{cb}|}
\newcommand{\vts}{|V_{ts}|}

\def\R1{\varepsilon_1}
\def\E8{\varepsilon_8}

\def\r#1{(\ref{#1})}
\def\eps{\varepsilon}
\def\epe{\varepsilon'/\varepsilon}
\def\as{\alpha_s}
\newcommand{\eqn}{\ref}
\def\Heff{{\cal H}_{\rm eff}}

\newcommand{\mt}{m_{\rm t}}
\newcommand{\mtb}{\overline{m}_{\rm t}}

\newcommand{\mc}{m_{\rm c}}
\newcommand{\ms}{m_{\rm s}}
\newcommand{\md}{m_{\rm d}}
\newcommand{\mb}{m_{\rm b}}
\newcommand{\mw}{M_{\rm W}}
\newcommand{\mz}{M_{\rm Z}}
\newcommand{\gev}{\, {\rm GeV}}
\newcommand{\mev}{\, {\rm MeV}}
\newcommand{\bsi}{B_6^{(1/2)}}
\newcommand{\bei}{B_8^{(3/2)}}
\newcommand{\Lms}{\Lambda_{\overline{\rm MS}}}

\newcommand{\bea}{\begin{eqnarray}}
\newcommand{\eea}{\end{eqnarray}}
\newcommand{\bd}{\begin{displaymath}}
\newcommand{\ed}{\end{displaymath}}
\newcommand{\aem}{\alpha}

\newcommand{\beq}{\begin{equation}}
\newcommand{\eeq}{\end{equation}}
\newcommand{\be}{\begin{equation}}
\newcommand{\ee}{\end{equation}}
\newcommand{\bi}{\begin{itemize}}
\newcommand{\ei}{\end{itemize}}
\newcommand{\ord}{{\cal O}}

\def\kpnn{$K^+\rightarrow\pi^+\nu\bar\nu$}
\def\kpn{K^+\rightarrow\pi^+\nu\bar\nu}
\def\klpn{K_{\rm L}\rightarrow\pi^0\nu\bar\nu}
\def\klpnn{$K_{\rm L}\rightarrow\pi^0\nu\bar\nu$}

\newcommand{\kppn}{K^+ \to \pi^+ \nu \bar \nu}
\newcommand{\kmm}{K_{\rm L} \to \mu^+ \mu^-}
\newcommand{\kpe}{K_{\rm L} \to \pi^0 e^+ e^-}

\def\aspi{\frac{\as}{4\pi}}

\newcommand{\imlt}{\IM\lambda_t}
\newcommand{\relt}{\RE\lambda_t}
\newcommand{\relc}{\RE\lambda_c}

\begin{document}
\thispagestyle{empty}
\phantom{xxx}
\vskip1truecm
\begin{flushright}
 TUM-HEP-349/99 \\
May 1999
\end{flushright}
\vskip1.8truecm
\centerline{\LARGE\bf CP Violation and Rare Decays of K and B Mesons}
   \vskip1truecm
\centerline{\Large\bf Andrzej J. Buras}
\bigskip
\centerline{\sl Technische Universit{\"a}t M{\"u}nchen}
\centerline{\sl Physik Department} 
\centerline{\sl D-85748 Garching, Germany}
\vskip1truecm
\centerline{\bf Abstract}
These lectures describe
CP violation and rare decays of K and B mesons and consist
of ten chapters:
i) Grand view of the field including CKM matrix and
the unitarity triangle, ii) General aspects of the theoretical
framework, iii) Particle-antiparticle mixing
and CP violation, iv) Standard analysis of the unitarity
triangle, v) The ratio $\epe$ including most recent developments,
vi) Rare decays $K^+\to\pi^+\nu\bar\nu$ and $K_L\to\pi^0\nu\bar\nu$,
vii) Express review of other rare decays, viii) Express review
of CP violation in B decays, ix) A brief look beyond the Standard
Model including connections between $\epe$ and CP violating
rare K decays, x) Final messages.
\vskip2truecm

\centerline{\it Lectures given at the 14th}
\centerline{\bf Lake Louise Winter Institute}
\centerline{\sl February 14-20, 1999}
%%% end title page %%%%%%%%%%%%%

\newpage

\thispagestyle{empty}

\mbox{}

\newpage

\pagenumbering{roman}

\tableofcontents

\newpage

\pagenumbering{arabic}

\setcounter{page}{1}

\section{Grand View}
\setcounter{equation}{0}
\subsection{Preface}
CP violation and rare decays of K and B mesons play an important
role in the tests of the Standard Model and of its extensions.
The prime examples, already observed experimentally, are
$K^0-\bar K^0$ and $B^0_d-\bar B^0_d$ mixings, CP violation
in $K_L\to\pi\pi$ and the rare decays $B\to X\gamma$,
$K_L\to\mu\bar\mu$ and $K^+\to \pi^+\nu\bar\nu$. In the coming
years CP violation in B decays, $B^0_s-\bar B^0_s$ mixing
and rare decays $K_L\to\pi^0\nu\bar\nu$, $K_L\to \pi^0e^+e^-$,
$B\to X_{s,d} l^+l^-$, $B_{d,s}\to l^+l^-$ and $B\to X_{s,d}\nu\bar\nu$
will hopefully be included in this list.

These lectures provide a non-technical description of this fascinating
field. There is unavoidably an overlap with my Les Houches lectures 
\cite{AJBLH} and with the reviews \cite{BBL} and \cite{BF97}.
On the other hand new developments are included and all numerical
results updated.
\subsection{Some Facts about the Standard Model}
Throughout these lectures we will dominantly work in the context of
the Standard Model with three generations of quarks and leptons and
the interactions described by the gauge group 
$ SU(3)_C\otimes SU(2)_L\otimes U(1)_Y$ spontaneously broken to
$SU(3)_C\otimes U(1)_Q$.
There are excellent text books on the dynamics of the Standard Model.
Let us therfore collect here only those ingredients of this model which
are fundamental for the subject of weak decays.

\bi
\item
The strong interactions are mediated by eight gluons $G_a$, the
electroweak interactions by $W^{\pm}$, $Z^0$ and $\gamma$.
\item
Concerning {\it Electroweak Interactions}, the left-handed leptons and
quarks are put into $SU(2)_L$ doublets:
\begin{equation}\label{2.31}
\left(\begin{array}{c}
\nu_e \\
e^-
\end{array}\right)_L\qquad
\left(\begin{array}{c}
\nu_\mu \\
\mu^-
\end{array}\right)_L\qquad
\left(\begin{array}{c}
\nu_\tau \\
\tau^-
\end{array}\right)_L
\end{equation}
\begin{equation}\label{2.66}
\left(\begin{array}{c}
u \\
d^\prime
\end{array}\right)_L\qquad
\left(\begin{array}{c}
c \\
s^\prime
\end{array}\right)_L\qquad
\left(\begin{array}{c}
t \\
b^\prime
\end{array}\right)_L       
\end{equation}
with the corresponding right-handed fields transforming as singlets
under $ SU(2)_L $. The primes in (\ref{2.66}) will be
discussed in a moment. 
\item
The charged current processes mediated by $W^{\pm}$ are
flavour violating with the strength of violation given by
the gauge coupling $g_2$  and effectively at low energies 
by the Fermi constant 
\begin{equation}\label{2.100}
\frac{G_{\rm F}}{\sqrt{2}}=\frac{g^2_2}{8 \mw^2}
\end{equation}
and a {\it unitary} $3\times3$
{\rm CKM} matrix. 
\item
The {\rm CKM} matrix \cite{CAB,KM} connects the {\it weak
eigenstates} $(d^\prime,s^\prime,b^\prime)$ and the corresponding {\it mass 
eigenstates} $d,s,b$ through
\begin{equation}\label{2.67}
\left(\begin{array}{c}
d^\prime \\ s^\prime \\ b^\prime
\end{array}\right)=
\left(\begin{array}{ccc}
V_{ud}&V_{us}&V_{ub}\\
V_{cd}&V_{cs}&V_{cb}\\
V_{td}&V_{ts}&V_{tb}
\end{array}\right)
\left(\begin{array}{c}
d \\ s \\ b
\end{array}\right)=\hat V_{\rm CKM}\left(\begin{array}{c}
d \\ s \\ b
\end{array}\right).
\end{equation}
In the leptonic sector the analogous mixing matrix is a unit matrix
due to the masslessness of neutrinos in the Standard Model.
\item
The unitarity of the CKM matrix assures the absence of
flavour changing neutral current transitions at the tree level.
This means that the
elementary vertices involving neutral gauge bosons ($G_a$, $Z^0$,
$\gamma$) are flavour conserving.
This property is known under the name of GIM mechanism \cite{GIM1}.
\item
The fact that the $V_{ij}$'s can a priori be complex
numbers allows  CP violation in the Standard Model \cite{KM}. 
\ei

\subsection{CKM Matrix}
\subsubsection{General Remarks}
We know from the text books that the CKM matrix can be
parametrized by
three angles and a single complex phase.
This phase leading to an
imaginary part of the CKM matrix is a necessary ingredient to describe
CP violation within the framework of the Standard Model.

Many parametrizations of the CKM
matrix have been proposed in the literature.  We will use
two parametrizations in these lectures: the standard parametrization 
\cite{CHAU} recommended by the Particle Data Group  \cite{PDG}  
and the Wolfenstein parametrization \cite{WO}.

\subsubsection{Standard Parametrization}
            \label{sec:sewm:stdparam}
With
$c_{ij}=\cos\theta_{ij}$ and $s_{ij}=\sin\theta_{ij}$ 
($i,j=1,2,3$), the standard parametrization is
given by:
\begin{equation}\label{2.72}
\hat V_{CKM}=
\left(\begin{array}{ccc}
c_{12}c_{13}&s_{12}c_{13}&s_{13}e^{-i\delta}\\ -s_{12}c_{23}
-c_{12}s_{23}s_{13}e^{i\delta}&c_{12}c_{23}-s_{12}s_{23}s_{13}e^{i\delta}&
s_{23}c_{13}\\ s_{12}s_{23}-c_{12}c_{23}s_{13}e^{i\delta}&-s_{23}c_{12}
-s_{12}c_{23}s_{13}e^{i\delta}&c_{23}c_{13}
\end{array}\right)\,,
\end{equation}
where $\delta$ is the phase necessary for {\rm CP} violation.
$c_{ij}$ and
$s_{ij}$ can all be chosen to be positive
and  $\delta$ may vary in the
range $0\le\delta\le 2\pi$. However, the measurements
of CP violation in $K$ decays force $\delta$ to be in the range
 $0<\delta<\pi$. 

From phenomenological applications we know that 
$s_{13}$ and $s_{23}$ are small numbers: $\ord(10^{-3})$ and ${\cal
O}(10^{-2})$,
respectively. Consequently to an excellent accuracy $c_{13}=c_{23}=1$
and the four independent parameters are given as 
\begin{equation}\label{2.73}
s_{12}=| V_{us}|, \quad s_{13}=| V_{ub}|, \quad s_{23}=|
V_{cb}|, \quad \delta.
\end{equation}

The first three can be extracted from tree level decays mediated
by the transitions $s \to u$, $b \to u$ and $b \to c$ respectively.
The phase $\delta$ can be extracted from CP violating transitions or 
loop processes sensitive to $| V_{td}|$. The latter fact is based
on the observation that
 for $0\le\delta\le\pi$, as required by the analysis of CP violation
in the $K$ system,
there is a one--to--one correspondence between $\delta$ and $|V_{td}|$
given by
\begin{equation}\label{10}
| V_{td}|=\sqrt{a^2+b^2-2 a b \cos\delta},
\qquad
a=| V_{cd} V_{cb}|,
\qquad
b=| V_{ud} V_{ub}|\,.
\end{equation} 

The main  phenomenological advantages of (\ref{2.72}) over other
parametrizations proposed in the literature are basically these
two: 
\begin{itemize}
\item
$s_{12}$, $s_{13}$ and $s_{23}$ being related in a very simple way
to $| V_{us}|$, $| V_{ub}|$ and $|V_{cb}|$ respectively, can be
measured independently in three decays.
\item
The CP violating phase is always multiplied by the very small
$s_{13}$. This shows clearly the suppression of CP violation
independently of the actual size of $\delta$.
\end{itemize}

For numerical evaluations the use of the standard parametrization
is strongly recommended. However once the four parameters in
(\ref{2.73}) have been determined it is often useful to make
a change of basic parameters in order to see the structure of
the result more transparently. This brings us to the Wolfenstein
parametrization \cite{WO} and its generalization given in 
\cite{BLO}.

\subsubsection{Wolfenstein Parameterization }\label{Wolf-Par}
 The Wolfenstein parametrization 
is an approximate parametrization of the CKM matrix in which
each element is expanded as a power series in the small parameter
$\lambda=| V_{us}|=0.22$,
\begin{equation}\label{2.75} 
\hat V=
\left(\begin{array}{ccc}
1-{\lambda^2\over 2}&\lambda&A\lambda^3(\varrho-i\eta)\\ -\lambda&
1-{\lambda^2\over 2}&A\lambda^2\\ A\lambda^3(1-\varrho-i\eta)&-A\lambda^2&
1\end{array}\right)
+\ord(\lambda^4)\,,
\end{equation}
and the set (\ref{2.73}) is replaced by
\begin{equation}\label{2.76}
\lambda, \qquad A, \qquad \varrho, \qquad \eta \, .
\end{equation}

Because of the
smallness of $\lambda$ and the fact that for each element 
the expansion parameter is actually
$\lambda^2$, it is sufficient to keep only the first few terms
in this expansion. 

The Wolfenstein parametrization is certainly more transparent than
the standard parametrization. However, if one requires sufficient 
level of accuracy, the higher order terms in $\lambda$ have to
be included in phenomenological applications.
This can be done in many ways.
The
point is that since (\ref{2.75}) is only an approximation the {\em exact}
definiton of the parameters in (\ref{2.76}) is not unique by terms of the 
neglected order
${\cal O}(\lambda^4)$. 
This situation is familiar from any perturbative expansion, where
different definitions of expansion parameters (coupling constants) 
are possible.
This is also the reason why in different papers in the
literature different ${\cal O}(\lambda^4)$ terms in (\ref{2.75})
 can be found. They simply
correspond to different definitions of the parameters in (\ref{2.76}).
Since the physics does not depend on a particular definition, it
is useful to make a choice for which the transparency of the original
Wolfenstein parametrization is not lost. Here we present one
way of achieving this.

\subsubsection{Wolfenstein Parametrization beyond LO}
An efficient and systematic way of finding higher order terms in $\lambda$
is to go back to the standard parametrization (\ref{2.72}) and to
 {\it define} the parameters $(\lambda,A,\varrho,\eta)$ through 
\cite{BLO,schubert}
\begin{equation}\label{2.77} 
s_{12}=\lambda\,,
\qquad
s_{23}=A \lambda^2\,,
\qquad
s_{13} e^{-i\delta}=A \lambda^3 (\varrho-i \eta)
\end{equation}
to {\it  all orders} in $\lambda$. 
It follows  that
\begin{equation}\label{2.84} 
\varrho=\frac{s_{13}}{s_{12}s_{23}}\cos\delta,
\qquad
\eta=\frac{s_{13}}{s_{12}s_{23}}\sin\delta.
\end{equation}
(\ref{2.77}) and (\ref{2.84}) represent simply
the change of variables from (\ref{2.73}) to (\ref{2.76}).
Making this change of variables in the standard parametrization 
(\ref{2.72}) we find the CKM matrix as a function of 
$(\lambda,A,\varrho,\eta)$ which satisfies unitarity exactly.
Expanding next each element in powers of $\lambda$ we recover the
matrix in (\ref{2.75}) and in addition find explicit corrections of
$\ord(\lambda^4)$ and higher order terms:. 

\be
V_{ud}=1-\frac{1}{2}\lambda^2-\frac{1}{8}\lambda^4 +\ord(\lambda^6)
\ee
\be
V_{us}=\lambda+\ord(\lambda^7),\qquad 
V_{ub}=A \lambda^3 (\varrho-i \eta)
\ee
\be
V_{cd}=-\lambda+\frac{1}{2} A^2\lambda^5 [1-2 (\varrho+i \eta)]+
\ord(\lambda^7)
\ee
\be
V_{cs}= 1-\frac{1}{2}\lambda^2-\frac{1}{8}\lambda^4(1+4 A^2) +\ord(\lambda^6)
\ee
\be
V_{cb}=A\lambda^2+\ord(\lambda^8), \qquad
V_{tb}=1-\frac{1}{2} A^2\lambda^4+\ord(\lambda^6)
\ee
\be
V_{td}=A\lambda^3 \left[ 1-(\varrho+i \eta)(1-\frac{1}{2}\lambda^2)\right]
+\ord (\lambda^7)
\ee
\begin{equation}\label{2.83d}
 V_{ts}= -A\lambda^2+\frac{1}{2}A(1-2 \varrho)\lambda^4
-i\eta A \lambda^4 +\ord(\lambda^6)
\end{equation}

We note that by definition
$V_{ub}$ remains unchanged and the
corrections to $V_{us}$ and $V_{cb}$ appear only at $\ord(\lambda^7)$ and
$\ord(\lambda^8)$, respectively.
Consequently to an 
 an excellent accuracy we have:
\begin{equation}\label{CKM1}
V_{us}=\lambda, \qquad V_{cb}=A\lambda^2,
\end{equation}
\begin{equation}\label{CKM2}
V_{ub}=A\lambda^3(\varrho-i\eta),
\qquad
V_{td}=A\lambda^3(1-\bar\varrho-i\bar\eta)
\end{equation}
with
\begin{equation}\label{2.88d}
\bar\varrho=\varrho (1-\frac{\lambda^2}{2}),
\qquad
\bar\eta=\eta (1-\frac{\lambda^2}{2}).
\end{equation}
The advantage of this generalization of the Wolfenstein parametrization
over other generalizations found in the literature is the absence of
relevant corrections to $V_{us}$, $V_{cb}$ and $V_{ub}$ and an elegant
change in $V_{td}$ which allows a simple generalization of the 
so-called unitarity triangle beyond LO.

Finally let us collect useful approximate analytic expressions
for $\lambda_i=V_{id}V^*_{is}$ with $i=c,t$:
\begin{equation}\label{2.51}
 \IM\lambda_t= -\IM\lambda_c=\eta A^2\lambda^5=
\mid V_{ub}\mid \mid V_{cb} \mid \sin\delta 
\end{equation}
\begin{equation}\label{2.52}
 \RE\lambda_c=-\lambda (1-\frac{\lambda^2}{2})
\end{equation}
\begin{equation}\label{2.53}
 \RE\lambda_t= -(1-\frac{\lambda^2}{2}) A^2\lambda^5 (1-\bar\varrho) \,.
\end{equation}
Expressions (\ref{2.51}) and (\ref{2.52}) represent to an accuracy of
0.2\% the exact formulae obtained using (\ref{2.72}). The expression
(\ref{2.53}) deviates by at most 2\% from the exact formula in the
full range of parameters considered. For $\varrho$ close to zero
this deviation is below 1\%. 
After inserting the expressions (\ref{2.51})--(\ref{2.53}) in the exact
formulae for quantities of interest, a further expansion in $\lambda$
should not be made. 
\subsubsection{Unitarity Triangle}
The unitarity of the CKM-matrix implies various relations between its
elements. In particular, we have
\begin{equation}\label{2.87h}
V_{ud}^{}V_{ub}^* + V_{cd}^{}V_{cb}^* + V_{td}^{}V_{tb}^* =0.
\end{equation}
Phenomenologically this relation is very interesting as it involves
simultaneously the elements $V_{ub}$, $V_{cb}$ and $V_{td}$ which are
under extensive discussion at present.

The relation (\ref{2.87h})  can be
represented as a ``unitarity'' triangle in the complex 
$(\bar\varrho,\bar\eta)$ plane. 
The invariance of (\ref{2.87h})  under any phase-transformations
implies that the  corresponding triangle
is rotated in the $(\bar\varrho,\bar\eta)$  plane under such transformations. 
Since the angles and the sides
(given by the moduli of the elements of the
mixing matrix)  in these triangles remain unchanged, they
 are phase convention independent and are  physical observables.
Consequently they can be measured directly in suitable experiments.  
The area of the unitarity triangle  is related to the measure of CP violation 
$J_{\rm CP}$ \cite{CJ,js}:
\begin{equation}
\mid J_{\rm CP} \mid = 2\cdot A_{\Delta},
\end{equation}
where $A_{\Delta}$ denotes the area of the unitarity triangle.

\begin{figure}[hbt]
\vspace{0.10in}
\centerline{
\epsfysize=2.1in
\epsffile{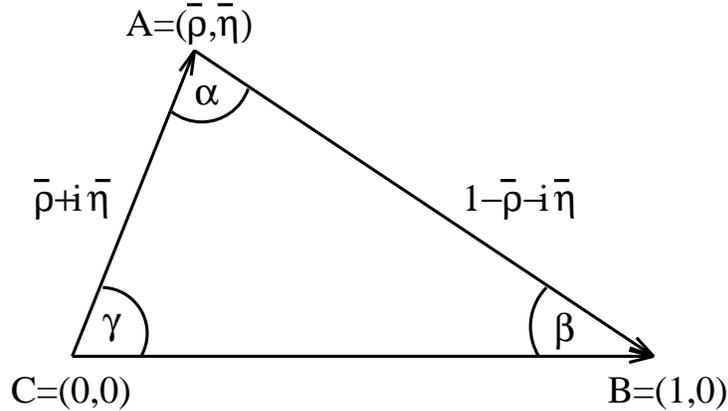}
}
\vspace{0.08in}
\caption{Unitarity Triangle.}\label{fig:utriangle}
\end{figure}

The construction of the unitarity triangle proceeds as follows:

\bi
\item
We note first that
\begin{equation}\label{2.88a}
V_{cd}^{}V_{cb}^*=-A\lambda^3+\ord(\lambda^7).
\end{equation}
Thus to an excellent accuracy $V_{cd}^{}V_{cb}^*$ is real with
$| V_{cd}^{}V_{cb}^*|=A\lambda^3$.
\item
Keeping $\ord(\lambda^5)$ corrections and rescaling all terms in
(\ref{2.87h})
by $A \lambda^3$ 
we find
\begin{equation}\label{2.88b}
 \frac{1}{A\lambda^3}V_{ud}^{}V_{ub}^*
=\bar\varrho+i\bar\eta,
\qquad
\qquad
 \frac{1}{A\lambda^3}V_{td}^{}V_{tb}^*
=1-(\bar\varrho+i\bar\eta)
\end{equation}
with $\bar\varrho$ and $\bar\eta$ defined in (\ref{2.88d}). 
\item
Thus we can represent (\ref{2.87h}) as the unitarity triangle 
in the complex $(\bar\varrho,\bar\eta)$ plane 
as shown in fig. \ref{fig:utriangle}.
\ei

Let us collect useful formulae related to this triangle:
\bi
\item
Using simple trigonometry one can express $\sin(2\phi_i$), $\phi_i=
\alpha, \beta, \gamma$, in terms of $(\bar\varrho,\bar\eta)$ as follows:
\begin{equation}\label{2.89}
\sin(2\alpha)=\frac{2\bar\eta(\bar\eta^2+\bar\varrho^2-\bar\varrho)}
  {(\bar\varrho^2+\bar\eta^2)((1-\bar\varrho)^2
  +\bar\eta^2)}  
\end{equation}
\begin{equation}\label{2.90}
\sin(2\beta)=\frac{2\bar\eta(1-\bar\varrho)}{(1-\bar\varrho)^2 + \bar\eta^2}
\end{equation}
 \begin{equation}\label{2.91}
\sin(2\gamma)=\frac{2\bar\varrho\bar\eta}{\bar\varrho^2+\bar\eta^2}=
\frac{2\varrho\eta}{\varrho^2+\eta^2}.
\end{equation}
\item
The lengths $CA$ and $BA$ in the
rescaled triangle  to be denoted by $R_b$ and $R_t$,
respectively, are given by
\begin{equation}\label{2.94}
R_b \equiv \frac{| V_{ud}^{}V^*_{ub}|}{| V_{cd}^{}V^*_{cb}|}
= \sqrt{\bar\varrho^2 +\bar\eta^2}
= (1-\frac{\lambda^2}{2})\frac{1}{\lambda}
\left| \frac{V_{ub}}{V_{cb}} \right|
\end{equation}
\begin{equation}\label{2.95}
R_t \equiv \frac{| V_{td}^{}V^*_{tb}|}{| V_{cd}^{}V^*_{cb}|} =
 \sqrt{(1-\bar\varrho)^2 +\bar\eta^2}
=\frac{1}{\lambda} \left| \frac{V_{td}}{V_{cb}} \right|.
\end{equation}
\item
The angles $\beta$ and $\gamma$ of the unitarity triangle are related
directly to the complex phases of the CKM-elements $V_{td}$ and
$V_{ub}$, respectively, through
\beq\label{e417}
V_{td}=|V_{td}|e^{-i\beta},\quad V_{ub}=|V_{ub}|e^{-i\gamma}.
\eeq
\item
The angle $\alpha$ can be obtained through the relation
\beq\label{e419}
\alpha+\beta+\gamma=180^\circ
\eeq
expressing the unitarity of the CKM-matrix.
\ei

The triangle depicted in fig. \ref{fig:utriangle} together with $|V_{us}|$ 
and $\vcb$ gives a full description of the CKM matrix. 
Looking at the expressions for $R_b$ and $R_t$, we observe that within
the Standard Model the measurements of four CP
{\it conserving } decays sensitive to $\mid V_{us}\mid$, $\mid V_{ub}\mid$,   
$\mid V_{cb}\mid $ and $\mid V_{td}\mid$ can tell us whether CP violation
($\eta \not= 0$) is predicted in the Standard Model. 
This is a very remarkable property of
the Kobayashi-Maskawa picture of CP violation: quark mixing and CP violation
are closely related to each other. 

\subsection{Grand Picture}
\begin{figure}[hbt]
\vspace{0.001in}
\centerline{
\epsfysize=4in
\rotate[r]{
\epsffile{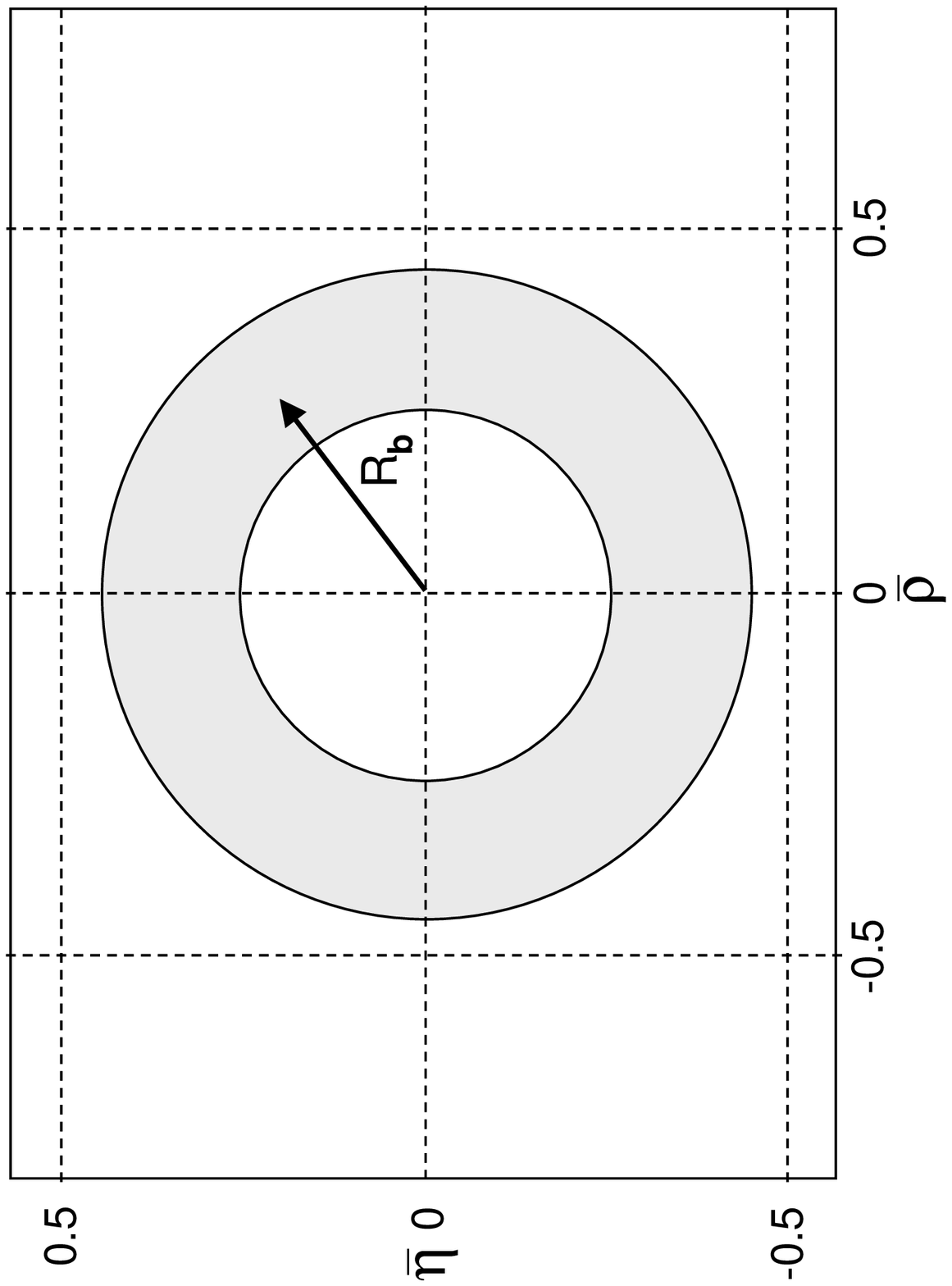}
}}
\vspace{-0.015in}
\caption[]{``Unitarity Clock".
\label{L:2}}
\end{figure}

\begin{figure}[hbt]
\vspace{0.10in}
\centerline{
\epsfysize=4.3in
\rotate[r]{
\epsffile{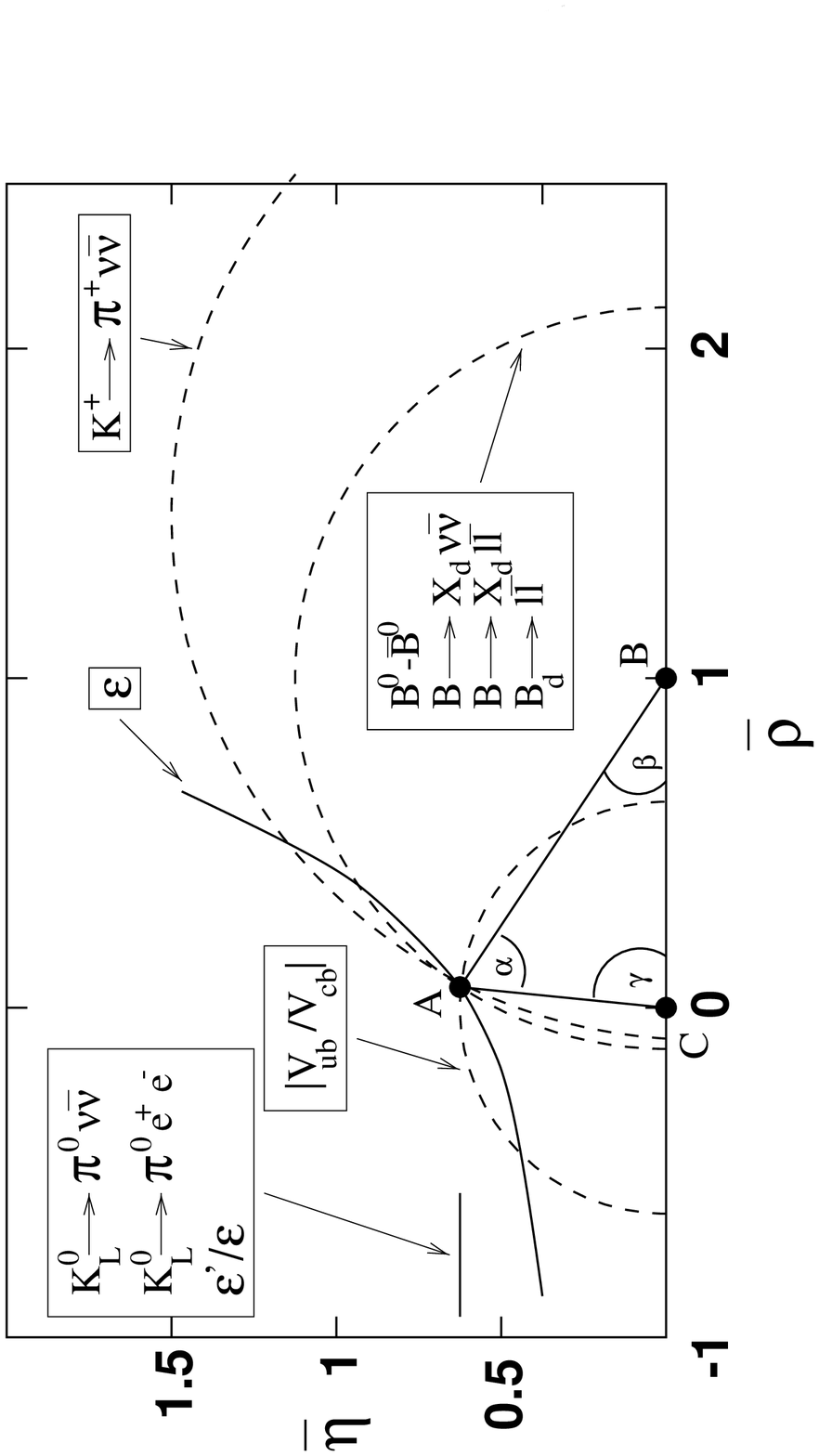}
}}
\vspace{0.08in}
\caption[]{
The ideal Unitarity Triangle. 
\label{fig:2011}}
\end{figure}

What do we know about the CKM matrix and the unitarity triangle on the
basis of {\it tree level} decays? A detailed answer to this question
can be found in the reports of
the Particle Data Group \cite{PDG} as well as other reviews 
\cite{UT99,STOCCHI},
where references to the relevant experiments and related theoretical
work can be found. In particular we have

\begin{equation}\label{vcb}
|V_{us}| = \lambda =  0.2205 \pm 0.0018\,
\quad\quad
\vcb=0.040\pm0.002,
\end{equation}
\begin{equation}\label{v13}
\frac{|V_{ub}|}{\vcb}=0.089\pm0.016, \quad\quad
|V_{ub}|=(3.56\pm0.56)\cdot 10^{-3}.
\end{equation}
Using (\ref{CKM1} and (\ref{2.94}) we find then
\be
 A=0.826\pm0.041,\qquad R_b=0.39\pm 0.07~.
\ee
This tells us only that the apex $A$ of the unitarity triangle lies
in the band shown in fig. \ \ref{L:2}. In order to answer the question where
the apex $A$ lies on this "unitarity clock'' we have to look at different
decays. Most promising in this respect are the so-called "loop induced''
decays and transitions which are the subject of several sections in these
lectures
and CP asymmetries in B-decays which will be briefly discussed in Section 8.
These two different routes for explorations of the CKM matrix
and of the related unitarity triangle may answer the important question, 
whether
the Kobayashi-Maskawa picture
of CP violation is correct and more generally whether the Standard
Model offers a correct description of weak decays of hadrons. Indeed,
in order
to answer these important questions it is essential to calculate as
many branching ratios as possible, measure them experimentally and
check if they all can be described by the same set of the parameters
$(\lambda,A,\varrho,\eta)$. In the language of the unitarity triangle
this means that the various curves in the $(\bar\varrho,\bar\eta)$ plane
extracted from different decays should cross each other at a single point
as shown in fig.~\ref{fig:2011}.
Moreover the angles $(\alpha,\beta,\gamma)$ in the
resulting triangle should agree with those extracted one day from
CP-asymmetries in B-decays. For artistic reasons the value of
$\bar\eta$ in fig.~\ref{fig:2011}
has been chosen to be higher than the fitted central value
$\bar\eta\approx 0.35.$

On the other hand if new physics contributes to weak decays
the different curves
based on the Standard Model expressions, will not cross each other 
at a single point
and the angles $(\alpha,\beta,\gamma)$ 
extracted one day from
CP-asymmetries in B-decays will disagree with the ones determined from
rare K and B decays.
Clearly the plot in fig. \ref{fig:2011} is highly idealized because in order
to extract such nice curves from various decays one needs perfect
experiments and perfect theory. 
One of the goals of these lectures is 
to identify those decays
for which at least the theory is under control. For such decays,
if they can be measured with a sufficient precision, the curves
in fig.~\ref{fig:2011} are  not fully unrealistic.
Let us then briefly discuss the theoretical framework for weak
decays.

\section{Theoretical Framework}
\setcounter{equation}{0}
\subsection{OPE and Renormalization Group}
The basis for any serious phenomenology of weak decays of
hadrons is the {\it Operator Product Expansion} (OPE) \cite{OPE,ZIMM},
which allows to write
the effective weak Hamiltonian simply as follows
\be\label{b1}
{\cal H}_{eff}=\frac{G_F}{\sqrt{2}}\sum_i V^i_{\rm CKM}C_i(\mu)Q_i~.
\ee
Here $G_F$ is the Fermi constant and $Q_i$ are the relevant local
operators which govern the decays in question. 
They are built out of quark and lepton fields.
The Cabibbo-Kobayashi-Maskawa
factors $V^i_{CKM}$ \cite{CAB,KM} 
and the Wilson coefficients $C_i$ \cite{OPE} describe the 
strength with which a given operator enters the Hamiltonian.
An amplitude for a decay of a given meson 
$M= K, B,..$ into a final state $F=\pi\nu\bar\nu,~\pi\pi,~DK$ is then
simply given by
\be\label{amp5}
A(M\to F)=\langle F|{\cal H}_{eff}|M\rangle
=\frac{G_F}{\sqrt{2}}\sum_i V^i_{CKM}C_i(\mu)\langle F|Q_i(\mu)|M\rangle,
\ee
where $\langle F|Q_i(\mu)|M\rangle$ 
are the hadronic matrix elements of $Q_i$ between M and F. 

The essential virtue of OPE is this one. It allows to separate the problem
of calculating the amplitude
$A(M\to F)$ into two distinct parts: the {\it short distance}
(perturbative) calculation of the coefficients $C_i(\mu)$ and 
the {\it long-distance} (generally non-perturbative) calculation of 
the matrix elements $\langle Q_i(\mu)\rangle$. The scale $\mu$
separates the physics contributions into short
distance contributions contained in $C_i(\mu)$ and the long distance 
contributions
contained in $\langle Q_i(\mu)\rangle$. 
Thus $C_i$ include the top quark contributions and
contributions from other heavy particles such as W, Z-bosons and charged
Higgs particles or supersymmetric particles in the supersymmetric extensions
of the Standard Model. 
Consequently $C_i(\mu)$ depend generally 
on $m_t$ and also on the masses of new particles if extensions of the 
Standard Model are considered. This dependence can be found by evaluating 
so-called {\it box} and {\it penguin} diagrams with full W-, Z-, top- and 
new particles exchanges and {\it properly} including short distance QCD 
effects. The latter govern the $\mu$-dependence of $C_i(\mu)$.

The value of $\mu$ can be chosen arbitrarily but the final result
must be $\mu$-independent.
Therefore 
the $\mu$-dependence of $C_i(\mu)$ has to cancel the 
$\mu$-dependence of $\langle Q_i(\mu)\rangle$. In other words it is a
matter of choice what exactly belongs to $C_i(\mu)$ and what to 
$\langle Q_i(\mu)\rangle$. This cancellation
of $\mu$-dependence involves generally several terms in the expansion 
in (\ref{amp5}).
The  coefficients $C_i(\mu)$ 
depend also
on the renormalization scheme.
This scheme dependence must also be cancelled
by the one of $\langle Q_i(\mu)\rangle$ so that the physical amplitudes are 
renormalization scheme independent. Again, as in the case of the 
$\mu$-dependence, the cancellation of
the renormalization scheme dependence involves generally several 
terms in the expansion (\ref{amp5}).

Although $\mu$ is in principle arbitrary,
it is customary 
to choose
$\mu$ to be of the order of the mass of the decaying hadron. 
This is $\ord (\mb)$ and $\ord(\mc)$ for B-decays and
D-decays respectively. In the case of K-decays the typical choice is
 $\mu=\ord(1-2~GeV)$
instead of $\ord(m_K)$, which is much too low for any perturbative 
calculation of the couplings $C_i$.
Now due to the fact that $\mu\ll  M_{W,Z},~ m_t$, large logarithms 
$\ln\mw/\mu$ compensate in the evaluation of
$C_i(\mu)$ the smallness of the QCD coupling constant $\alpha_s$ and 
terms $\alpha^n_s (\ln\mw/\mu)^n$, $\alpha^n_s (\ln\mw/\mu)^{n-1}$ 
etc. have to be resummed to all orders in $\alpha_s$ before a reliable 
result for $C_i$ can be obtained.
This can be done very efficiently by means of the renormalization group
methods. 
The resulting {\it renormalization group improved} perturbative
expansion for $C_i(\mu)$ in terms of the effective coupling constant 
$\alpha_s(\mu)$ does not involve large logarithms and is more reliable.

All this looks rather formal but in fact should be familiar.
Indeed,
in the simplest case of the $\beta$-decay, ${\cal H}_{eff}$ takes 
the familiar form
\be\label{beta}
{\cal H}^{(\beta)}_{eff}=\frac{G_F}{\sqrt{2}}
\cos\theta_c[\bar u\gamma_\mu(1-\gamma_5)d \otimes
\bar e \gamma^\mu (1-\gamma_5)\nu_e]~,
\ee
where $V_{ud}$ has been expressed in terms of the Cabibbo angle. In this
particular case the Wilson coefficient is equal unity and the local
operator, the object between the square brackets, is given by a product 
of two $V-A$ currents. 
Equation (\ref{beta}) represents the Fermi theory for $\beta$-decays 
as formulated by Sudarshan and
Marshak \cite{SUMA} and Feynman and Gell-Mann \cite{GF} forty years ago, 
except that in (\ref{beta})
the quark language has been used and following Cabibbo a small departure of
$V_{ud}$ from unity has been incorporated. In this context the basic 
formula (\ref{b1})
can be regarded as a generalization of the Fermi Theory to include all known
quarks and leptons as well as their strong and electroweak interactions as
summarized by the Standard Model. 

Due to the interplay of electroweak 
and strong interactions the structure of the local operators is 
much richer than in the case of the $\beta$-decay. They can be classified 
with respect to the Dirac structure, colour structure and the type of quarks 
and leptons relevant for a given decay. Of particular interest are the 
operators involving quarks only. In the case of the $\Delta S=1$
transitions the relevant set of operators is given as follows:
 
{\bf Current--Current :}
\begin{equation}\label{OS1} 
Q_1 = (\bar s_{\alpha} u_{\beta})_{V-A}\;(\bar u_{\beta} d_{\alpha})_{V-A}
~~~~~~Q_2 = (\bar s u)_{V-A}\;(\bar u d)_{V-A} 
\end{equation}

{\bf QCD--Penguins :}
\begin{equation}\label{OS2}
Q_3 = (\bar s d)_{V-A}\sum_{q=u,d,s}(\bar qq)_{V-A}~~~~~~   
 Q_4 = (\bar s_{\alpha} d_{\beta})_{V-A}\sum_{q=u,d,s}(\bar q_{\beta} 
       q_{\alpha})_{V-A} 
\end{equation}
\begin{equation}\label{OS3}
 Q_5 = (\bar s d)_{V-A} \sum_{q=u,d,s}(\bar qq)_{V+A}~~~~~  
 Q_6 = (\bar s_{\alpha} d_{\beta})_{V-A}\sum_{q=u,d,s}
       (\bar q_{\beta} q_{\alpha})_{V+A} 
\end{equation}

{\bf Electroweak--Penguins :}
\begin{equation}\label{OS4} 
Q_7 = {3\over 2}\;(\bar s d)_{V-A}\sum_{q=u,d,s}e_q\;(\bar qq)_{V+A} 
~~~~~ Q_8 = {3\over2}\;(\bar s_{\alpha} d_{\beta})_{V-A}\sum_{q=u,d,s}e_q
        (\bar q_{\beta} q_{\alpha})_{V+A}
\end{equation}
\begin{equation}\label{OS5} 
 Q_9 = {3\over 2}\;(\bar s d)_{V-A}\sum_{q=u,d,s}e_q(\bar q q)_{V-A}
~~~~~Q_{10} ={3\over 2}\;
(\bar s_{\alpha} d_{\beta})_{V-A}\sum_{q=u,d,s}e_q\;
       (\bar q_{\beta}q_{\alpha})_{V-A} \,.
\end{equation}
Here, $e_q$ denotes the electric quark charges reflecting the
electroweak origin of $Q_7,\ldots,Q_{10}$. 

Clearly, in order to calculate the amplitude $A(M\to F)$, the matrix 
elements $\langle Q_i(\mu)\rangle$ have to be evaluated. 
Since they involve long distance contributions one is forced in
this case to use non-perturbative methods such as lattice calculations, the
1/N expansion (N is the number of colours), QCD sum rules, hadronic sum rules,
chiral perturbation theory and so on. In the case of certain B-meson decays,
the {\it Heavy Quark Effective Theory} (HQET) also turns out to be a 
useful tool.
Needless to say, all these non-perturbative methods have some limitations.
Consequently the dominant theoretical uncertainties in the decay amplitudes
reside in the matrix elements $\langle Q_i(\mu)\rangle$.

The fact that in most cases the matrix elements $\langle Q_i(\mu)\rangle$
 cannot be reliably
calculated at present, is very unfortunate. One of the main goals of the
experimental studies of weak decays is the determination of the CKM factors 
$V_{\rm CKM}$
and the search for the physics beyond the Standard Model. Without a reliable
estimate of $\langle Q_i(\mu)\rangle$ this goal cannot be achieved unless 
these matrix elements can be determined experimentally or removed from the 
final measurable quantities
by taking the ratios or suitable combinations of amplitudes or branching
ratios. However, this can be achieved only in a handful of decays and
generally one has to face directly the calculation of 
$\langle Q_i(\mu)\rangle$. We will discuss these issues later on.

\subsection{Inclusive Decays}
So far I have discussed only  {\it exclusive} decays. It turns out that
in the case of {\it inclusive} decays of heavy mesons, like B-mesons,
things turn out to be easier. In an inclusive decay one sums over all 
(or over
a special class) of accessible final states so that the amplitude for an
inclusive decay takes the form:
\be\label{ampi}
A(B\to X)
=\frac{G_F}{\sqrt{2}}\sum_{f\in X} 
V^i_{\rm CKM}C_i(\mu)\langle f|Q_i(\mu)|B\rangle~.
\ee
At first sight things look as complicated as in the case of exclusive decays.
It turns out, however, that the resulting branching ratio can be calculated
in the expansion in inverse powers of $\mb$ with the leading term 
described by the spectator model
in which the B-meson decay is modelled by the decay of the $b$-quark:
\be\label{hqe}
{\rm Br}(B\to X)={\rm Br}(b\to q) +\ord(\frac{1}{\mb^2})~. 
\ee
This formula is known under the name of the Heavy Quark Expansion (HQE)
\cite{HQE1}.
Since the leading term in this expansion represents the decay of the quark,
it can be calculated in perturbation theory or more correctly in the
renormalization group improved perturbation theory. It should be realized
that also here the basic starting point is the effective Hamiltonian 
 (\ref{b1})
and that the knowledge of $C_i(\mu)$ is essential for 
the evaluation of
the leading term in (\ref{hqe}). But there is an important difference 
relative to the
exclusive case: the matrix elements of the operators $Q_i$ can be 
"effectively"
evaluated in perturbation theory. 
This means, in particular, that their $\mu$ and renormalization scheme
dependences can be evaluated and the cancellation of these dependences by
those present in $C_i(\mu)$ can be investigated.

Clearly in order to complete the evaluation of $Br(B\to X)$ also the 
remaining terms in
(\ref{hqe}) have to be considered. These terms are of a non-perturbative 
origin, but
fortunately they are suppressed by at least two powers of $m_b$. 
They have been
studied by several authors in the literature with the result that they affect
various branching ratios by less then $10\%$ and often by only a few percent.
Consequently the inclusive decays give generally more precise theoretical
predictions at present than the exclusive decays. On the other hand their
measurements are harder. There are of course some important theoretical
issues related to the validity of HQE in (\ref{hqe}) which appear in the 
literature under the name of quark-hadron duality. 
Since these matters are rather involved I will not discuss them
here. 

\subsection{Status of NLO Calculations}
In order to achieve
sufficient precision for the theoretical predictions it is desirable to have
accurate values of $C_i(\mu)$. Indeed it has been realized at the end of
the eighties
that the leading term (LO) in the renormalization group improved perturbation
theory, in which the terms $\alpha^n_s (\ln\mw/\mu)^n$ are summed, is 
generally insufficient and the
inclusion of next-to-leading corrections  (NLO) which correspond to summing
the terms $\alpha^n_s (\ln\mw/\mu)^{n-1}$ is necessary. 
In particular, unphysical left-over $\mu$-dependences
in the decay amplitudes and branching ratios resulting from the truncation of
the perturbative series are considerably reduced by including NLO
corrections. These corrections are known by now for the most important and
interesting decays and will be taken into account in these lectures. 
The review of all 
existing NLO calculations 
can be found  in  \cite{BBL,ZIMG}.

\begin{figure}[hbt]
\centerline{
\epsfysize=4.5in
\epsffile{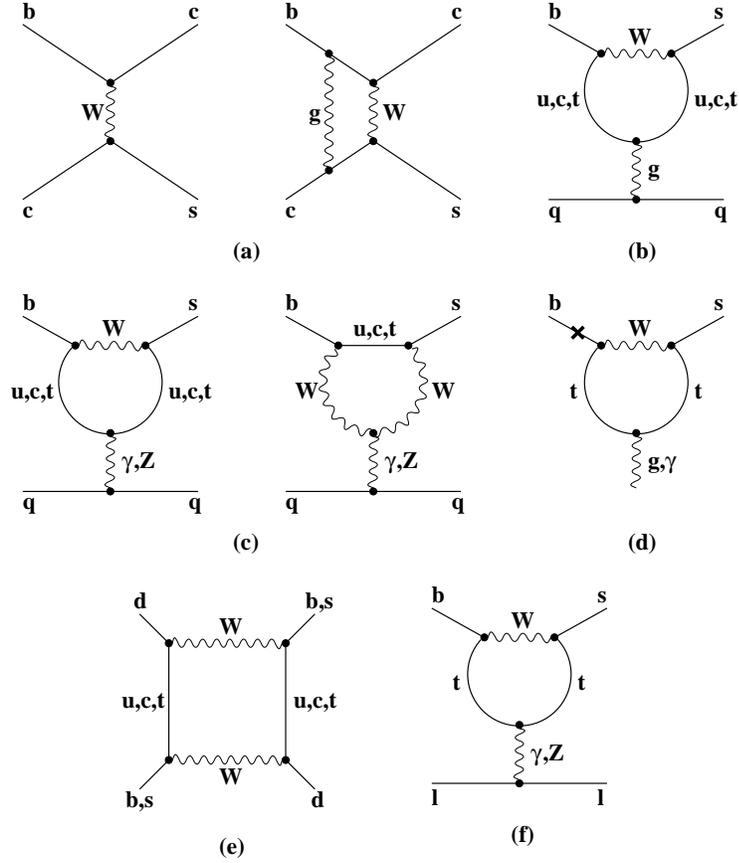}
}
\caption{Typical Penguin and Box Diagrams.}
\label{fig:fdia}
\end{figure}

\subsection{Penguin--Box Expansion}
The rare and CP violating decays of K and B mesons are governed 
by various penguin and box diagrams with internal top quark and charm quark
exchanges. Some examples are shown in fig.~\ref{fig:fdia}. 
Evaluating these diagrams one finds
a set of basic universal (process independent) 
$\mt$-dependent functions $F_r(x_t)$ \cite{IL} where $x_t=\mt^2/\mw^2$. 
Explicit expressions for these
functions will be given below. 

It is useful to express the OPE formula (\ref{amp5}) directly in terms
of the functions $F_r(x_t)$ \cite{PBE0}:
\begin{equation}
A({M\to F}) = P_0(M\to F) + \sum_r P_r(M\to F) \, F_r(x_t),
\label{generalPBE1}
\end{equation}
where the sum runs over all possible functions contributing to a given
amplitude. 
$P_0$  summarizes contributions stemming from internal quarks
other than the top, in particular the charm quark. 
The coefficients $P_0$ and $P_r$ are process dependent and
include QCD corrections. They depend also on hadronic matrix
elements of local operators and the relevant CKM factors.
I would like to call (\ref{generalPBE1}) {\it Penguin-Box Expansion} 
(PBE).
We will encounter many
examples of PBE in the course of 
these lectures.

Originally  PBE was designed to expose the $\mt$-dependence
of FCNC processes \cite{PBE0}. After the top quark mass has been
measured precisely this role of PBE is less important.
On the other hand,
PBE is very well suited for the study of the extentions of the
Standard Model in which new particles are exchanged in the loops.
If there are no new local operators
the mere change is to modify the functions $F_r(x_t)$ which now
acquire the dependence on the masses of new particles such as
charged Higgs particles and supersymmetric particles. The process
dependent coefficients $P_0$ and $P_r$ remain unchanged. 
The effects of new physics can be then transparently seen.
However, if
new effective operators with different Dirac and colour structures
are present the values of $P_0$ and $P_r$ are modified. 

Let us denote by $B_0$, $C_0$ and $D_0$ the functions $F_r(x_t)$
resulting from $\Delta F=1$ ($F$ stands for flavour) box diagram,
$Z^0$-penguin and $\gamma$-penguin diagram respectively. These
diagrams are gauge dependent and it is useful to introduce
gauge independent combinations \cite{PBE0}
\be\label{XYZ}
X_0=C_0-4 B_0,\qquad Y_0=C_0-B_0,\qquad Z_0=C_0+\frac{1}{4}D_0
\ee
Then the set of gauge independent basic functions which govern
the FCNC processes in the Standard Model is given to a very good
approximation as follows:
\begin{equation}\label{S0}
 S_0(x_t)=2.46~\left(\frac{\mt}{170\gev}\right)^{1.52},
\quad\quad S_0(x_c)=x_c
\ee
\begin{equation}\label{BFF}
S_0(x_c, x_t)=x_c\left[\ln\frac{x_t}{x_c}-\frac{3x_t}{4(1-x_t)}-
 \frac{3 x^2_t\ln x_t}{4(1-x_t)^2}\right].
\end{equation}
\be\label{XA0}
X_0(x_t)=1.57~\left(\frac{\mt}{170\gev}\right)^{1.15},
\quad\quad
Y_0(x_t)=1.02~\left(\frac{\mt}{170\gev}\right)^{1.56},
\end{equation}
\begin{equation}
 Z_0(x_t)=0.71~\left(\frac{\mt}{170\gev}\right)^{1.86},\quad\quad
   E_0(x_t)= 0.26~\left(\frac{\mt}{170\gev}\right)^{-1.02},
\end{equation}
\begin{equation}
 D'_0(x_t)=0.38~\left(\frac{\mt}{170\gev}\right)^{0.60}, \quad\quad 
E'_0(x_t)=0.19~\left(\frac{\mt}{170\gev}\right)^{0.38}.
\end{equation}
The first three functions correspond to
$\Delta F=2$ box diagrams with $(t,t)$, $(c,c)$ and $(t,c)$ exchanges.
$E_0$ results from QCD penguin diagram with off-shell gluon, 
$D'_0$ and $E'_0$ from $\gamma$ and QCD penguins with on-shell
photons and gluons respectively. 
The subscript ``$0$'' indicates that 
these functions
do not include QCD corrections to the relevant penguin and box diagrams.

In the range $150\gev \le \mt \le 200\gev$ 
these approximations reproduce the
exact expressions to an accuracy better than 1\%. 
These formulae will allow us to exhibit elegantly the $\mt$ dependence
of various branching ratios in the phenomenological sections of
these lectures.
Exact expressions for all functions can be found in \cite{AJBLH}.
 
Generally, several basic functions contribute to a given decay,
although decays exist which depend only on a single function.
We have the following correspondence between the most interesting FCNC
processes and the basic functions:
\begin{center}
\begin{tabular}{lcl}
$K^0-\bar K^0$-mixing &\qquad\qquad& $S_0(x_t)$, $S_0(x_c,x_t)$, 
$S_0(x_c)$ \\
$B^0-\bar B^0$-mixing &\qquad\qquad& $S_0(x_t)$ \\
$K \to \pi \nu \bar\nu$, $B \to X_{d,s} \nu \bar\nu$ 
&\qquad\qquad& $X_0(x_t)$ \\
$K_{\rm L}\to \mu \bar\mu$, $B \to l\bar l$ &\qquad\qquad& $Y_0(x_t)$ \\
$K_{\rm L} \to \pi^0 e^+ e^-$ &\qquad\qquad& $Y_0(x_t)$, $Z_0(x_t)$, 
$E_0(x_t)$ \\
$\varepsilon'$ &\qquad\qquad& $X_0(x_t)$, $Y_0(x_t)$, $Z_0(x_t)$,
$E_0(x_t)$ \\
$B \to X_s \gamma$ &\qquad\qquad& $D'_0(x_t)$, $E'_0(x_t)$ \\
$B \to X_s \mu^+ \mu^-$ &\qquad\qquad&
$Y_0(x_t)$, $Z_0(x_t)$, $E_0(x_t)$, $D'_0(x_t)$, $E'_0(x_t)$
\end{tabular}
\end{center}
 
\section{Particle-Antiparticle Mixing and CP
Violation}
        \label{sec:epsBBUT}
\setcounter{equation}{0}
\subsection{Preliminaries}
Let us next discuss particle--antiparticle mixing which in the past
 has been of fundamental
importance in testing the Standard Model and often has proven to be an
undefeatable challenge for suggested extensions of this model.
Let us just recall that
 from the calculation of the
$K_{\rm L}-K_{\rm S}$ mass difference, Gaillard and Lee \cite{GALE} 
were able to estimate the
value of the charm quark mass before charm discovery. On the
other hand $B_d^0-\bar B_d^0$ mixing \cite{ARGUS} gave the first 
indication of a large top quark mass. 
Finally, particle--antiparticle mixing in the $K^0-\bar K^0$ system
offers within the Standard Model a plausible description of
CP violation in $K_L\to\pi\pi$ discovered in 1964 \cite{CRONIN}. 

In this section we will predominantly discuss  the parameter 
$\varepsilon$ 
describing the {\it indirect} CP violation in the $K$ system and  
the mass differences $\Delta M_{d,s}$  which
describe the size of $B_{d,s}^0-\bar B^0_{d,s}$ mixings. 
In the Standard Model these phenomena
appear first at the one--loop level and as such they are
sensitive measures of the top quark couplings $V_{ti}(i=d,s,b)$ and 
and in particular of the phase $\delta=\gamma$.
They allow then to construct the unitarity triangle.

Let us next enter some details. The following subsection borrows a lot 
from \cite{CHAU83,BSSII}. A nice review of CP violation can also
be found in \cite{NIRSLAC}.

\begin{figure}[hbt]
\vspace{0.10in}
\centerline{
\epsfysize=1.5in
%\rotate[r]{
\epsffile{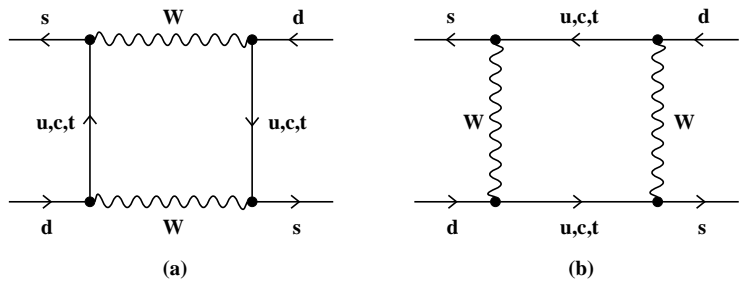}
}%}
\vspace{0.08in}
\caption[]{Box diagrams contributing to $K^0-\bar K^0$ mixing
in the Standard Model.
\label{L:9}}
\end{figure}

\subsection{Express Review of $K^0-\bar K^0$ Mixing}
$K^0=(\bar s d)$ and $\bar K^0=(s\bar d)$ are flavour eigenstates which 
in the Standard Model
may mix via weak interactions through the box diagrams in fig.
\ref{L:9}.
We will choose the phase conventions so that 
\be
CP|K^0\rangle=-|\bar K^0\rangle, \qquad   CP|\bar K^0\rangle=-|K^0\rangle.
\ee

In the absence of mixing the time evolution of $|K^0(t)\rangle$ is
given by
\be
|K^0(t)\rangle=|K^0(0)\rangle \exp(-iHt)~, 
\qquad H=M-i\frac{\Gamma}{2}~,
\ee
where $M$ is the mass and $\Gamma$ the width of $K^0$. Similar formula
for $\bar K^0$ exists.

On the other hand, in the presence of flavour mixing the time evolution 
of the $K^0-\bar K^0$ system is described by
\be
i\frac{d\psi(t)}{dt}=\hat H \psi(t) \qquad  
\psi(t)=
\left(\begin{array}{c}
|K^0(t)\rangle\\
|\bar K^0(t)\rangle
\end{array}\right)
\ee
where
\be
\hat H=\hat M-i\frac{\hat\Gamma}{2}
= \left(\begin{array}{cc} 
M_{11}-i\frac{\Gamma_{11}}{2} & M_{12}-i\frac{\Gamma_{12}}{2} \\
M_{21}-i\frac{\Gamma_{21}}{2}  & M_{22}-i\frac{\Gamma_{22}}{2}
    \end{array}\right)
\ee
with $\hat M$ and $\hat\Gamma$ being hermitian matrices having positive
(real) eigenvalues in analogy with $M$ and $\Gamma$. $M_{ij}$ and
$\Gamma_{ij}$ are the transition matrix elements from virtual and physical
intermediate states respectively.
Using
\be
M_{21}=M^*_{12}~, \qquad 
\Gamma_{21}=\Gamma_{12}^*~,\quad\quad {\rm (hermiticity)}
\ee
\be
M_{11}=M_{22}\equiv M~, \qquad \Gamma_{11}=\Gamma_{22}\equiv\Gamma~,
\quad {\rm (CPT)}
\ee
we have
\be\label{MM12}
\hat H=
 \left(\begin{array}{cc} 
M-i\frac{\Gamma}{2} & M_{12}-i\frac{\Gamma_{12}}{2} \\
M^*_{12}-i\frac{\Gamma^*_{12}}{2}  & M-i\frac{\Gamma}{2}
    \end{array}\right)~.
\ee

We can next diagonalize the system to find:

{\bf Eigenstates:}
\be\label{KLS}
K_{L,S}=\frac{(1+\bar\varepsilon)K^0\pm (1-\bar\varepsilon)\bar K^0}
        {\sqrt{2(1+\mid\bar\varepsilon\mid^2)}}
\ee
where $\bar\varepsilon$ is a small complex parameter given by
\be\label{bare3}
\frac{1-\bar\varepsilon}{1+\bar\varepsilon}=
\sqrt{\frac{M^*_{12}-i\frac{1}{2}\Gamma^*_{12}}
{M_{12}-i\frac{1}{2}\Gamma_{12}}}=
\frac{\Delta M-i\frac{1}{2}\Delta\Gamma}
{2 M_{12}-i\Gamma_{12}}\equiv r\exp(i\kappa)~.
\ee
with $\Delta\Gamma$ and $\Delta M$ given below.

{\bf Eigenvalues:}
\be
M_{L,S}=M\pm \RE Q  \qquad \Gamma_{L,S}=\Gamma\mp 2 \IM Q
\ee
where
\be
Q=\sqrt{(M_{12}-i\frac{1}{2}\Gamma_{12})(M^*_{12}-i\frac{1}{2}\Gamma^*_{12})}.
\ee
Consequently we have
\be\label{deltak}
\Delta M= M_L-M_S = 2\RE Q
\quad\quad
\Delta\Gamma=\Gamma_L-\Gamma_S=-4 \IM Q.
\ee

It should be noted that the mass eigenstates $K_S$ and $K_L$ differ from 
CP eigenstates
\begin{equation}
K_1={1\over{\sqrt 2}}(K^0-\bar K^0),
  \qquad\qquad CP|K_1\rangle=|K_1\rangle~,
\end{equation}
\begin{equation}
K_2={1\over{\sqrt 2}}(K^0+\bar K^0),
  \qquad\qquad CP|K_2\rangle=-|K_2\rangle~,
\end{equation}
by 
a small admixture of the
other CP eigenstate:
\begin{equation}
K_{\rm S}={{K_1+\bar\varepsilon K_2}
\over{\sqrt{1+\mid\bar\varepsilon\mid^2}}},
\qquad
K_{\rm L}={{K_2+\bar\varepsilon K_1}
\over{\sqrt{1+\mid\bar\varepsilon\mid^2}}}\,.
\end{equation}

It should be stressed that
the small parameter $\bar\varepsilon$  depends on the 
phase convention
chosen for $K^0$ and $\bar K^0$. Therefore it may not 
be taken as a physical measure of CP violation.
On the other hand $\RE\bar\varepsilon$ and $r$ are independent of
phase conventions. In particular the departure of $r$ from 1
measures CP violation in the $K^0-\bar K^0$ mixing:
\be
r=1+\frac{2 |\Gamma_{12}|^2}{4 |M_{12}|^2+|\Gamma_{12}|^2}
    \IM\left(\frac{M_{12}}{\Gamma_{12}}\right)~.
\ee
Since $\bar\varepsilon$ is $\ord(10^{-3})$, one has
 to a very good approximation:
\be\label{deltak1}
\Delta M_K = 2 \RE M_{12}, \qquad \Delta\Gamma_K=2 \RE \Gamma_{12}~,
\ee
where we have introduced the subscript K to stress that these formulae apply
only to the $K^0-\bar K^0$ system.

The 
$K_{\rm L}-K_{\rm S}$
mass difference is experimentally measured to be \cite{PDG}
\begin{equation}\label{DMEXP}
\Delta M_K=M(K_{\rm L})-M(K_{\rm S}) = 
(3.489\pm 0.009) \cdot 10^{-15} \gev\,.
\end{equation}
In the Standard Model roughly $70\%$ of the measured $\Delta M_K$
is described by the real parts of the box diagrams with charm quark
and top quark exchanges, whereby the contribution of the charm exchanges
is by far dominant. This is related to the smallness of the real parts
of the CKM top quark couplings compared with the corresponding charm
quark couplings. 
Some non-negligible contribution comes from the box diagrams with
simultaneous charm and top exchanges.
The remaining $20 \%$ of the measured $\Delta M_K$ is attributed to long 
distance contributions which are difficult to estimate \cite{GERAR}.
Further information with the relevant references can be found in 
\cite{HNa}.

The situation with $\Delta \Gamma_K$ is rather different.
It is fully dominated by long distance effects. Experimentally
one has
$\Delta\Gamma_K\approx-2 \Delta M_K$.

\subsection{The First Look at $\varepsilon$ and $\varepsilon'$}
Since a two pion final state is CP even while a three pion final state is CP
odd, $K_{\rm S}$ and $K_{\rm L}$ preferably decay to $2\pi$ and $3\pi$, 
respectively
via the following CP conserving decay modes:
\begin{equation}
K_{\rm L}\to 3\pi {\rm ~~(via~K_2),}\qquad K_{\rm S}\to 2 
\pi {\rm ~~(via~K_1).}
\end{equation}
This difference is responsible for the large disparity in their
life-times. A factor of 579.
However, $K_{\rm L}$ and $K_{\rm S}$ are not CP eigenstates and 
may decay with small branching fractions as follows:
\begin{equation}
K_{\rm L}\to 2\pi {\rm ~~(via~K_1),}\qquad K_{\rm S}\to 3 
\pi {\rm ~~(via~K_2).}
\end{equation}
This violation of CP is called {\it indirect} as it
proceeds not via explicit breaking of the CP symmetry in 
the decay itself but via the admixture of the CP state with opposite 
CP parity to the dominant one.
 The measure for this
indirect CP violation is defined as
\begin{equation}\label{ek}
\varepsilon
={{A(K_{\rm L}\rightarrow(\pi\pi)_{I=0}})\over{A(K_{\rm 
S}\rightarrow(\pi\pi)_{I=0})}}.
\end{equation}

Following the derivation in \cite{CHAU83} one finds
\begin{equation}
\eps = \frac{\exp(i \pi/4)}{\sqrt{2} \Delta M_K} \,
\left( \IM M_{12} + 2 \xi \RE M_{12} \right),
\quad\quad
\xi = \frac{\IM A_0}{\RE A_0}.
\label{eq:epsdef}
\end{equation}
where the term involving $\IM M_{12}$ represents $\bar\varepsilon$
 defined in (\ref{bare3}).
The phase convention dependence of the term involving $\xi$ cancells
the convention dependence of $\bar\varepsilon$ so that $\varepsilon$
is free from this dependence. 

\begin{figure}[hbt]
\centerline{
\epsfysize=1.5in
\epsffile{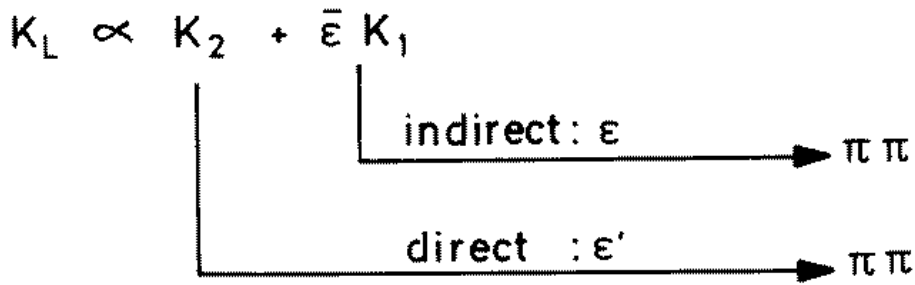}
}
\caption[]{
Indirect versus direct CP violation in $K_L \to \pi\pi$.
\label{fig:14}}
\end{figure}

While {\it indirect} CP violation reflects the fact that the mass
eigenstates are not CP eigenstates, so-called {\it direct}
CP violation is realized via a 
direct transition of a CP odd to a CP even state or vice versa (see
fig.~\ref{fig:14}). 
A measure of such a direct CP violation in $K_L\to \pi\pi$ is characterized
by a complex parameter $\varepsilon'$  defined as
\be\label{eprime}
\varepsilon'=\frac{1}{\sqrt{2}}\IM\left(\frac{A_2}{A_0}\right)
              \exp(i\Phi_{\varepsilon'}), 
   \qquad \Phi_{\varepsilon'}=\frac{\pi}{2}+\delta_2-\delta_0, 
\ee
where
the isospin amplitudes $A_I$ in $K\to\pi\pi$
decays are introduced through
\begin{equation} 
A(K^+\rightarrow\pi^+\pi^0)=\sqrt{3\over 2} A_2 e^{i\delta_2}
\end{equation}
\begin{equation} 
A(K^0\rightarrow\pi^+\pi^-)=\sqrt{2\over 3} A_0 e^{i\delta_0}+ \sqrt{1\over
3} A_2 e^{i\delta_2}
\end{equation}
\begin{equation}
A(K^0\rightarrow\pi^0\pi^0)=\sqrt{2\over 3} A_0 e^{i\delta_0}-2\sqrt{1\over
3} A_2 e^{i\delta_2}\,.
\end{equation} 
Here the subscript $I=0,2$ denotes states with isospin $0,2$
equivalent to $\Delta I=1/2$ and $\Delta I = 3/2$ transitions,
respectively, and $\delta_{0,2}$ are the corresponding strong phases. 
The weak CKM phases are contained in $A_0$ and $A_2$.
The isospin amplitudes $A_I$ are complex quantities which depend on
phase conventions. On the other hand, $\varepsilon'$ measures the 
difference between the phases of $A_2$ and $A_0$ and is a physical
quantity.
The strong phases $\delta_{0,2}$ can be extracted from $\pi\pi$ scattering. 
Then $\Phi_{\varepsilon'}\approx \pi/4$.

Experimentally $\varepsilon$ and $\varepsilon'$
can be found by measuring the ratios
\begin{equation}
\eta_{00}={{A(K_{\rm L}\to\pi^0\pi^0)}\over{A(K_{\rm S}\to\pi^0\pi^0)}},
            \qquad
  \eta_{+-}={{A(K_{\rm L}\to\pi^+\pi^-)}\over{A(K_{\rm S}\to\pi^+\pi^-)}}.
\end{equation}
Indeed, assuming $\varepsilon$ and $\varepsilon'$ to be small numbers one
finds
\be
\eta_{00}=\varepsilon-{{2\varepsilon'}\over{1-\sqrt{\omega}}}
            \simeq \varepsilon-2\varepsilon',~~~~
  \eta_{+-}=\varepsilon+{{\varepsilon'}\over{1+\omega/\sqrt{2}}}
            \simeq \varepsilon+ \varepsilon'
\end{equation}
where experimentally $\omega=\RE A_2/\RE A_0=0.045$.

In the absence of direct CP violation $\eta_{00}=\eta_{+-}$.
The ratio ${\varepsilon'}/{\varepsilon}$  can then be measured through
\begin{equation}
\left|{{\eta_{00}}\over{\eta_{+-}}}\right|^2\simeq 1 -6\; 
\RE(\frac{\varepsilon'}{\varepsilon})\,.
\end{equation}
\subsection{Basic Formula for $\eps$}
            \label{subsec:epsformula}
With all this information at hand let us derive a formula for $\varepsilon$
which can be efficiently used in pheneomenological applications.
The off-diagonal 
element $M_{12}$ in
the neutral $K$-meson mass matrix representing $K^0$-$\bar K^0$
mixing is given by
\begin{equation}
2 m_K M^*_{12} = \langle \bar K^0| \Heff(\Delta S=2) |K^0\rangle\,,
\label{eq:M12Kdef}
\end{equation}
where $\Heff(\Delta S=2)$ is the effective Hamiltonian for the 
$\Delta S=2$ transitions.
That $ M^*_{12}$ and not $ M_{12}$ stands on the l.h.s of this formula,
is evident from (\ref{MM12}). The factor $2 m_K$ reflects our normalization
of external states.

To lowest order in electroweak interactions $\Delta S=2$ transitions 
are induced
through the box diagrams of fig. \ref{L:9}. Including
 QCD corrections one has 
\begin{eqnarray}\label{hds2}
{\cal H}^{\Delta S=2}_{\rm eff}&=&\frac{G^2_{\rm F}}{16\pi^2}M^2_W
 \left[\lambda^2_c\eta_1 S_0(x_c)+\lambda^2_t \eta_2 S_0(x_t)+
 2\lambda_c\lambda_t \eta_3 S_0(x_c, x_t)\right] \times
\nonumber\\
& & \times \left[\as^{(3)}(\mu)\right]^{-2/9}\left[
  1 + \frac{\as^{(3)}(\mu)}{4\pi} J_3\right]  Q(\Delta S=2) + h. c.
\end{eqnarray}
where
$\lambda_i = V_{is}^* V_{id}^{}$,
$\mu<\mu_c=\ord(m_c)$ and $\as^{(3)}$ is the strong coupling constant
in an effective three flavour theory.
In (\ref{hds2}),
the relevant operator
\begin{equation}\label{qsdsd}
Q(\Delta S=2)=(\bar sd)_{V-A}(\bar sd)_{V-A},
\end{equation}
is multiplied by the corresponding coefficient function.
This function is decomposed into a
charm-, a top- and a mixed charm-top contribution.
The functions $S_0$  are given in (\ref{S0})and (\ref{BFF}).

Short-distance QCD effects are described through the correction
factors $\eta_1$, $\eta_2$, $\eta_3$ and the explicitly
$\alpha_s$-dependent terms in (\ref{hds2}). 
The NLO values of $\eta_i$ are given as follows \cite{HNa,BJW90,HNb}:
\begin{equation}
\eta_1=1.38\pm 0.20,\qquad
\eta_2=0.57\pm 0.01,\qquad
  \eta_3=0.47\pm0.04~.
\end{equation}
The quoted errors reflect the remaining theoretical uncertainties due to
leftover $\mu$-dependences at $\ord(\as^2)$ and $\Lambda_{\overline{MS}}$,
the scale in the QCD running coupling.

Defining  the renormalization group 
invariant parameter $\hat B_K$ by
\begin{equation}
\hat B_K = B_K(\mu) \left[ \alpha_s^{(3)}(\mu) \right]^{-2/9} \,
\left[ 1 + \frac{\alpha_s^{(3)}(\mu)}{4\pi} J_3 \right]
\label{eq:BKrenorm}
\end{equation}
\begin{equation}
\langle \bar K^0| (\bar s d)_{V-A} (\bar s d)_{V-A} |K^0\rangle
\equiv \frac{8}{3} B_K(\mu) F_K^2 m_K^2
\label{eq:KbarK}
\end{equation}
and using (\eqn{hds2}) one finds
\begin{equation}
M_{12} = \frac{G_{\rm F}^2}{12 \pi^2} F_K^2 \hat B_K m_K \mw^2
\left[ {\lambda_c^*}^2 \eta_1 S_0(x_c) + {\lambda_t^*}^2 \eta_2 S_0(x_t) +
2 {\lambda_c^*} {\lambda_t^*} \eta_3 S_0(x_c, x_t) \right],
\label{eq:M12K}
\end{equation}
where $F_K$ is the $K$-meson decay constant and $m_K$
the $K$-meson mass. 

To proceed further we neglect the last term in (\eqn{eq:epsdef}) as it
 constitutes at most a 2\,\% correction to $\eps$. This is justified
in view of other uncertainties, in particular those connected with
$\hat B_K$.
Inserting (\eqn{eq:M12K}) into (\eqn{eq:epsdef}) we find
\begin{equation}
\eps=C_{\eps} \hat B_K \IM\lambda_t \left\{
\RE\lambda_c \left[ \eta_1 S_0(x_c) - \eta_3 S_0(x_c, x_t) \right] -
\RE\lambda_t \eta_2 S_0(x_t) \right\} \exp(i \pi/4)\,,
\label{eq:epsformula}
\end{equation}
where we have used the unitarity relation $\IM\lambda_c^* = {\rm
Im}\lambda_t$ and  have neglected $\RE\lambda_t/\RE\lambda_c
 = \ord(\lambda^4)$ in evaluating $\IM(\lambda_c^* \lambda_t^*)$.
The numerical constant $C_\eps$ is given by
\begin{equation}
C_\eps = \frac{G_{\rm F}^2 F_K^2 m_K \mw^2}{6 \sqrt{2} \pi^2 \Delta M_K}
       = 3.84 \cdot 10^4 \, .
\label{eq:Ceps}
\end{equation}
To this end we have used the experimental value of $\Delta M_K$ 
in (\ref{DMEXP}). 

Using the standard parametrization of (\eqn{2.72}) to evaluate ${\rm
Im}\lambda_i$ and $\RE\lambda_i$, setting the values for $s_{12}$,
$s_{13}$, $s_{23}$ and $\mt$ in accordance with experiment
 and taking a value for $\hat B_K$ (see below), one can
determine the phase $\delta$ by comparing (\eqn{eq:epsformula}) with the
experimental value for $\eps$
\begin{equation}\label{eexp}
\varepsilon_{exp}
=(2.280\pm0.013)\cdot10^{-3}\;\exp{i\Phi_{\varepsilon}},
\qquad \Phi_{\varepsilon}={\pi\over 4}.
\end{equation}

Once $\delta$ has been determined in this manner one can find the
apex $(\bar\varrho,\bar\eta)$ of the unitarity triangle
in fig. \ref{fig:utriangle}   by using 
\begin{equation}\label{2.84a} 
\varrho=\frac{s_{13}}{s_{12}s_{23}}\cos\delta,
\qquad
\eta=\frac{s_{13}}{s_{12}s_{23}}\sin\delta
\end{equation}
and
\begin{equation}\label{2.88da}
\bar\varrho=\varrho (1-\frac{\lambda^2}{2}),
\qquad
\bar\eta=\eta (1-\frac{\lambda^2}{2}).
\end{equation}

For a given set ($s_{12}$, $s_{13}$, $s_{23}$,
$\mt$, $\hat B_K$) there are two solutions for $\delta$ and consequently two
solutions for $(\bar\varrho,\bar\eta)$. 
This will be evident from the analysis of the unitarity triangle discussed
in detail below.

Finally we have to say a few words about the non-perturbative
parameter $\hat B_K$, the main uncertainty in this analysis. 
Reviews are given in \cite{GUPTA98,EP99}. Here we only collect
in table~\ref{TAB9} values for $\hat B_K$ obtained
in various non-perturbative approaches.
In our numerical analysis presented 
below we will use 
\begin{equation}\label{BKT}
\hat B_K=0.80\pm 0.15 \,
\end{equation}
which is in the ball park of various lattice and large-N estimates.

\begin{table}[thb]
\caption[]{$\hat B_K$ obtained using various methods. WA stands
for recent world avarage.  
 \label{TAB9}}
\vspace{0.4cm}
\begin{center}
\begin{tabular}{|c||c||c|}\hline
{\bf Method} & {\bf $\hat B_K$} & {\bf Reference} \\ \hline
Chiral QM &$1.1\pm0.2$ & \cite{BERT97} \\ \hline
Lattice (APE) &$0.93\pm 0.16$ & \cite{APE}  \\ \hline
Lattice (JLQCD) &$0.86\pm0.06$ & \cite{JLQCD}  \\ \hline
Lattice (GKS) &$0.85\pm 0.05$ & \cite{GKS} \\ \hline
Lattice (WA) &$0.89\pm0.13$ & \cite{LL} \\ \hline
Large-N &$0.70\pm0.10$ & \cite{BBG0,Bijnens} \\ \hline
Large-N &$0.4-0.7 $ & \cite{DORT99} \\ \hline
QCDS &$0.5-0.6$ & \cite{BKQCD} \\ \hline
CHPTH &$0.42\pm0.06 $ & \cite{Bruno} \\ \hline
QCD HD &$0.39\pm0.10 $ & \cite{Prades} \\ \hline
SU(3)+PCAC &$0.33 $ & \cite{Donoghue} \\ \hline
\end{tabular}
\end{center}
\end{table}

\subsection{Basic Formula for $B^0$-$\bar B^0$ Mixing}
            \label{subsec:BBformula}
The strength of the $B^0_{d,s}-\bar B^0_{d,s}$ mixings
is described by the mass differences
\begin{equation}
\Delta M_{d,s}= M_H^{d,s}-M_L^{d,s}
\end{equation}
with ``H'' and ``L'' denoting {\it Heavy} and {\it Light} respectively. 
In contrast to $\Delta M_K$, in this case the long distance contributions
are estimated to be very small and $\Delta M_{d,s}$ is very well
approximated by the relevant box diagrams. 
Moreover, due $m_{u,c}\ll m_t$ 
only the top sector can contribute significantly to 
$B_{d,s}^0-\bar B_{d,s}^0$ mixings.
The charm sector and the mixed top-charm contributions are
entirely negligible.

 $\Delta M_{d,s}$ can be expressed
in terms of the off-diagonal element in the neutral B-meson mass matrix
by using the formulae developed previously for the K-meson system.
One finds
\begin{equation}
\Delta M_q= 2 |M_{12}^{(q)}|, \qquad q=d,s.
\label{eq:xdsdef}
\end{equation}
This formula differs from $\Delta M_K=2 \RE M_{12}$ because in the
B-system $\Gamma_{12}\ll M_{12}$.

The off-diagonal
term $M_{12}$ in the neutral $B$-meson mass matrix is then given by
a formula analogous to (\ref{eq:M12Kdef})
\begin{equation}
2 m_{B_q} |M_{12}^{(q)}| = 
|\langle \bar B^0_q| \Heff(\Delta B=2) |B^0_q\rangle|,
\label{eq:M12Bdef}
\end{equation}
where 
in the case of $B_d^0-\bar B_d^0$
mixing 
\begin{eqnarray}\label{hdb2}
{\cal H}^{\Delta B=2}_{\rm eff}&=&\frac{G^2_{\rm F}}{16\pi^2}M^2_W
 \left(V^\ast_{tb}V_{td}\right)^2 \eta_{B}
 S_0(x_t)\times
\nonumber\\
& &\times \left[\alpha^{(5)}_s(\mu_b)\right]^{-6/23}\left[
  1 + \frac{\alpha^{(5)}_s(\mu_b)}{4\pi} J_5\right]  Q(\Delta B=2) + h. c.
\end{eqnarray}
Here $\mu_b=\ord(m_b)$,
\begin{equation}\label{qbdbd}
Q(\Delta B=2)=(\bar bd)_{V-A}(\bar bd)_{V-A}
\end{equation}
and \cite{BJW90}
\begin{equation}
\eta_B=0.55\pm0.01.
\end{equation}
In the case of  $B_s^0-\bar B_s^0$ mixing one should simply replace
$d\to s$ in (\ref{hdb2}) and (\ref{qbdbd}) with all other quantities
unchanged.

Defining the renormalization group invariant parameters $\hat B_q$
in analogy to (\ref{eq:BKrenorm}) and (\ref{eq:KbarK})
one finds
 using (\ref{hdb2}) 
\begin{equation}
\Delta M_q = \frac{G_{\rm F}^2}{6 \pi^2} \eta_B m_{B_q} 
(\hat B_{B_q} F_{B_q}^2 ) \mw^2 S_0(x_t) |V_{tq}|^2,
\label{eq:xds}
\end{equation}
where $F_{B_q}$ is the $B_q$-meson decay constant.
This implies two useful formulae
\begin{equation}\label{DMD}
\Delta M_d=
0.50/{\rm ps}\cdot\left[ 
\frac{\sqrt{\hat B_{B_d}}F_{B_d}}{200\mev}\right]^2
\left[\frac{\mtb(\mt)}{170\gev}\right]^{1.52} 
\left[\frac{\vtd}{8.8\cdot10^{-3}} \right]^2 
\left[\frac{\eta_B}{0.55}\right]  
\end{equation}
and
\begin{equation}\label{DMS}
\Delta M_{s}=
15.1/{\rm ps}\cdot\left[ 
\frac{\sqrt{\hat B_{B_s}}F_{B_s}}{240\mev}\right]^2
\left[\frac{\mtb(\mt)}{170\gev}\right]^{1.52} 
\left[\frac{\vts}{0.040} \right]^2
\left[\frac{\eta_B}{0.55}\right] \,.
\end{equation}

There is a vast literature on the calculations of $F_{B_d}$ and
$\hat B_d$. The most recent lattice results are summarized in 
\cite{BF}.
They are compatible with the results obtained 
with the help of QCD sum rules   \cite{QCDSF}.
In our numerical analysis we will use the value for
$F_{B_d}\sqrt{B_{B_d}}$ given in table~\ref{tab:inputparams}.
The experimental situation on
$\Delta M_{d,s}$  is also given there.  
\section{Standard Analysis of the Unitarity Triangle}\label{UT-Det}
\subsection{Basic Procedure}
With all these formulae at hand we can now summarize the standard
analysis of the unitarity triangle in fig. \ref{fig:utriangle}. 
It proceeds in five steps.

{\bf Step 1:}

{}From  $b\to c$ transition in inclusive and exclusive 
leading $B$ meson decays
one finds $\vcb$ and consequently the scale of the unitarity triangle:
\begin{equation}
\vcb\quad \Longrightarrow\quad\lambda \vcb= \lambda^3 A
\end{equation}

{\bf Step 2:}

{}From  $b\to u$ transition in inclusive and exclusive $B$ meson decays
one finds $\vub$ and consequently the side $CA=R_b$ of the unitarity
triangle:
\begin{equation}\label{rb}
\left| \frac{V_{ub}}{V_{cb}} \right|
 \quad\Longrightarrow \quad R_b=\sqrt{\bar\varrho^2+\bar\eta^2}=
4.44 \cdot \left| \frac{V_{ub}}{V_{cb}} \right|
\end{equation}

{\bf Step 3:}

\begin{figure}[hbt]
\vspace{0.010in}
\centerline{
\epsfysize=4in
\rotate[r]{
\epsffile{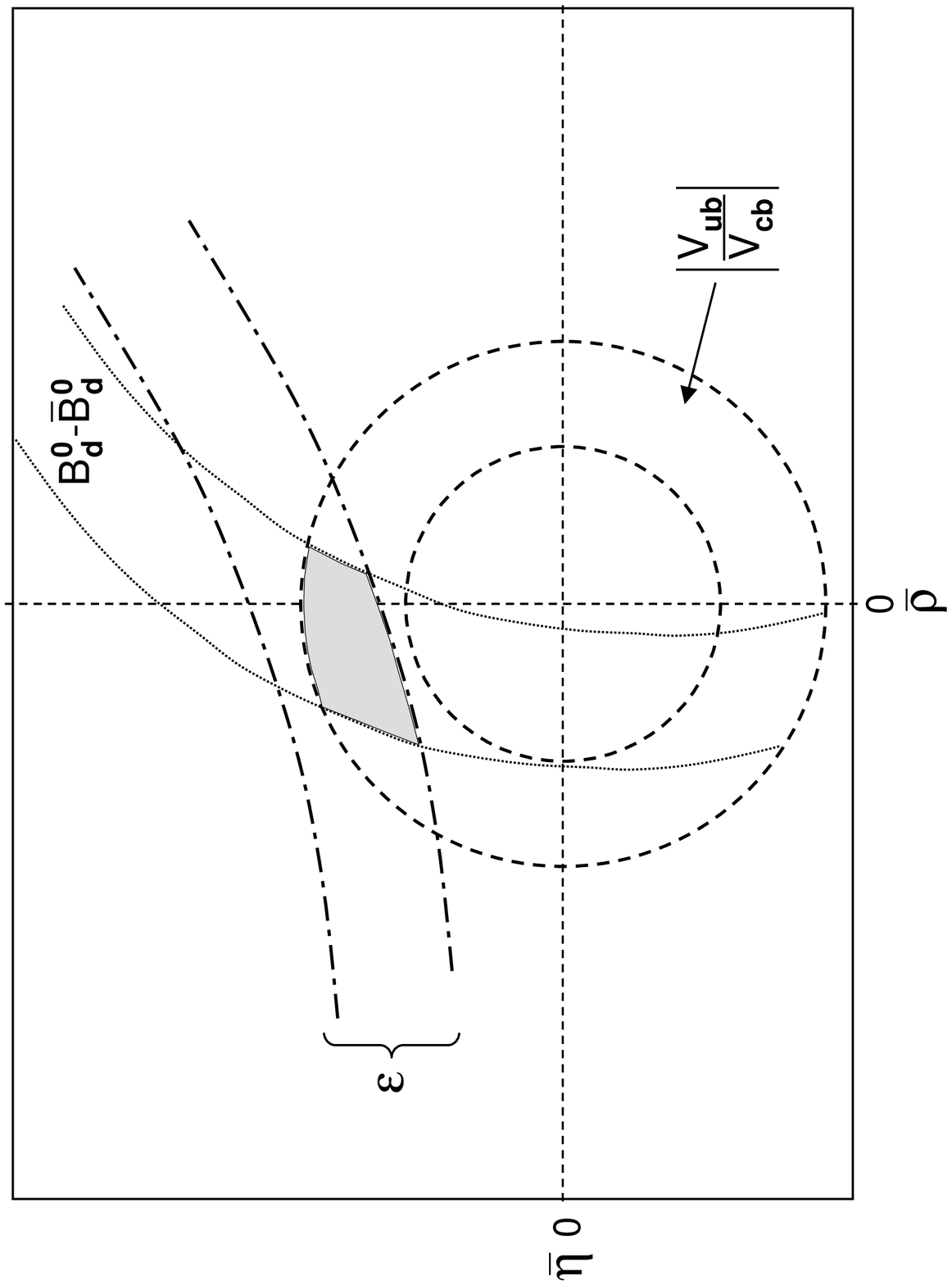}
}}
\vspace{0.0108in}
\caption[]{Schematic determination of the Unitarity Triangle.
\label{L:10}}
\end{figure}

{}From the experimental value of $\varepsilon$ (\ref{eexp}) 
and the formula (\ref{eq:epsformula}) one 
derives, using the approximations (\ref{2.51})--(\ref{2.53}), 
the constraint
\begin{equation}\label{100}
\bar\eta \left[(1-\bar\varrho) A^2 \eta_2 S_0(x_t)
+ P_0(\varepsilon) \right] A^2 \hat B_K = 0.224,
\end{equation}
where
\begin{equation}\label{102}
P_0(\varepsilon) = 
\left[ \eta_3 S_0(x_c,x_t) - \eta_1 x_c \right] \frac{1}{\lambda^4},
\qquad
x_t=\frac{\mt^2}{\mw^2}.
\end{equation}
 $P_0(\varepsilon)=0.31\pm0.05$ summarizes the contributions
of box diagrams with two charm quark exchanges and the mixed 
charm-top exchanges. 
The main uncertainties in the constraint (\ref{100}) reside in
$\hat B_K$ and to some extent in $A^4$ which multiplies the leading term.
Equation (\ref{100}) specifies 
a hyperbola in the $(\bar \varrho, \bar\eta)$
plane.
This hyperbola intersects the circle found in step 2
in two points which correspond to the two solutions for
$\delta$ mentioned earlier. This is illustrated in fig. \ref{L:10}.
The position of the hyperbola (\ref{100}) in the
$(\bar\varrho,\bar\eta)$ plane depends on $\mt$, $|V_{cb}|=A \lambda^2$
and $\hat B_K$. With decreasing $\mt$, $|V_{cb}|$ and $\hat B_K$ the
$\eps$-hyperbola moves away from the origin of the
$(\bar\varrho,\bar\eta)$ plane. 

{\bf Step 4:}
{}From the observed $B^0_d-\bar B^0_d$ mixing parametrized by $\Delta M_d$ 
the side $BA=R_t$ of the unitarity triangle can be determined:
\begin{equation}\label{106}
 R_t= \frac{1}{\lambda}\frac{|V_{td}|}{\vcb} = 1.0 \cdot
\left[\frac{|V_{td}|}{8.8\cdot 10^{-3}} \right] 
\left[ \frac{0.040}{\vcb} \right]
\end{equation}
with
\begin{equation}\label{VT}
\vtd=
8.8\cdot 10^{-3}\left[ 
\frac{200\mev}{\sqrt{B_{B_d}}F_{B_d}}\right]
\left[\frac{170~GeV}{\mtb(\mt)} \right]^{0.76} 
\left[\frac{\Delta M_d}{0.50/{\rm ps}} \right ]^{0.5} 
\sqrt{\frac{0.55}{\eta_B}}.
\end{equation}

\begin{table}[thb]
\caption[]{Collection of input parameters.
\label{tab:inputparams}}
\vspace{0.4cm}
\begin{center}
\begin{tabular}{|c|c|c|c|}
\hline
{\bf Quantity} & {\bf Central} & {\bf Error} & {\bf Reference} \\
\hline
$|V_{cb}|$ & 0.040 & $\pm 0.002$ & \cite{PDG}     \\
$|V_{ub}|$ & $3.56\cdot 10^{-3}$ & $\pm 0.56\cdot 10^{-3} $ &
\cite{STOCCHI}  \\
$\hat B_K$ & 0.80 & $\pm 0.15$ & See Text  \\
$\sqrt{B_d} F_{B_{d}}$ & $200\mev$ & $\pm 40\mev$ & \cite{BF} \\
$\mt$ & $165\gev$ & $\pm 5\gev$ & \cite{CDFD0} \\
$\Delta M_d$ & $0.471~\mbox{ps}^{-1}$ & $\pm 0.016~\mbox{ps}^{-1}$ 
& \cite{LEPB}\\
$\Delta M_s$ & $>12.4~\mbox{ps}^{-1}$ & $ 95\% {\rm C.L.}$ 
& \cite{LEPB}\\
$\xi$ & $1.14$ & $\pm 0.08$ 
& \cite{BF,NAR} \\
\hline
\end{tabular}
\end{center}
\end{table}

Since $\mt$, $\Delta M_d$ and $\eta_B$ are already rather precisely
known, the main uncertainty in the determination of $\vtd$ from
$B_d^0-\bar B_d^0$ mixing comes from $F_{B_d}\sqrt{B_{B_d}}$.
Note that $R_t$ suffers from additional uncertainty in $\vcb$,
which is absent in the determination of $\vtd$ this way. 
The constraint in the $(\bar\varrho,\bar\eta)$ plane coming from
this step is illustrated in fig.~\ref{L:10}.

{\bf Step 5:}

{}The measurement of $B^0_s-\bar B^0_s$ mixing parametrized by $\Delta M_s$
together with $\Delta M_d$  allows to determine $R_t$ in a different
way. Using (\ref{eq:xds}) one finds 
\begin{equation}\label{107b}
\frac{\vtd}{|V_{ts}|}= 
\xi\sqrt{\frac{m_{B_s}}{m_{B_d}}}
\sqrt{\frac{\Delta M_d}{\Delta M_s}},
\qquad
\xi = 
\frac{F_{B_s} \sqrt{B_{B_s}}}{F_{B_d} \sqrt{B_{B_d}}}.
\end{equation}
Using next $\Delta M^{{\rm max}}_d= 0.487/
\mbox{ps}$ and 
$|V_{ts}/V_{cb}|^{{\rm max}}=0.991$  one finds a useful 
approximate formula
\begin{equation}\label{107bu}
(R_t)_{\rm max} = 1.0 \cdot \xi 
\sqrt{\frac{10.2/ps}{(\Delta M_s)_{\rm min}}},
\end{equation}
One should 
note that $\mt$ and $|V_{cb}|$ dependences have been eliminated this way
 and that $\xi$ should in principle 
contain much smaller theoretical
uncertainties than the hadronic matrix elements in $\Delta M_d$ and 
$\Delta M_s$ separately.  
The most recent values relevant for (\ref{107bu}) are summarized
in table~\ref{tab:inputparams}.

\subsection{Numerical Results}\label{sec:standard}
\subsubsection{Input Parameters}

The input parameters needed to perform the
standard analysis of the unitarity triangle
are given in table \ref{tab:inputparams}, where
 $\mt$ 
 refers
to the running current top quark mass defined at $\mu=\mt^{Pole}$.
It corresponds to 
$\mt^{Pole}=174.3\pm 5.1\gev$ measured by CDF and D0 \cite{CDFD0}.

\begin{figure}[thb]
\vspace{0.10in}
\centerline{
\epsfysize=3.4in
\rotate[r]{\epsffile{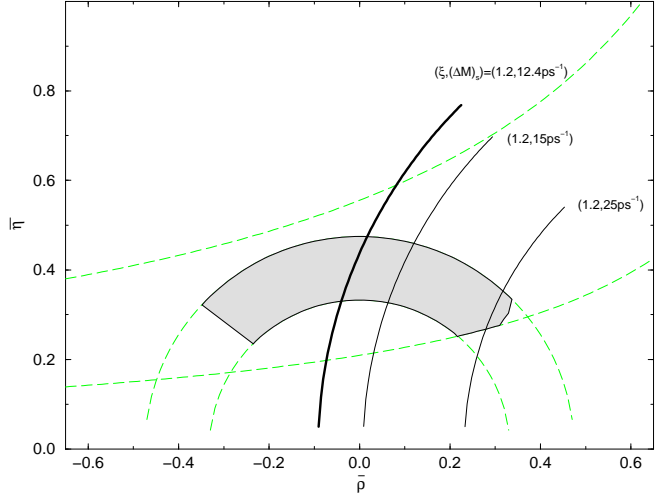}}
}
\vspace{0.08in}
\caption[]{
Unitarity Triangle 1999.
\label{fig:utdata}}
\end{figure}

\subsubsection{Output of the Standard Analysis}

In what follows we will present two types of numerical analyses
\cite{EP99,MGO}:

\begin{itemize}
\item
Method 1: The experimentally measured numbers  are used with Gaussian errors
 and for the theoretical input parameters we take a flat distribution
in the ranges given in
table~\ref{tab:inputparams}.
\item
Method 2: Both the experimentally measured numbers and the theoretical input
parameters are scanned independently within the ranges given in
table~\ref{tab:inputparams}.
\end{itemize}

The results are shown in table \ref{TAB2}.
The allowed region for $(\bar\varrho,\bar\eta)$ is presented in fig. 
\ref{fig:utdata}.
It is the shaded area on the right hand side of the solid circle
which represents 
the upper bound for $(\Delta M)_d/(\Delta M)_s$. The hyperbolas
give the constraint from $\varepsilon$ and the two circles centered
at $(0,0)$ the constraint from $\vub$. The white areas between the
lower $\varepsilon$-hyperbola and the shaded region are excluded
by $B^0_{d}-\bar B^0_{d}$ mixing. We observe that the region
$\bar\varrho<0$ is practically excluded. 
 The results in fig. \ref{fig:utdata} 
correspond to a simple independent 
scanning of all parameters within one standard deviation.
We find that whereas the angle $\beta$ is rather
constrained, the uncertainties in 
$\alpha$ and $\gamma$ are  substantially larger: 
\be\label{apr}
66^\circ\le \alpha \le 113^\circ~,
\quad
17^\circ\le \beta \le 29^\circ~,
\quad
44^\circ\le \gamma \le 97^\circ~.
\ee
The result for $\sin2\beta$ is consistent with the recent
measurement of CP asymmetry in $B\to\psi K_S$ by CDF \cite{CDF99},
although the large experimental error precludes any definite 
conclusion.

Other studies of the unitarity triangle can be found in
\cite{UT99,STOCCHI,FRENCH,PAGA}.

\begin{table}[thb]
\caption[]{Output of the Standard Analysis. 
 $\lambda_t=V^*_{ts} V_{td}$.\label{TAB2}}
\vspace{0.4cm}
\begin{center}
\begin{tabular}{|c||c||c|}\hline
{\bf Quantity} & {\bf Scanning} & {\bf Gaussian} \\ \hline
$\mid V_{td}\mid/10^{-3}$ &$7.0 - 9.8$ &$ 8.1\pm 0.6$ \\ \hline
$\mid V_{ts}/V_{cb}\mid$ &$0.975 - 0.991$ &$0.984\pm 0.004$  \\ \hline
$\mid V_{td}/V_{ts}\mid$ &$0.17 - 0.24$ &$0.201\pm 0.017$  \\ \hline
$\sin(2\beta)$ &$0.57 - 0.84$ &$ 0.73\pm0.09 $ \\ \hline
$\sin(2\alpha)$ &$-0.72 - 1.0$ &$- 0.03\pm 0.36 $ \\ \hline
$\sin(\gamma)$ &$0.69 - 1.0 $ &$ 0.89\pm0.08 $ \\ \hline
$\IM \lambda_t/10^{-4}$ &$1.04 - 1.63 $ &$ 1.33\pm 0.14 $ \\ \hline
\end{tabular}
\end{center}
\end{table}

\subsection{Final Remarks}
In this section we have completed the determination of the CKM matrix.
It is given by the values of $|V_{us}|$, $\vcb$ and $|V_{ub}|$ in
(\ref{vcb}) and (\ref{v13}), the results in table~\ref{TAB2} and
the unitarity triangle shown in fig.~\ref{fig:utdata}. Clearly
the accuracy of this determination is not yet impressive. We should
stress, however, that in a few years from now the standard analysis
may give much more accurate results.
In particular a single precise measurement of $\Delta M_s$ 
will have a very important impact on the allowed area in the 
$(\bar\varrho,\bar\eta)$ plane. Such a measurement should come from
SLD and later from LHC.

Having the values of CKM parameters at hand, we can use them to predict
various branching ratios for rare and CP-violating decays.
This we will do in the subsequent sections. 

%\clearpage
\section{$\epe$ in the Standard Model}\label{EpsilonPrime}
\setcounter{equation}{0}
\subsection{Preliminaries}
Direct CP violation remains one of the important targets 
of contemporary particle physics. 
In the case of $K\to\pi\pi$,
a non-vanishing value of the ratio Re($\epe$) defined in (\ref{eprime}) 
would give the first
signal for direct CP violation ruling out superweak models
\cite{wolfenstein:64}.
Until recently the experimental situation on $\epe$ was rather unclear:
\begin{equation}\label{eprime2}
\RE(\varepsilon'/\varepsilon) =\left\{ \begin{array}{ll}
(23 \pm 7)\cdot 10^{-4} & {\rm (NA31)}~\cite{barr:93} \\
(7.4 \pm 5.9)\cdot 10^{-4} &{\rm (E731)}~ \cite{gibbons:93}.
\end{array} \right.
\end{equation}
While the result of the NA31 collaboration at CERN 
\cite{barr:93}
clearly indicated direct CP violation, the value of E731 at Fermilab
\cite{gibbons:93}, was compatible with superweak theories
\cite{wolfenstein:64} in which $\varepsilon'/\varepsilon = 0$.
This controversy is now settled with the very recent measurement
by KTeV at Fermilab \cite{KTeV}
\begin{equation}\label{eprime1}
\RE(\epe) =
(28.0 \pm 4.1)\cdot 10^{-4}~~{\rm (KTeV)} 
\end{equation}
which together with the NA31 result confidently establishes direct 
CP violation in nature.
The grand average including NA31, E731 and KTeV results reads
\cite{KTeV} 
\be
\RE(\epe) = (21.8\pm 3.0)\cdot 10^{-4}
\label{ga}
\ee
very close to the NA31 result but with a smaller error. The error 
should be further reduced once the first data from 
NA48 collaboration at
CERN are available and complete data from both collaborations have 
been analyzed. It is also 
of great interest to see what value for $\epe$ will be measured by KLOE at 
Frascati, which uses a different experimental technique than KTeV and NA48.

There is a long history of calculations of $\epe$ in the Standard
Model.
The first calculation of $\epe$ for $\mt \ll \mw$ without the inclusion
of renormalization group effects can be found in \cite{EGN}. 
Renormalization group effects
in the leading
logarithmic approximation have been first presented in \cite{GW79}. 
For $\mt \ll \mw$ only QCD
penguins play a substantial role. 
First extensive phenomenological analyses in this approximation
can be found in
\cite{BSS}.
Over the eighties these calculations
were refined through the inclusion of QED penguin effects 
for $\mt \ll \mw$ \cite{BW84,donoghueetal:86,burasgerard:87},
the inclusion of isospin breaking in the
quark masses \cite{donoghueetal:86,burasgerard:87,lusignoli:89},
and through improved estimates of hadronic matrix elements in
the framework of the $1/N$ approach \cite{bardeen:87}. 
This era of $\epe$ culminated
in the analyses in \cite{flynn:89,buchallaetal:90}, where QCD
penguins, electroweak penguins ($\gamma$ and $Z^0$ penguins)
and the relevant box diagrams were included for arbitrary
top quark masses. The strong cancellation between QCD penguins
and electroweak penguins for $m_t > 150~\gev$ found in these
papers was confirmed by other authors \cite{PW91}.

During the nineties considerable progress has been made by
calculating complete NLO corrections to $\varepsilon'$
\cite{BJLW1}--\cite{ROMA2}. Together with the NLO
corrections to $\varepsilon$ and $B^0-\bar B^0$ mixing
\cite{HNa,BJW90,HNb}, this allowed
a complete NLO analysis of $\varepsilon'/\varepsilon$ including
constraints from the observed indirect CP violation ($\varepsilon$)
and  $B_{d,s}^0-\bar B_{d,s}^0$ mixings ($\Delta M_{d,s}$). The improved
determination of the $V_{ub}$ and $V_{cb}$ elements of the CKM matrix,
the improved estimates of hadronic matrix elements using the lattice 
approach as well as other non-perturbative approaches 
and in particular the determination of the top quark mass
$\mt$ had of course also an important impact on
$\varepsilon'/\varepsilon$. 

Now, $\epe$ is given by (\ref{eq:epe1}) where
in a crude approximation (not to be used for any serious analysis)
\be\label{ap}
F_{\varepsilon'}\approx 13\cdot 
\left[\frac{110\mev}{\ms(2~\gev)}\right]^2
\left[\bsi(1-\OEE)-0.4\cdot \bei\left(\frac{\mt}{165\gev}\right)^{2.5}\right]
\left(\frac{\Lms^{(4)}}{340~\mev}\right)~.
\ee
Here $\OEE\approx 0.25$ represents isospin breaking corrections and $B_i$ 
are hadronic parameters which we will 
define later on. 
This formula exhibits very clearly the dominant uncertainties in 
$F_{\varepsilon'}$ which 
reside in the values of $\ms$, $\bsi$, $\bei$, $\Lms^{(4)}$ and $\OEE$.
Moreover, the partial 
cancellation between QCD penguin ($\bsi$) and electroweak 
penguin ($\bei$) contributions requires 
accurate 
values of $\bsi$ and $\bei$ for an acceptable estimate of $\epe$. 
Because of the accurate value $\mt(\mt)=165\pm 5~\gev$, the uncertainty 
in $\epe$ due to the top quark mass amounts only to a few percent. 
A more accurate formula 
for $F_{\varepsilon'}$ will be given below.

Now, it has been known for some time that for central values of the 
input parameters the size of $\epe$
 in the Standard Model is well below the NA31 value of 
$(23.0\pm6.5)\cdot 10^{-4}$. Indeed, 
extensive NLO analyses with lattice and large--N estimates of 
$\bsi\approx 1$ and 
$\bei\approx 1$ performed 
first in \cite{BJLW,ROMA1} and after the top discovery in
\cite{ciuchini:95,BJL96a,ciuchini:96} have found $\epe$ in 
the ball park of 
$(3-7)\cdot 10^{-4}$ for $\ms(2~\gev)\approx 130~\mev$. 
 On the other hand 
it has been stressed repeatedly in \cite{AJBLH,BJL96a}  that 
for extreme values of $\bsi$, $\bei$ and $\ms$ still consistent with 
lattice, QCD sum rules and large--N estimates as well as 
sufficiently high values of $\IM\lambda_t$ and $\Lms^{(4)}$, 
a ratio $\epe$ as high as $(2-3)\cdot 10^{-3}$ could be obtained within 
the Standard Model. Yet, it has also been admitted that such simultaneously 
extreme values of all input parameters and consequently values of $\epe$
 close to the NA31 result 
are rather improbable in the Standard Model. 
Different conclusions have been reached in \cite{paschos:96}, where
values  $(1-2)\cdot 10^{-3}$ for $\epe$ can be found.
Also the Trieste group \cite{BERT98}, which calculated the parameters 
$\bsi$ and $\bei$
 in the chiral quark model, found $\epe=(1.7\pm 1.4)\cdot 10^{-3}$.
On the other hand using an  effective chiral
lagrangian approach, the authors in \cite{BELKOV} found $\epe$
consistent with zero.

After these general remarks let us discuss 
$\epe$ in explicit terms. Other reviews of $\epe$ can be found
in \cite{WW,BERT98}.
\subsection{Basic Formulae}
           \label{subsec:epeformulae}
The parameter $\varepsilon'$ is given in terms of the isospin amplitudes
$A_I$ in (\ref{eprime}). Applying OPE to these amplitudes one finds
\begin{equation}
\frac{\varepsilon'}{\varepsilon} = 
\IM \lambda_t\cdot F_{\varepsilon'},
\label{eq:epe1}
\end{equation}
where
\begin{equation}
F_{\varepsilon'} = 
\left[ P^{(1/2)} - P^{(3/2)} \right] \exp(i\Phi),
\label{eq:epe2}
\end{equation}
with
\begin{eqnarray}
P^{(1/2)} & = & r \sum y_i \langle Q_i\rangle_0
(1-\Omega_{\eta+\eta'})~,
\label{eq:P12} \\
P^{(3/2)} & = &\frac{r}{\omega}
\sum y_i \langle Q_i\rangle_2~.~~~~~~
\label{eq:P32}
\end{eqnarray}
Here
\begin{equation}
r = \frac{G_{\rm F} \omega}{2 |\eps| \RE A_0}~, 
\qquad
\langle Q_i\rangle_I \equiv \langle (\pi\pi)_I | Q_i | K \rangle~,
\qquad
\omega = \frac{\RE A_2}{\RE A_0}.
\label{eq:repe}
\end{equation}
Since
\begin{equation}
\Phi=\Phi_{\varepsilon'}-\Phi_\varepsilon \approx 0,
\label{Phi}
\end{equation}
$F_{\varepsilon'}$ and $\epe$
are real  to an excellent approximation.
The operators $Q_i$ have been given already in (\ref{OS1})-(\ref{OS5}).
The Wilson coefficient functions $ y_i(\mu)$
were calculated including
the complete next-to-leading order (NLO) corrections in
\cite{BJLW1}--\cite{ROMA2}. The details
of these calculations can be found there and in the review
\cite{BBL}. 
Their numerical values for $\Lms^{(4)}$ corresponding to
$\alpha_{\overline{MS}}^{(5)}(\mz)=0.119\pm 0.003$
and two renormalization schemes (NDR and HV)
are given in table 
\ref{tab:wc10smu13} \cite{EP99}.

\begin{table}[htb]
\caption[]{$\Delta S=1 $ Wilson coefficients at $\mu=\mc=1.3\gev$ for
$\mt=165\gev$ and $f=3$ effective flavours.
$y_1 = y_2 \equiv 0$.
\label{tab:wc10smu13}}
\begin{center}
\begin{tabular}{|c|c|c||c|c||c|c|}
\hline
& \multicolumn{2}{c||}{$\Lms^{(4)}=290\mev$} &
  \multicolumn{2}{c||}{$\Lms^{(4)}=340\mev$} &
  \multicolumn{2}{c| }{$\Lms^{(4)}=390\mev$} \\
\hline
Scheme & NDR & HV & NDR & HV & NDR & HV \\
\hline
$y_3$ & 0.027 & 0.030 & 0.030 & 0.034 & 0.033 & 0.038 \\
$y_4$ & --0.054 & --0.056 & --0.059 & --0.061 & --0.064 & --0.067 \\
$y_5$ & 0.006 & 0.015 & 0.005 & 0.016 & 0.003 & 0.017 \\
$y_6$ & --0.082 & --0.074 & --0.092 & --0.083 & --0.105 & --0.093 \\
\hline
$y_7/\aem$ & --0.038 & --0.037 & --0.037 & --0.036 & --0.037 & --0.034 \\
$y_8/\aem$ & 0.118 & 0.127 & 0.134 & 0.143 & 0.152 & 0.161 \\
$y_9/\aem$ & --1.410 & --1.410 & --1.437 & --1.437 & --1.466 & --1.466 \\
$y_{10}/\aem$ & 0.496 & 0.502 & 0.539 & 0.546 & 0.585 & 0.593 \\
\hline
\end{tabular}
\end{center}
\end{table}

It is customary in phenomenological
applications to take $\RE A_0$ and $\omega$ from
experiment, i.e.
\begin{equation}
\RE A_0 = 3.33 \cdot 10^{-7}\gev,
\qquad
\omega = 0.045,
\label{eq:ReA0data}
\end{equation}
where the last relation reflects the so-called $\Delta I=1/2$ rule.
This strategy avoids to a large extent the hadronic uncertainties 
in the real parts of the isospin amplitudes $A_I$.

The sum in (\ref{eq:P12}) and (\ref{eq:P32}) runs over all contributing
operators. $P^{(3/2)}$ is fully dominated by electroweak penguin
contributions. $P^{(1/2)}$ on the other hand is governed by QCD penguin
contributions which are suppressed by isospin breaking in the quark
masses ($m_u \not= m_d$). The latter effect is described by

\begin{equation}
\Omega_{\eta+\eta'} = \frac{1}{\omega} \frac{(\IM A_2)_{\rm
I.B.}}{\IM A_0}\,.
\label{eq:Omegaeta}
\end{equation}
For $\Omega_{\eta+\eta'}$ we will first set
\begin{equation}
\Omega_{\eta+\eta'} = 0.25\,,
\label{eq:Omegaetadata}
\end{equation}
which is in the ball park of the values obtained in the $1/N$ approach
\cite{burasgerard:87} and in chiral perturbation theory
\cite{donoghueetal:86,lusignoli:89}. $\Omega_{\eta+\eta'}$ is
independent of $\mt$. We will investigate the sensitivity of $\epe$
to $\OEE$ later on.

\subsection{Hadronic Matrix Elements}
The main source of uncertainty in the calculation of
$\epe$ are the hadronic matrix elements $\langle Q_i \rangle_I$.
They generally depend
on the renormalization scale $\mu$ and on the scheme used to
renormalize the operators $Q_i$. These two dependences are canceled by
those present in the Wilson coefficients $y_i(\mu)$ so that the
resulting physical $\epe$ does not (in principle) depend on $\mu$ and on the
renormalization scheme of the operators.  Unfortunately, the accuracy of
the present non-perturbative methods used to evalutate $\langle Q_i
\rangle_I$  is not
sufficient to have the $\mu$ and scheme dependences of
$\langle Q_i \rangle_I$ fully under control. 
We believe that this situation will change once the lattice calculations
and QCD sum rule calculations improve.
A brief review of the existing methods 
including most recent developments will be given below.

In view of this situation it has been suggested in \cite{BJLW} to
determine as many matrix elements $\langle Q_i \rangle_I$ as possible
from the leading CP conserving $K \to \pi\pi$ decays, for which the
experimental data is summarized in (\ref{eq:ReA0data}). 
To this end it turned out to be very convenient to determine $\langle
Q_i \rangle_I$ in the three-flavour effective theory at a scale $\mu
\approx m_c$. 
The details of this approach will not be discussed here.
It sufficies to say that
this method allows to determine only the matrix
elements of the $(V-A)\otimes(V-A)$ operators. 
For the central value of $\IM\lambda_t$
these operators give a negative contribution to $\epe$ 
of about $-2.5\cdot 10^{-4}$. This shows that these
operators are only relevant if  $\epe$ is below $1 \cdot 10^{-3}$.
Unfortunately the matrix elements of the dominant $(V-A)\otimes(V+A)$
operators cannot be  determined by the CP conserving data and
one has to use  non-perturbative methods to estimate them.

Concerning the $(V-A)\otimes(V+A)$ operators $Q_5-Q_8$, it 
is customary to express their matrix elements
$\langle Q_i \rangle_I$ in terms of non-perturbative parameters
$B_i^{(1/2)}$ and $B_i^{(3/2)}$ as follows:
\begin{equation}
\langle Q_i \rangle_0 \equiv B_i^{(1/2)} \, \langle Q_i
\rangle_0^{\rm (vac)}\,,
\qquad
\langle Q_i\rangle_2 \equiv B_i^{(3/2)} \, \langle Q_i
\rangle_2^{\rm (vac)} \,.
\label{eq:1}
\end{equation}
The label ``vac'' stands for the vacuum
insertion estimate of the hadronic matrix elements in question 
for
which $B_i^{(1/2)}=B_i^{(3/2)}=1$.

As the numerical analysis in \cite{BJLW} shows $\epe$ is only
weakly sensitive to the values of the parameters
$B_3^{(1/2)}$, $B_5^{(1/2)}$, $B_7^{(1/2)}$, $B_8^{(1/2)}$
and $B_7^{(3/2)}$ as long as their absolute values are not
substantially larger than 1.
As in \cite{BJLW} our strategy
is to set
\begin{equation}
B_{3,7,8}^{(1/2)}(\mc) = 1,
\qquad
B_5^{(1/2)}(\mc) = B_6^{(1/2)}(\mc),
\qquad
B_7^{(3/2)}(\mc) = B_8^{(3/2)}(\mc)
\label{eq:B1278mc}
\end{equation}
and to treat $B_6^{(1/2)}(\mc)$ and $B_8^{(3/2)}(\mc)$ as free
parameters.

The approach in \cite{BJLW} allows then in a  good approximation
to express $\epe$ or equivalently $F_{\varepsilon'}$ in terms of
$\Lms^{(4)}$, $\mt$, $\ms$ and the two non-perturbative parameters 
$\bsi\equiv B_6^{(1/2)}(\mc)$ and $\bei\equiv B_8^{(3/2)}(\mc)$ 
which cannot be fixed by the CP conserving data.

\subsection{An Analytic Formula for $\epe$}
           \label{subsec:epeanalytic}
As shown in \cite{buraslauten:93}, it is possible to cast the formal
expressions for $\epe$ in (\ref{eq:epe1})--(\ref{eq:P32})
into an analytic formula which exhibits the $\mt$ dependence
together with the dependence on $\ms$, $\Lms^{(4)}$,
$B_6^{(1/2)}$ and $B_8^{(3/2)}$. 
To this end the approach for hadronic matrix elements presented
above is used and $\OEE$ is set to $0.25$.
The analytic formula given below, while being rather accurate, 
exhibits
various features which are not transparent in a pure numerical
analysis. It can be used in phenomenological applications if
one is satisfied with a few percent accuracy. Needless to say, 
in the numerical analysis \cite{EP99} presented below
 we have used exact expressions.

In this formulation
the function $F_{\varepsilon'}$
is given simply as follows ($x_t=\mt^2/\mw^2$):
\begin{equation}
F_{\varepsilon'} =
P_0 + P_X \, X_0(x_t) + P_Y \, Y_0(x_t) + P_Z \, Z_0(x_t) 
+ P_E \, E_0(x_t). 
\label{eq:3b}
\end{equation}
with the $\mt$-dependent functions given in subsection 2.4. 

The coefficients $P_i$ are given in terms of $B_6^{(1/2)} \equiv
B_6^{(1/2)}(\mc)$, $B_8^{(3/2)} \equiv B_8^{(3/2)}(\mc)$ and $\ms(\mc)$
as follows:
\begin{equation}
P_i = r_i^{(0)} + 
r_i^{(6)} R_6 + r_i^{(8)} R_8 \, .
\label{eq:pbePi}
\end{equation}
where
\be\label{RS}
R_6\equiv \bsi\left[ \frac{137\mev}{\ms(\mc)+\md(\mc)} \right]^2,
\qquad
R_8\equiv \bei\left[ \frac{137\mev}{\ms(\mc)+\md(\mc)} \right]^2.
\ee
The $P_i$ are renormalization scale and scheme independent. They depend,
however, on $\Lms^{(4)}$. In table~\ref{tab:pbendr} we give the numerical
values of $r_i^{(0)}$, $r_i^{(6)}$ and $r_i^{(8)}$ for different values
of $\Lms^{(4)}$ at $\mu=\mc$ in the NDR renormalization scheme
\cite{EP99}. Actually at NLO only $r_0$ coefficients are renormalization
scheme dependent. The last row gives them in the HV scheme.
The inspection of table~\ref{tab:pbendr} shows
that the terms involving $r_0^{(6)}$ and $r_Z^{(8)}$ dominate the ratio
$\epe$. Moreover, the function $Z_0(x_t)$ representing a gauge invariant
combination of $Z^0$- and $\gamma$-penguins grows rapidly with $\mt$
and due to $r_Z^{(8)} < 0$ these contributions suppress $\epe$ strongly
for large $\mt$ \cite{flynn:89,buchallaetal:90}.

\begin{table}[thb]
\caption[]{Coefficients in the formula (\ref{eq:pbePi})
 for various $\Lms^{(4)}$ in 
the NDR scheme.
The last row gives the $r_0$ coefficients in the HV scheme.
\label{tab:pbendr}}
\begin{center}
\begin{tabular}{|c||c|c|c||c|c|c||c|c|c|}
\hline
& \multicolumn{3}{c||}{$\Lms^{(4)}=290\mev$} &
  \multicolumn{3}{c||}{$\Lms^{(4)}=340\mev$} &
  \multicolumn{3}{c| }{$\Lms^{(4)}=390\mev$} \\
\hline
$i$ & $r_i^{(0)}$ & $r_i^{(6)}$ & $r_i^{(8)}$ &
      $r_i^{(0)}$ & $r_i^{(6)}$ & $r_i^{(8)}$ &
      $r_i^{(0)}$ & $r_i^{(6)}$ & $r_i^{(8)}$ \\
\hline
0 &
   --2.771 &   9.779 &   1.429 &
   --2.811 &  11.127 &   1.267 &
   --2.849 &  12.691 &   1.081 \\
$X_0$ &
     0.532 &   0.017 &       0 &
     0.518 &   0.021 &       0 &
     0.506 &   0.024 &       0 \\
$Y_0$ &
     0.396 &   0.072 &       0 &
     0.381 &   0.079 &       0 &
     0.367 &   0.087 &       0 \\
$Z_0$ &
     0.354 &  --0.013 &   --9.404 &
     0.409 &  --0.015 &  --10.230 &
     0.470 &  --0.017 &  --11.164 \\
$E_0$ &
     0.182 &  --1.144 &   0.411 &
     0.167 &  --1.254 &   0.461 &
     0.153 &  --1.375 &   0.517 \\
\hline
0 &
   --2.749 &   8.596 &   1.050 &
   --2.788 &   9.638 &   0.871 &
   --2.825 &  10.813 &   0.669 \\
\hline
\end{tabular}
\end{center}
\end{table}

\subsection{The Status of $\ms$, $\bsi$, $\bei$, $\OEE$ and $\Lms^{(4)}$}

The present status of these parameters has been recently
reviewed in details in \cite{EP99}. Therefore our
presentation will be very brief.
\subsubsection{$\ms$}
The present values for $\ms(2\gev)$ extracted from lattice calculations
and QCD sum rules are
\begin{equation}\label{ms}
\ms(2\gev) =\left\{ \begin{array}{ll}
(110\pm20)\;\mev & {\rm (Lattice)}~\cite{GUPTA98,kenway98} \\
(124\pm22)\;\mev & {\rm (QCDS)}~\cite{QCDS} \end{array} \right.
\end{equation}

The value for QCD sum rules is an average over the results given in
\cite{QCDS}.
QCD sum rules also allow to
derive lower bounds on the strange quark mass. It was found that generally
$\ms(2\gev)\gsim 100\,\mev$ \cite{MSBOUND}. If these bounds
hold, they would rule out the very low strange mass values found in
unquenched lattice QCD simulations given above.

Finally, one should also mention the very recent determination of the
strange mass from the hadronic $\tau$-spectral
  function \cite{PP,aleph:99}:
$\ms(2\gev)=(170^{+44}_{-55})\,\mev$.
We observe that the central value is much
larger than the corresponding results given above
although the error  is still large. 
In the future, however, improved
experimental statistics and a better understanding of perturbative QCD
corrections should make the determination of $\ms$ from the $\tau$-spectral
function competitive to the other methods.
On the other hand a very recent estimate using new $\tau$-like
$\phi$-meson sum rules gives $\ms(2\gev)=(136\pm16)\,\mev$ \cite{NAR99}.

We conclude that the error on $\ms$ is still rather large. In our 
numerical analysis of $\epe$, where $\ms$ is evaluated at the scale $\mc$,
we will set
\begin{equation}\label{msvalues}
\ms(\mc)=(130\pm25)\;\mev \,,
\end{equation}
roughly corresponding to $\ms(2~\gev)$ obtained in the lattice
approach. 
\subsubsection{$\bsi$ and $\bei$}
The values for $\bsi$ and $\bei$ obtained in various approaches
are collected in table~\ref{tab:317}. 
The lattice results have been obtained at $\mu=2\gev$.
The results in the large--N approach and the chiral quark model
correspond to scales below $1\gev$.
However,
 as a detailed numerical analysis in \cite{BJLW}
showed, $\bsi$ and $\bei$ are only weakly dependent on $\mu$.
Consequently
the  comparison
of these parameters
obtained in different approaches at different $\mu$ is meaningful.

Next, the values coming from lattice and
chiral quark model are given in the NDR renormalization 
scheme. The
corresponding values in the HV scheme can be found
using approximate relations \cite{EP99} 
\be\label{NDRHV}
(\bsi)_{\rm HV}\approx 1.2 (\bsi)_{\rm NDR},
\qquad
 (\bei)_{\rm HV}\approx 1.2 (\bei)_{\rm NDR}.
\ee
The results in the large-N approach are unfortunately
not sensitive to the renormalization scheme. 

Concerning the
lattice results for $B^{(1/2)}_{6}$,
the old results read
$B^{(1/2)}_{5,6}(2~\gev)=1.0 \pm 0.2$ \cite{kilcup:91,sharpe:91}.
More accurate estimates for $B^{(1/2)}_{6}$ have been given
in \cite{kilcup:98}: 
$B^{(1/2)}_{6}(2~\gev)=0.67 \pm 0.04\pm 0.05$
(quenched) and $B^{(1/2)}_{6}(2~\gev)=0.76 \pm 0.03\pm0.05$
($f=2$). However,
a recent work in \cite{kilcup:99} shows
that lattice calculations of $\bsi$ are very uncertain 
and one has
to conclude that there are no solid predictions for
$B^{(1/2)}_{6}$ from the lattice at present. 

\begin{table}[thb]
\caption[]{ Results for $\bsi$ and $\bei$ obtained
in various approaches. 
\label{tab:317}}
\begin{center}
\begin{tabular}{|c|c|c|}\hline
  { Method}& $\bsi$& $B^{(3/2)}_8$  \\
 \hline
Lattice\cite{GKS,G67,APE}&$-$ &$0.69-1.06$  \\
Large$-$N\cite{DORT98,DORT99}& $0.72-1.10$ &$0.42-0.64$ \\
ChQM\cite{BERT98}& $1.07-1.58$ &$0.75-0.79$  \\
\hline
\end{tabular}
\end{center}
\end{table}

We observe that most non-perturbative approaches discussed
above found $\bei$ below unity. The suppression of $\bei$
below unity is rather modest (at most $20\%$) in the lattice 
approaches and in the chiral quark model. In the $1/N$ approach
$\bei$ is rather strongly suppressed and can be as low as 0.5.

Concerning $\bsi$ the situation is worse. As we stated above
there is no solid prediction for this parameter in the lattice
approach. On the other hand while the average value of $\bsi$
in the $1/N$ approach is close to $1.0$, the chiral
quark model gives in the NDR scheme
the value for $\bsi$
as high as $1.33\pm 0.25$.
Interestingly both approaches give the ratio
$\bsi/\bei$ in the ball park of 1.7.

Guided by the results presented above and biased to some
extent by the results from the large-N approach and lattice
calculations, we will use
in our numerical analysis below $\bsi$ and $\bei$ in
the ranges:
\be\label{bbb}
\bsi=1.0\pm0.3,
\qquad
\bei=0.8\pm 0.2
\ee
keeping always $\bsi\ge \bei$.

\subsubsection{$\OEE$ and $\Lms^{(4)}$}
The dependence of $\epe$ on $\OEE$ can be studied numerically by 
using the formula (\ref{eq:P12}) or incorporated approximately
into the analytic formula (\ref{eq:3b}) by simply replacing
$\bsi$ with an effective parameter
\be\label{eff}
(\bsi)_{\rm eff}=\bsi\frac{(1-0.9~\OEE)}{0.775}
\ee
A numerical analysis shows that using $(1-\OEE)$
overestimates the role of $\OEE$. In our numerical analysis
we have incorporated the uncertainty in $\OEE$ by increasing
the error in $\bsi$ from $\pm 0.2$ to $\pm 0.3$.

The last estimates of $\OEE$ have been done more than ten years
ago \cite{donoghueetal:86}-\cite{lusignoli:89} 
and it is desirable to update these analyses which
 can be summarized by
\be
\OEE=0.25\pm 0.08~.
\ee

In table~\ref{tab:inputp} we summarize the input parameters
used in the numerical analysis of $\epe$ below.
The range for $\Lms^{(4)}$ in table~\ref{tab:inputp} 
corresponds roughly to $\alpha_s(\mz)=0.119\pm 0.003$.

\begin{table}[thb]
\caption[]{Collection of input parameters.
We impose $\bsi\ge\bei$.
\label{tab:inputp}}
\vspace{0.4cm}
\begin{center}
\begin{tabular}{|c|c|c|c|}
\hline
{\bf Quantity} & {\bf Central} & {\bf Error} & {\bf Reference} \\
\hline
$\Lms^{(4)}$ & $340 \mev$ & $\pm 50\mev$ & \cite{PDG,BETKE} \\
$\ms(\mc)$ & $130\mev$    & $\pm 25\mev$ & See Text\\
$\bsi $ & 1.0 & $\pm 0.3$ & See Text\\
$\bei $ & 0.8 & $\pm 0.2$ & See Text\\
\hline
\end{tabular}
\end{center}
\end{table}
 
\subsection{Numerical Results for $\epe$}
 
In order to make predictions for $\epe$ we need the value of
$\IM \lambda_t$. This can be obtained from the standard analysis
of the unitarity triangle as discussed in section 4.

In what follows we will present two types of  numerical analyses of
$\epe$ which use the methods 1 and 2 discussed already in section 4.
This analysis is based on \cite{EP99}.

\begin{figure}
\begin{center}
% GNUPLOT: LaTeX picture with Postscript
\setlength{\unitlength}{0.1bp}
\begin{picture}(4539,2808)(0,0)
\special{psfile=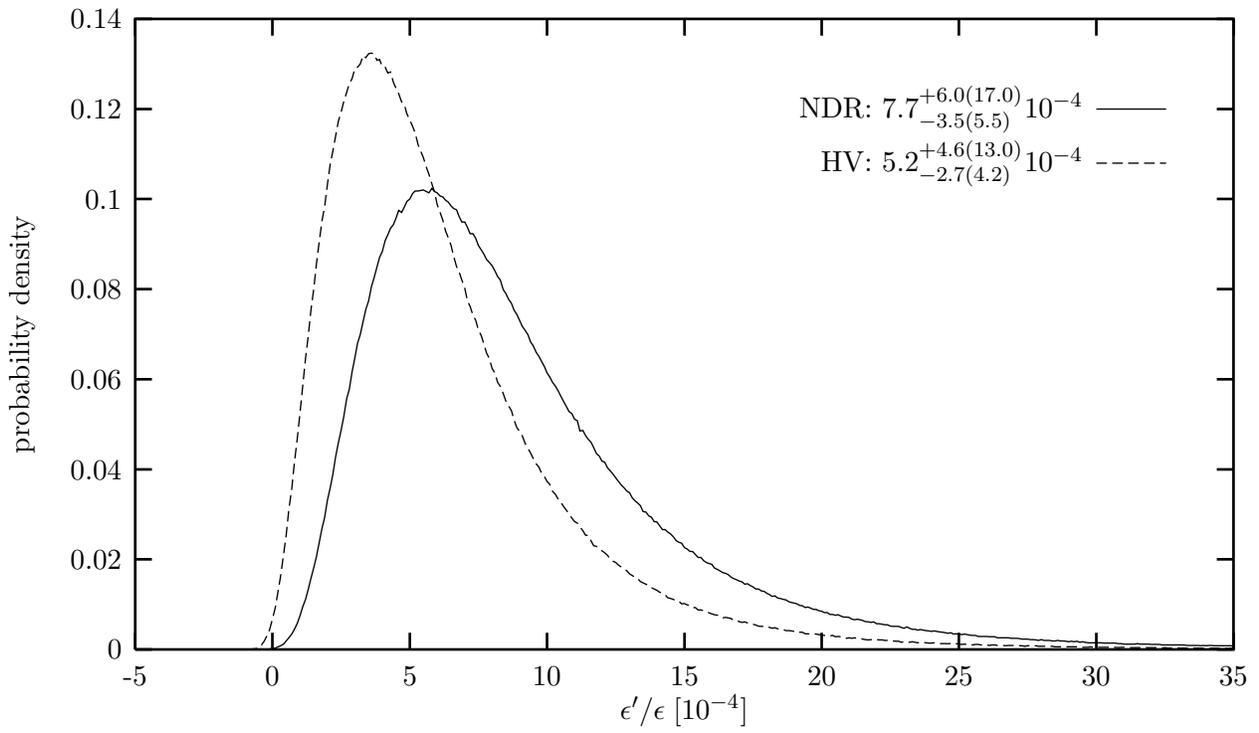 llx=0 lly=0 urx=908 ury=655 rwi=9080}
\put(3972,2164){\makebox(0,0)[r]{HV: $5.2^{+4.6(13.0)}_{-2.7(4.2)} 10^{-4}$}}
\put(3972,2368){\makebox(0,0)[r]{NDR: $7.7^{+6.0(17.0)}_{-3.5(5.5)} 10^{-4}$}}
\put(2469,100){\makebox(0,0){$\epsilon'/\epsilon \;[10^{-4}]$}}
\put(0,1519){%
\special{ps: gsave currentpoint currentpoint translate
270 rotate neg exch neg exch translate}%
\makebox(0,0)[b]{\shortstack{probability density}}%
\special{ps: currentpoint grestore moveto}%
}
\put(4539,230){\makebox(0,0){35}}
\put(4022,230){\makebox(0,0){30}}
\put(3504,230){\makebox(0,0){25}}
\put(2987,230){\makebox(0,0){20}}
\put(2470,230){\makebox(0,0){15}}
\put(1952,230){\makebox(0,0){10}}
\put(1435,230){\makebox(0,0){5}}
\put(917,230){\makebox(0,0){0}}
\put(400,230){\makebox(0,0){-5}}
\put(350,2708){\makebox(0,0)[r]{0.14}}
\put(350,2368){\makebox(0,0)[r]{0.12}}
\put(350,2029){\makebox(0,0)[r]{0.1}}
\put(350,1689){\makebox(0,0)[r]{0.08}}
\put(350,1349){\makebox(0,0)[r]{0.06}}
\put(350,1009){\makebox(0,0)[r]{0.04}}
\put(350,670){\makebox(0,0)[r]{0.02}}
\put(350,330){\makebox(0,0)[r]{0}}
\end{picture}
\end{center}
\vspace{-6mm}
\caption{Probability density distributions for $\epe$ in NDR and HV
schemes.}
\label{g1}
\end{figure}

Using the first method we find the probability density 
distributions for
$\epe$ in fig.~ \ref{g1}.
From this distribution
we deduce the following results:
\begin{equation}\label{eq:eperangefinal}
\epe =\left\{ \begin{array}{ll}
( 7.7^{~+6.0}_{~-3.5}) \cdot 10^{-4} & {\rm (NDR)} \\
( 5.2^{~+4.6}_{~-2.7}) \cdot 10^{-4} & {\rm (HV)} \end{array} \right.
\end{equation}
The difference between these two results indicates the left over
renormalization scheme dependence.
Since, the resulting probability density distributions for
$\epsilon'/\epsilon$ are very asymmetric with  very long
tails towards large values we quote the medians
and the $68\%(95\%)$ confidence level intervals. This means that 
$68\%(95\%)$
of our data can be found inside the corresponding error interval and
that $50\%$ of our data has smaller $\epsilon'/\epsilon$ than our
median.

We observe that negative values of $\epsilon'/\epsilon$ can be
excluded at $95\%$ C.L. For completeness we quote the mean and the
standard deviation for $\epsilon'/\epsilon$:
\begin{equation}\label{mean}
\epe =\left\{ \begin{array}{ll}
( 9.1\pm 6.2) \cdot 10^{-4} & {\rm (NDR)} \\
( 6.3 \pm 4.8) \cdot 10^{-4} & {\rm (HV)} \end{array} \right.
\end{equation}

Using the second method and the parameters in table~\ref{tab:inputparams} 
we find :
\begin{equation}
~~~~~1.05 \cdot 10^{-4} \le \epe \le 28.8 \cdot 10^{-4}\qquad {\rm (NDR)}.
\label{eq:eperangenew}
\end{equation}
and
\begin{equation}
~~~~~0.26 \cdot 10^{-4} \le \epe \le 22.0 \cdot 10^{-4}\qquad {\rm (HV)}.
\label{hv:eperangenew}
\end{equation}

We observe that
$\epe$ is generally lower in the HV scheme if the same values for
$B_6^{(1/2)}$ and $B_8^{(3/2)}$ are used in both schemes. 
Since the present non-perturbative methods do not have renormalization
scheme dependence fully under control we think that such treatment of
$B_6^{(1/2)}$ and $B_8^{(3/2)}$ is the proper way of estimating
scheme dependences at present.
Assuming, on the other hand, that the values in (\ref{bbb}) correspond
to the NDR scheme and using the relation (\ref{NDRHV}), we find for
the HV scheme the range 
$0.58 \cdot 10^{-4} \le \epe \le 26.9 \cdot 10^{-4}$ which is
much closer to the NDR result in (\ref{eq:eperangenew}). 
This exercise shows that it is very desirable to have the
scheme dependence under control.

We observe that the most probable values for $\epe$ in the NDR scheme
 are in the
ball park of $1 \cdot 10^{-3}$. They are lower by roughly $30\%$ in the
HV scheme if the same values for $(\bsi,\bei)$ are used.
On the other hand the ranges in (\ref{eq:eperangenew}) and
(\ref{hv:eperangenew}) show that for
particular choices of the input parameters, values for $\epe$ as high as
$(2-3)\cdot 10^{-3}$ cannot be excluded at present. Let us study 
this in  more detail.

In table~ \ref{tab:31731}, taken from \cite{EP99}, we show the values 
 of $\epe$ in units of $10^{-4}$ 
for specific values of $\bsi$, $\bei$ and $\ms(\mc)$ as calculated
in the NDR scheme. The corresponding values in the HV scheme
are lower as discussed above.
The fourth column shows the results for central values of all remaining
parameters. The comparison of the the fourth and the fifth column
 demonstrates
how $\epe$ is increased when $\Lms^{(4)}$ is raised from $340~\mev$
to $390~\mev$. As stated in (\ref{ap}) $\epe$ is roughly proportional
to $\Lms^{(4)}$. Finally, in the last column maximal values of $\epe$
are given.
To this end we have scanned all parameters relevant for
the analysis of $\IM\lambda_t$ within one
standard deviation and have chosen the highest value of
$\Lms^{(4)}=390\mev$. Comparison of the last two columns demonstrates
the impact of the increase of $\IM\lambda_t$ from its
central to its maximal value and of the variation of $\mt$.

Table~\ref{tab:31731} gives a good insight in the dependence of
$\epe$ on various parameters which is roughly described
by (\ref{ap}). We observe the following hierarchies:

\begin{itemize}
\item
The largest uncertainties reside in $\ms$, $\bsi$ and $\bei$.
$\epe$ increases universally by roughly a factor of 2.3 when
$\ms(\mc)$ is changed from $155 \mev$ to $105 \mev$. The increase
of $\bsi$ from 1.0 to 1.3 increases $\epe$ by $(55\pm 10)\%$,
depending on $\ms$ and $\bei$. The corresponding changes
due to $\bei$ are approximately $(40\pm 15)\%$.
\item
The combined uncertainty due to $\IM\lambda_t$ and $\mt$,
present both in $\IM\lambda_t$ and $F_{\varepsilon'}$,
is approximately $\pm 25\%$. The uncertainty due to
$\mt$ alone is only $\pm 5\%$.
\item
The uncertainty due to $\Lms^{(4)}$ is approximately $\pm 16\%$.
\item
The uncertainty due to $\OEE$ is approximately $\pm 12\%$.
\end{itemize}

The large sensitivity of $\epe$ to $\ms$ has been known 
since the analyses in the eighties. In the context of
the KTeV result this issue has been analyzed in \cite{Nierste}.
It has been found that provided $2\bsi-\bei\le 2$ the
consistency of the Standard Model with the KTeV result
requires the $2\sigma$ bound $\ms(2\gev)\le 110\mev$.
Our analysis is compatible with these findings.

\begin{table}[thb]
\caption[]{ Values of $\epe$ in units of $10^{-4}$ 
for specific values of $\bsi$, $\bei$ and $\ms(\mc)$
and other parameters
as explained in the text.
\label{tab:31731}}
\begin{center}
\begin{tabular}{|c|c|c|c|c|c|}\hline
 $B^{(1/2)}_6$& $B^{(3/2)}_8$ & $\ms(\mc)[\mev]$ &
  Central & $\Lms^{(4)}=390\mev $ & Maximal \\ \hline
      &     & $105$ &  20.2 & 23.3 & 28.8\\
 $1.3$&$0.6$& $130$ &  12.8 & 14.8 & 18.3\\
      &     & $155$ &   8.5 &  9.9 & 12.3 \\
 \hline
      &     & $105$ &  18.1 & 20.8 & 26.0\\
 $1.3$&$0.8$ & $130$ & 11.3 & 13.1 & 16.4\\
      &     & $155$ &   7.5 &  8.7 & 10.9\\
 \hline
      &     & $105$ &   15.9 & 18.3 & 23.2\\
 $1.3$&$1.0$ & $130$ &  9.9 &  11.5 & 14.5\\
      &     & $155$ &   6.5  &  7.6 &  9.6\\
 \hline\hline
      &     & $105$ &   13.7 & 15.8 & 19.7\\
 $1.0$&$0.6$ & $130$ &  8.4 &  9.8& 12.2 \\
      &     & $155$ &   5.4 &  6.4 & 7.9 \\
 \hline
      &     & $105$ &   11.5 & 13.3 & 16.9\\
$1.0$&$0.8$ & $130$  &  7.0 &   8.1 & 10.4\\
     &     & $155$ &   4.4 &    5.2 &  6.6\\
 \hline
     &     & $105$ &   9.4 &   10.9 & 14.1 \\
$1.0$&$1.0$ & $130$  &  5.5 &   6.5 &  8.5 \\
     &     & $155$ &   3.3 &    4.0 &  5.2\\
 \hline
\end{tabular}
\end{center}
\end{table}

\subsection{Summary}
 As we have seen,
the estimates of $\epe$ in the Standard Model are typically below 
the experimental data. However,
as our scanning analysis shows,
for suitably chosen parameters, $\epe$ in the Standard 
Model can be made consistent with data. However, this happens only if all 
relevant parameters are simultaneously close to their extreme values. 
This is clearly seen in table~\ref{tab:31731}.
Moreover, the probability density distributions for $\epe$ in
fig.~\ref{g1} indicates that values of $\epe$ in the ball park of
NA31 and KTeV results are rather improbable.

Unfortunately, in view of very large hadronic and substantial parametric 
uncertainties, it 
is impossible to conclude at present whether new physics contributions are 
indeed required to fit the data. 
Similarly it is difficult to conclude what is precisely the impact of 
the $\epe$-data on the CKM matrix. However, as analyzed in \cite{EP99}
there are indications  
that the lower limit on 
$\IM\lambda_t$ is improved. The same applies to the lower limits for the 
branching ratios for $K_L\to\pi^0\nu\bar\nu$ and 
$K_L\to\pi^0 e^+ e^-$ decays discussed in the following 
sections.

It is also clear that the $\epe$ data puts models in which there 
are new positive contributions to $\eps$ and negative contibutions to 
$\varepsilon'$ in serious difficulties. In particular 
as analyzed in \cite{EP99} the two Higgs Doublet Model II
\cite{Abbott} can either be ruled out  with improved 
hadronic matrix elements or a  
powerful lower bound on $\tan\beta$ can 
be obtained from $\epe$.
In the Minimal Supersymmetric Standard Model, in addition to charged
Higgs exchanges in loop diagrams, also charginos contribute.
For suitable  choice of the
supersymmetric parameters, the chargino contribution
can enhance $\epe$  with respect to the Standard
Model expectations \cite{GG95}.
Yet, generally
the most conspicuous effect of
minimal supersymmetry is a depletion of $\epe$.
The situation can be different in more general
models in which there are more parameters than
in the two Higgs doublet model II and in the MSSM, in particular
new CP violating phases.
As an example, in more general supersymmetric models
$\epe$ can be made consistent with experimental
findings \cite{GMS,MM99}. Unfortunately, in view of the large number
of free  parameters
such models are not very predictive.

The future of $\epe$ in the Standard Model and in its extensions depends on 
the progress in the reduction of parametric and hadronic uncertainties. 
In any case $\epe$ already played a decisive role in establishing direct 
CP violation in nature and its rather large value gives additional strong 
motivation for searching for this phenomenon in 
cleaner K decays like 
$K_L\to\pi^0\nu\bar\nu$ and 
$K_L\to\pi^0 e^+ e^-$, in B decays, in D decays and elsewhere.
We now turn to discuss some of these topics.

\section{ The Decays $K^+\to\pi^+\nu\bar\nu$ and
$K_{\rm L}\to\pi^0\nu\bar\nu$}
         \label{sec:HeffRareKB}
\setcounter{equation}{0}
\subsection{General Remarks}
            \label{sec:HeffRareKB:overview}
We will now move to discuss
the semileptonic rare FCNC
transitions $\kpn$ and $K_{\rm L}\to\pi^0\nu\bar\nu$.
Within the Standard Model these decays are loop-induced
semileptonic FCNC processes determined only 
by $Z^0$-penguin and box diagrams
and  are governed by the single
function $X_0(x_t)$ given in (\ref{XA0}).

A particular and very important virtue of $K\to\pi\nu\bar\nu$
is their clean theoretical character.
This is related to the fact that
the low energy hadronic
matrix elements required are just the matrix elements of quark currents
between hadron states, which can be extracted from the leading
(non-rare) semileptonic decays. Other long-distance contributions
are negligibly small \cite{RS,GBGI}. As a consequence of these features,
the scale ambiguities, inherent to perturbative QCD, essentially
constitute  the only theoretical uncertainties 
present in the analysis of these decays.
These theoretical uncertainties have been considerably reduced
through the inclusion of
the next-to-leading QCD corrections 
 \cite{BB1}--\cite{BB98}. 

The investigation of these low energy rare decay processes in
conjunction with their theoretical cleanliness, allows to probe,
albeit indirectly, high energy scales of the theory and in particular
to measure $V_{td}$ and $\IM\lambda_t= \IM V^*_{ts} V_{td}$
from $K^+\to\pi^+\nu\bar\nu$ and $K_{\rm L}\to\pi^0\nu\bar\nu$
respectively.
However, the very fact
that these processes are based on higher order electroweak effects
implies
that their branching ratios are expected to be very small and not easy to
access experimentally.

\subsection{The Decay \kpnn}
            \label{sec:HeffRareKB:kpnn}
\subsubsection{The effective Hamiltonian}
The effective Hamiltonian for $\kpn$  can
be written as
\begin{equation}\label{hkpn} 
{\cal H}_{\rm eff}={G_{\rm F} \over{\sqrt 2}}{\alpha\over 2\pi 
\sin^2\Theta_{\rm W}}
 \sum_{l=e,\mu,\tau}\left( V^{\ast}_{cs}V_{cd} X^l_{\rm NL}+
V^{\ast}_{ts}V_{td} X(x_t)\right)
 (\bar sd)_{V-A}(\bar\nu_l\nu_l)_{V-A} \, .
\end{equation}
The index $l$=$e$, $\mu$, $\tau$ denotes the lepton flavour.
The dependence on the charged lepton mass resulting from the box-graph
is negligible for the top contribution. In the charm sector this is the
case only for the electron and the muon but not for the $\tau$-lepton.

The function $X(x_t)$ relevant for the top part is given by
\begin{equation}\label{xx9} 
X(x_t)=X_0(x_t)+\aspi X_1(x_t) 
=\eta_X\cdot X_0(x_t), \qquad\quad \eta_X=0.994,
\end{equation}
with the QCD correction \cite{BB2,MU98,BB98}
\begin{equation}\label{xx1}
X_1(x_t)=\tilde X_1(x_t)+
8x_t{\partial X_0(x_t)\over\partial x_t}\ln x_\mu\,.
\end{equation}
Here $x_\mu=\mu_t^2/M^2_W$ with $\mu_t=\ord(m_t)$ and
$\tilde X_1(x_t)$ is a complicated function given in 
\cite{BB2,MU98,BB98}.
The $\mu_t$-dependence of the last term in (\ref{xx1}) cancels to the
considered order the $\mu_t$-dependence of the leading term 
$X_0(x_t(\mu))$.
The leftover $\mu_t$-dependence in $X(x_t)$ is below $1\%$.
The factor $\eta_X$ summarizes the NLO 
corrections represented by the second
term in (\ref{xx9}).
With $\mt\equiv \mtb(\mt)$ the QCD factor $\eta_X$
is practically independent of $m_t$ and $\Lambda_{\overline{MS}}$
and is very close to unity.

The expression corresponding to $X(x_t)$ in the charm sector is the function
$X^l_{\rm NL}$. It results from the NLO calculation \cite{BB3} and is given
explicitly in \cite{BB98}.
The inclusion of NLO corrections reduced considerably the large
$\mu_c$ dependence
(with $\mu_c={\cal O}(m_c)$) present in the leading order expressions
for the charm contribution
 \cite{novikovetal:77}.
Varying $\mu_c$ in the range $1\gev\le\mu_c\le 3\gev$ changes $X_{\rm NL}$
by roughly $24\%$ after the inclusion of NLO corrections to be compared
with $56\%$ in the leading order. Further details can be found in
\cite{BB3,BBL}. The impact of the $\mu_c$ uncertainties on the
resulting branching ratio $Br(\kpn)$ is discussed below.

The
numerical values for $X^l_{\rm NL}$ for $\mu = \mc$ and several values of
$\Lms^{(4)}$ and $\mc(\mc)$ can be found in \cite{BB98}. 
The net effect of QCD corrections is to suppress the charm contribution
by roughly $30\%$. For our purposes we need only
\begin{equation}\label{p0k}
P_0(X)=\frac{1}{\lambda^4}\left[\frac{2}{3} X^e_{\rm NL}+\frac{1}{3}
 X^\tau_{\rm NL}\right]=0.42\pm0.06
\end{equation}
where the error results from the variation of $\Lms^{(4)}$ and $\mc(\mc)$.

\subsubsection{Deriving the Branching Ratio}
The relevant hadronic
matrix element of the weak current $(\bar sd)_{V-A}$ in (\ref{hkpn}) 
can be extracted
with the help of isospin symmetry from
the leading decay $K^+\to\pi^0e^+\nu$.
Consequently the resulting theoretical
expression for  the branching fraction $Br(K^+\to\pi^+\nu\bar\nu)$ can
be related to the experimentally well known quantity
$Br(K^+\to\pi^0e^+\nu)$. Let us demonstrate this.

The effective Hamiltonian for the tree level decay $K^+\to\pi^0 e^+\nu$
is given by
\begin{equation}\label{kp0} 
{\cal H}_{\rm eff}(K^+\to\pi^0 e^+\nu)
={G_{\rm F} \over{\sqrt 2}}
 V^{\ast}_{us}
 (\bar su)_{V-A}(\bar\nu_e e)_{V-A} \, .
\end{equation}
Using isospin symmetry we have
\be\label{iso1}
\langle \pi^+|(\bar sd)_{V-A}|K^+\rangle=\sqrt{2}
\langle \pi^0|(\bar su)_{V-A}|K^+\rangle.
\ee
Consequently neglecting differences in the phase space of these two decays,
due to $m_{\pi^+}\not=m_{\pi^0}$ and $m_e\not=0$, we find 
\be\label{br1}
\frac{Br(\kpn)}{Br(K^+\to\pi^0 e^+\nu)}=
{\alpha^2\over |V_{us}|^2 2\pi^2 
\sin^4\Theta_{\rm W}}
 \sum_{l=e,\mu,\tau}\left| V^{\ast}_{cs}V_{cd} X^l_{\rm NL}+
V^{\ast}_{ts}V_{td} X(x_t)\right|^2~.
\end{equation}
\subsubsection{Basic Phenomenology}
Using (\ref{br1}) 
and including isospin breaking corrections one finds
\begin{equation}\label{bkpn}
Br(\kpn)=\kappa_+\cdot\left[\left({\imlt\over\lambda^5}X(x_t)\right)^2+
\left({\relc\over\lambda}P_0(X)+{\relt\over\lambda^5}X(x_t)\right)^2
\right]~,
\end{equation}
\begin{equation}\label{kapp}
\kappa_+=r_{K^+}{3\alpha^2 Br(K^+\to\pi^0e^+\nu)\over 2\pi^2
\sin^4\Theta_{\rm W}}
 \lambda^8=4.11\cdot 10^{-11}\,,
\end{equation}
where we have used
\begin{equation}\label{alsinbr}
\alpha=\frac{1}{129},\qquad \sin^2\Theta_{\rm W}=0.23, \qquad
Br(K^+\to\pi^0e^+\nu)=4.82\cdot 10^{-2}\,.
\end{equation}
Here $\lambda_i=V^\ast_{is}V_{id}$ with $\lambda_c$ being
real to a very high accuracy. $r_{K^+}=0.901$ summarizes isospin
breaking corrections in relating $\kpn$ to $K^+\to\pi^0e^+\nu$.
These isospin breaking corrections are due to quark mass effects and 
electroweak radiative corrections and have been calculated in
\cite{MP}. Finally $P_0(X)$ is given in (\ref{p0k}).

Using the improved Wolfenstein parametrization and the approximate
formulae (\ref{2.51}) -- (\ref{2.53}) we can next put 
(\ref{bkpn}) into a more transparent form \cite{BLO}:
\begin{equation}\label{108}
Br(K^{+} \to \pi^{+} \nu \bar\nu) = 4.11 \cdot 10^{-11} A^4 X^2(x_t)
\frac{1}{\sigma} \left[ (\sigma \bar\eta)^2 +
\left(\varrho_0 - \bar\varrho \right)^2 \right]\,,
\end{equation}
where
\begin{equation}\label{109}
\sigma = \left( \frac{1}{1- \frac{\lambda^2}{2}} \right)^2\,.
\end{equation}

The measured value of $Br(K^{+} \to \pi^{+} \nu \bar\nu)$ then
determines  an ellipse in the $(\bar\varrho,\bar\eta)$ plane  centered at
$(\varrho_0,0)$ with 
\begin{equation}\label{110}
\varrho_0 = 1 + \frac{P_0(X)}{A^2 X(x_t)}
\end{equation}
and having the squared axes
\begin{equation}\label{110a}
\bar\varrho_1^2 = r^2_0, \qquad \bar\eta_1^2 = \left( \frac{r_0}{\sigma}
\right)^2
\end{equation}
where
\begin{equation}\label{111}
r^2_0 = \frac{1}{A^4 X^2(x_t)} \left[
\frac{\sigma \cdot Br(K^{+} \to \pi^{+} \nu \bar\nu)}
{4.11 \cdot 10^{-11}} \right]\,.
\end{equation}
Note that $r_0$ depends only on the top contribution.
The departure of $\varrho_0$ from unity measures the relative importance
of the internal charm contributions.

The ellipse defined by $r_0$, $\varrho_0$ and $\sigma$ given above
intersects with the circle (\ref{2.94}).  This allows to determine
$\bar\varrho$ and $\bar\eta$  with 
\begin{equation}\label{113}
\bar\varrho = \frac{1}{1-\sigma^2} \left( \varrho_0 - \sqrt{\sigma^2
\varrho_0^2 +(1-\sigma^2)(r_0^2-\sigma^2 R_b^2)} \right), \qquad
\bar\eta = \sqrt{R_b^2 -\bar\varrho^2}
\end{equation}
and consequently
\begin{equation}\label{113aa}
R_t^2 = 1+R_b^2 - 2 \bar\varrho,
\end{equation}
where $\bar\eta$ is assumed to be positive.
Given $\bar\varrho$ and $\bar\eta$ one can determine $V_{td}$:
\begin{equation}\label{vtdrhoeta}
V_{td}=A \lambda^3(1-\bar\varrho-i\bar\eta),\qquad
|V_{td}|=A \lambda^3 R_t.
\end{equation}
The determination of $|V_{td}|$ and of the unitarity triangle requires
the knowledge of $V_{cb}$ (or $A$) and of $|V_{ub}/V_{cb}|$. Both
values are subject to theoretical uncertainties present in the existing
analyses of tree level decays. Whereas the dependence on
$|V_{ub}/V_{cb}|$ is rather weak, the very strong dependence of
$Br(\kpn)$ on $A$ or $V_{cb}$ makes a precise prediction for this
branching ratio difficult at present. We will return to this below.
The dependence of $Br(\kpn)$ on $\mt$ is also strong. However $\mt$
is known already  within $\pm 4\%$ and
consequently the related uncertainty in 
$Br(\kpn)$ is substantialy smaller than the corresponding uncertainty 
due to $V_{cb}$.

\subsubsection{Numerical Analysis of \kpnn}
\label{sec:Kpnn:NumericalKp}
The uncertainties 
in the prediction for $Br(\kpn)$ and in the determination of  $|V_{td}|$
related to the choice of the renormalization scales $\mu_t$
and $\mu_c$ in the top part and the charm part, respectively
have been inestigated in \cite{BBL}.
To this end the scales $\mu_c$ and $\mu_t$ entering $m_c(\mu_c)$
and $m_t(\mu_t)$, respectively, have been varied in the ranges
$1\gev\leq\mu_c\leq 3\gev$ and $100\gev\leq\mu_t\leq 300\gev$.
It has been found that including
the full next-to-leading corrections reduces the uncertainty in the
determination of $|V_{td}|$ from $\pm 14\%$ (LO) to $\pm 4.6\%$ (NLO).
The main bulk of this theoretical error stems
from the charm sector. 
In the case of $Br(\kpn)$,
the theoretical uncertainty
due to $\mu_{c,t}$ is reduced from $\pm 22\%$ (LO) to $\pm 7\%$ (NLO).

Scanning the input parameters of table \ref{tab:inputparams}
we find 
\begin{equation}\label{kpnr}
Br(\kpn)=
(7.9 \pm 3.1)\cdot 10^{-11} 
\end{equation}
where the error comes dominantly from the uncertainties in the CKM
parameters.

It is possible to derive an upper bound on $Br(\kpn)$ \cite{BB98}:
\be\label{boundk}
Br(\kpn)_{\rm max}=\frac{\kappa_+}{\sigma}
\left[ P_0(X)+A^2 X(x_t)\frac{r_{sd}}{\lambda}
\sqrt{\frac{\Delta M_d}{\Delta M_s}}\right]^2
\ee
where $r_{ds}=\xi\sqrt{m_{B_s}/m_{B_d}}$. This equation translates
 a lower bound on $\Delta M_s$ into an upper bound on $Br(\kpn)$.
This bound is very clean and does not involve theoretical
hadronic uncertainties except for $r_{sd}$. Using
\be
\sqrt{\frac{\Delta M_d}{\Delta M_s}}<0.2~,~\quad
A<0.87~,~\quad P_0(X)<0.48~,~\quad X(x_t)<1.56~,~\quad
r_{sd}<1.2
\ee
we find
\be
Br(\kpn)_{\rm max}=12.2 \cdot 10^{-11}~.
\ee
This limit could be further strengthened with improved input.
However, this bound is strong enough to indicate a clear
conflict with the Standard Model if $Br(\kpn)$ should
be measured at $2\cdot 10^{-10}$.

\subsubsection{$\vtd$ from $K^+\to\pi^+\nu\bar\nu$}
Once $Br(K^+\to\pi^+\nu\bar\nu)\equiv Br(K^+)$ is measured, $\vtd$ can be
extracted subject to various uncertainties:
\be\label{vtda}
\frac{\sigma(\vtd)}{\vtd}=\pm 0.04_{scale}\pm \frac{\sigma(\vcb)}{\vcb}
\pm 0.7 \frac{\sigma(\bar\mc)}{\bar\mc}
\pm 0.65 \frac{\sigma( Br(K^+))}{Br(K^+)}~.
\ee
Taking $\sigma(\vcb)=0.002$, $\sigma(\bar\mc)=100\mev$ and
$\sigma( Br(K^+))=10\%$ and adding the errors in quadrature we find that
$\vtd$ can be determined with an accuracy of $\pm 10\%$.
This number
is increased to $\pm 11\%$ once the uncertainties due to $\mt$,
$\alpha_s$ and $|V_{ub}|/\vcb$ are taken into account. Clearly this
determination can be improved although a determination of $\vtd$ with
an accuracy better than $\pm 5\%$ seems rather unrealistic.

\subsubsection{Summary and Outlook}
The accuracy of the Standard Model prediction for $Br(\kpn)$ has
improved considerably during the last five years. 
This progress can be traced back to the
improved values of $\mt$ and $\vcb$ and to the inclusion of NLO
QCD corrections which considerably reduced the scale uncertainties
in the charm sector. 

Now, what about the experimental status of this decay ?
One of the high-lights of 97 was the observation by BNL787
collaboration at Brookhaven \cite{Adler97} 
of one event consistent with the signature expected for this decay.
The branching ratio:
\be\label{kp97}
Br(K^+ \rightarrow \pi^+ \nu \bar{\nu})=
(4.2^{+9.7}_{-3.5})\cdot 10^{-10}
\end{equation}
has the central value  by a factor of 5 above the Standard Model
expectation but in view of large errors the result is compatible with the
Standard Model. 
The analysis of additional
data on $K^+\to \pi^+\nu\bar\nu$ present on tape at BNL787 should narrow
this range in the near future considerably.
In view of the clean character of this decay a measurement of its
branching ratio at the level of $ 2 \cdot 10^{-10}$ 
would signal the presence of physics
beyond the Standard Model. The Standard Model sensitivity is
expected to be reached at AGS around the year 2000 \cite{AGS2}.
Also Fermilab with the Main Injector 
could measure this decay \cite{Cooper}.
\subsection{The Decay $K_{\rm L}\to\pi^0\nu\bar\nu$}
            \label{sec:HeffRareKB:klpinn1}
\subsubsection{The effective Hamiltonian}
The effective
Hamiltonian for $K_{\rm L}\to\pi^0\nu\bar\nu$
is given as follows:
\begin{equation}\label{hxnu}
{\cal H}_{\rm eff} = {G_{\rm F}\over \sqrt 2} {\alpha \over
2\pi \sin^2 \Theta_{\rm W}} V^\ast_{ts} V_{td}
X (x_t) (\bar sd)_{V-A} (\bar\nu\nu)_{V-A} + h.c.\,,   
\end{equation}
where the function $X(x_t)$, present already in $\kpn$,
includes NLO corrections and is given in (\ref{xx9}). 

As we will demonstrate shortly, $\klpn$  proceeds in the Standard Model 
almost
entirely through direct CP violation \cite{littenberg:89}. It
is completely dominated by short-distance loop diagrams with top quark
exchanges. The charm contribution can be fully
neglected and the theoretical uncertainties present in $\kpn$ due to
$m_c$, $\mu_c$ and $\Lambda_{\overline{MS}}$ are absent here. 
Consequently the rare decay $\klpn$ is even cleaner than $\kpn$
and is very well suited for the determination of 
the Wolfenstein parameter $\eta$ and in particular $\imlt$.

 It is usually stated in the literature that the
decay $\klpn$ is dominated by {\it direct} CP violation. Now
the standard definition of the direct CP violation 
requires the presence of strong phases which are
completely negligible in $\klpn$. Consequently the violation of
CP symmetry in $\klpn$ arises through the interference between
$K^0-\bar K^0$ mixing and the decay amplitude. This type of CP
violation is often called {\it mixing-induced} CP violation.
However, as already pointed out by Littenberg \cite{littenberg:89}
and demonstrated explictly in a moment,
the contribution of CP violation to $\klpn$ via $K^0-\bar K^0$ mixing 
alone is tiny. It gives $Br(\klpn) \approx 2\cdot 10^{-15}$.
Consequently, in this sence,  CP violation in $\klpn$ with
$Br(\klpn) = {\cal O}(10^{-11})$ is a manifestation of CP violation
in the decay and as such deserves the name of {\it direct} CP violation.
In other words the difference in the magnitude of CP violation in
$K_{\rm L}\to\pi\pi~(\varepsilon)$ and $\klpn$ is a signal of direct
CP violation and measuring $\klpn$ at the expected level would
be another signal of this phenomenon. More details on this
issue can be found in \cite{NIR96,BUCH96,BB96}.
\subsubsection{Deriving the Branching Ratio}
Let us derive the basic formula for $Br(\klpn)$ in a manner analogous
to the one for  $Br(K^+ \to \pi^+ \nu \bar\nu)$. To this end we
consider one neutrino flavour and define the complex function:
\begin{equation}\label{hxnu1}
F = {G_{\rm F}\over \sqrt 2} {\alpha \over
2\pi \sin^2 \Theta_{\rm W}} V^\ast_{ts} V_{td}
X (x_t).   
\end{equation}
Then the effective Hamiltonian in (\ref{hxnu}) can be written as
\begin{equation}\label{hxnu2}
{\cal H}_{\rm eff} =  F (\bar sd)_{V-A} (\bar\nu\nu)_{V-A}+
F^\ast (\bar ds)_{V-A} (\bar\nu\nu)_{V-A}~.
\end{equation}
Now, from (\ref{KLS}) we have
\be\label{KLS1}
K_L=\frac{1}{\sqrt{2}}
[(1+\bar\varepsilon)K^0+ (1-\bar\varepsilon)\bar K^0]
\ee
where we have neglected
$\mid\bar\varepsilon\mid^2\ll 1$. Thus the amplitude
for $K_L\to\pi^0\nu\bar\nu$ is given by
\be\label{ampkl0}
A(K_L\to\pi^0\nu\bar\nu)=
\frac{1}{\sqrt{2}}
\left[F(1+\bar\varepsilon) \langle \pi^0|(\bar sd)_{V-A}|K^0\rangle
+ 
F^\ast (1-\bar\varepsilon) \langle \pi^0|(\bar ds)_{V-A}|\bar K^0\rangle
 \right] (\bar\nu\nu)_{V-A}.
\ee
Recalling
\be\label{DEF}
CP|K^0\rangle = - |\bar K^0\rangle, \quad\quad
C|K^0\rangle =  |\bar K^0\rangle
\ee
we have
\be
\langle \pi^0|(\bar ds)_{V-A}|\bar K^0\rangle=-
\langle \pi^0|(\bar sd)_{V-A}|K^0\rangle,
\ee
where the minus sign is crucial for the subsequent steps.

Thus we can write
\be\label{bmpkl0}
A(K_L\to\pi^0\nu\bar\nu)=
\frac{1}{\sqrt{2}}
\left[F(1+\bar\varepsilon) -F^\ast (1-\bar\varepsilon)\right]
 \langle \pi^0|(\bar sd)_{V-A}| K^0\rangle
 (\bar\nu\nu)_{V-A}.
\ee
Now the terms $\bar\varepsilon$ can be safely neglected in comparision
with unity, which implies that the indirect CP violation
(CP violation in the $K^0-\bar K^0$ mixing) is negligible in this decay.
We have then
\be
F(1+\bar\varepsilon) -F^\ast (1-\bar\varepsilon)=
{G_{\rm F}\over \sqrt 2} {\alpha \over
\pi \sin^2 \Theta_{\rm W}} \IM (V^\ast_{ts} V_{td})
\cdot X(x_t).   
\end{equation}
Consequently using isospin relation
\be
\langle \pi^0|(\bar ds)_{V-A}|\bar K^0\rangle=
\langle \pi^0|(\bar su)_{V-A}|K^+\rangle
\ee
together with (\ref{kp0}) and taking into account the difference
in the lifetimes of $K_L$ and $K^+$ we have after summation over three
neutrino flavours
\be\label{br2}
\frac{Br(K_L\to\pi^0\nu\bar\nu)}{Br(K^+\to\pi^0 e^+\nu)}=
3\frac{\tau(K_L)}{\tau(K^+)}
{\alpha^2\over |V_{us}|^2 2\pi^2 
\sin^4\Theta_{\rm W}}
 \left[\IM \lambda_t \cdot X(x_t)\right]^2
\end{equation}
where $\lambda_t=V^{\ast}_{ts}V_{td}$.
\subsubsection{Master Formulae for $Br(\klpn)$}
\label{sec:Kpnn:MasterKL}
Using (\ref{br2}) we can write $Br(\klpn)$ simply as
follows
\begin{equation}\label{bklpn}
Br(K_{\rm L}\to\pi^0\nu\bar\nu)=\kappa_{\rm L}\cdot
\left({\imlt\over\lambda^5}X(x_t)\right)^2
\end{equation}
\begin{equation}\label{kapl}
\kappa_{\rm L}=\frac{r_{K_{\rm L}}}{r_{K^+}}
 {\tau(K_{\rm L})\over\tau(K^+)}\kappa_+ =1.80\cdot 10^{-10}
\end{equation}
with $\kappa_+$ given in (\ref{kapp}) and
$r_{K_{\rm L}}=0.944$ summarizing isospin
breaking corrections in relating $\klpn$ to $K^+\to\pi^0e^+\nu$
\cite{MP}.

Using the Wolfenstein
parametrization and (\ref{xx9}) we can rewrite (\ref{bklpn}) as
\begin{equation}
Br(K_{\rm L}\to\pi^0\nu\bar\nu)=
3.0\cdot 10^{-11}
\left [ \frac{\eta}{0.39}\right ]^2
\left [\frac{\mtb(\mt)}{170~GeV} \right ]^{2.3} 
\left [\frac{\mid V_{cb}\mid}{0.040} \right ]^4 \,.
\label{bklpn1}
\end{equation}

The determination of $\eta$ using $Br(\klpn)$ requires the knowledge
of $V_{cb}$ and $\mt$. The very strong dependence on $V_{cb}$ or $A$
makes a precise prediction for this branching ratio difficult at
present.

On the other hand inverting (\ref{bklpn}) and using (\ref{xx9})
 one finds \cite{BB96}:
\begin{equation}\label{imlta}
\IM\lambda_t=1.36\cdot 10^{-4} 
\left[\frac{170\gev}{\mtb(\mt)}\right]^{1.15}
\left[\frac{Br(\klpn)}{3\cdot 10^{-11}}\right]^{1/2}\,.
\end{equation}
without any uncertainty in $\vcb$.
(\ref{imlta}) offers
 the cleanest method to measure $\IM\lambda_t$;
even better than the CP asymmetries
in $B$ decays discussed briefly in section 8.
\subsubsection{Numerical Analysis of \klpnn}
\label{sec:Kpnn:NumericalKL}
The $\mu_t$-uncertainties present in the function $X(x_t)$ have 
already been
discussed in connection with $\kpn$. After the inclusion of NLO
corrections they are so small that they can be neglected for all
practical purposes. 
Scanning the input parameters of table \ref{tab:inputparams}
we find 
\begin{equation}\label{klpnr4}
Br(\klpn)=
(2.8 \pm 1.1)\cdot 10^{-11} 
\end{equation}
where the error comes dominantly from the uncertainties in the CKM
parameters. 
\subsubsection{Summary and Outlook}
The accuracy of the Standard Model prediction for $Br(\klpn)$ has
improved considerably during the last five years. 
This progress can be traced back mainly to the
improved values of $\mt$ and $\vcb$ and to some extent to 
the inclusion of NLO QCD corrections.

The present upper bound on $Br(K_{\rm L}\to \pi^0\nu\bar\nu)$ from
FNAL experiment E799 \cite{XX98} is 
\begin{equation}\label{KLD}
Br(\klpn)<1.6 \cdot 10^{-6}\,.
\end{equation}
This is about five orders of magnitude above the Standard Model expectation
(\ref{klpnr4}).
Moreover this bound is substantially weaker than the 
{\it model independent} bound \cite{NIR96}
from isospin symmetry:
\begin{equation}
Br(\klpn) < 4.4 \cdot Br(\kpn)
\end{equation}
which through (\ref{kp97})  gives
\begin{equation}\label{B108}
Br(\klpn) < 6.1 \cdot 10^{-9}
\end{equation}

Now FNAL-E799 expects to reach
the accuracy ${\cal O}(10^{-8})$ and
a very interesting new experiment
at Brookhaven (BNL E926) \cite{AGS2} 
expects to reach the single event sensitivity $2\cdot 10^{-12}$
allowing a $10\%$ measurement of the expected branching ratio. 
There are furthermore plans
to measure this gold-plated  decay with comparable sensitivity
at Fermilab \cite{FNALKL} and KEK \cite{KEKKL}.
\begin{figure}[hbt]
\vspace{0.10in}
\centerline{
\epsfysize=2.7in
\epsffile{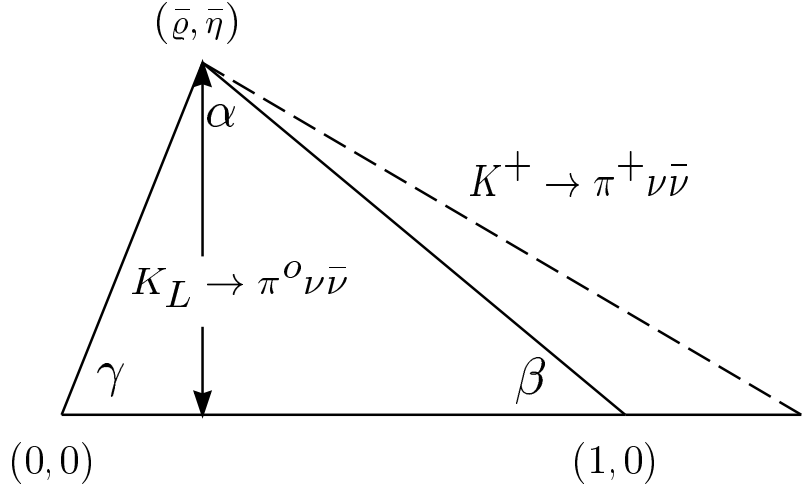}
}
\vspace{0.08in}
\caption{Unitarity triangle from $K\to\pi\nu\bar\nu$.}\label{fig:KPKL}
\end{figure}

\subsection{Unitarity Triangle and $\sin 2\beta$ from $K\to\pi\nu\bar\nu$}
\label{sec:Kpnn:Triangle}
The measurement of $Br(\kpn)$ and $Br(\klpn)$ can determine the
unitarity triangle completely, (see fig.~\ref{fig:KPKL}), 
provided $\mt$ and $V_{cb}$ are known \cite{BB4}.
Using these two branching ratios simultaneously allows to eliminate
$|V_{ub}/V_{cb}|$ from the analysis which removes a considerable
uncertainty. Indeed it is evident from (\ref{bkpn}) and
(\ref{bklpn}) that, given $Br(\kpn)$ and $Br(\klpn)$, one can extract
both $\imlt$ and $\relt$. One finds \cite{BB4,BBL}
\begin{equation}\label{imre}
\imlt=\lambda^5{\sqrt{B_2}\over X(x_t)}\qquad
\relt=-\lambda^5{{\relc\over\lambda}P_0(X)+\sqrt{B_1-B_2}\over X(x_t)}\,,
\end{equation}
where we have defined the ``reduced'' branching ratios
\begin{equation}\label{b1b2}
B_1={Br(\kpn)\over 4.11\cdot 10^{-11}}\qquad
B_2={Br(\klpn)\over 1.80\cdot 10^{-10}}\,.
\end{equation}
Using next the expressions for $\imlt$, $\relt$ and $\relc$ given
in (\ref{2.51})--(\ref{2.53}) we find
\begin{equation}\label{rhetb}
\bar\varrho=1+{P_0(X)-\sqrt{\sigma(B_1-B_2)}\over A^2 X(x_t)}\,,\qquad
\bar\eta={\sqrt{B_2}\over\sqrt{\sigma} A^2 X(x_t)}
\end{equation}
with $\sigma$ defined in (\ref{109}). An exact treatment of the CKM
matrix shows that the formulae (\ref{rhetb}) are rather precise
\cite{BB4}. 

Using (\ref{rhetb}) one finds subsequently \cite{BB4}
\begin{equation}\label{sin}
r_s=r_s(B_1, B_2)\equiv{1-\bar\varrho\over\bar\eta}=\cot\beta\,, \qquad
\sin 2\beta=\frac{2 r_s}{1+r^2_s}
\end{equation}
with
\begin{equation}\label{cbb}
r_s(B_1, B_2)=\sqrt{\sigma}{\sqrt{\sigma(B_1-B_2)}-P_0(X)\over\sqrt{B_2}}\,.
\end{equation}
Thus within the approximation of (\ref{rhetb}) $\sin 2\beta$ is
independent of $V_{cb}$ (or $A$) and $m_t$.

It should be stressed that $\sin 2\beta$ determined this way depends
only on two measurable branching ratios and on the function
$P_0(X)$ which is completely calculable in perturbation theory.
Consequently this determination is free from any hadronic
uncertainties and its accuracy can be estimated with a high degree
of confidence. 

An extensive numerical analysis of the formulae above has been presented
in \cite{BB4,BB96}. 
Assuming that the branching ratios are known to within $\pm 10\%$
and $\mt$ within $\pm 3~\gev$ one finds the results in table~\ref{tabkb2}
\cite{BB96}. 
We observe that respectable determinations of all considered 
quantities except for 
$\bar\varrho$ can be obtained.
Of particular interest are the accurate determinations of
$\sin 2\beta$ and of ${\rm Im}\lambda_t$.
The latter quantity as seen in (\ref{imlta}) 
can be obtained from
$K_{\rm L}\to\pi^0\nu\bar\nu$ alone and does not require knowledge
of $V_{cb}$.
The importance of measuring accurately  ${\rm Im}\lambda_t$ is evident.
It plays a central role in the phenomenology of CP violation
in $K$ decays and is furthermore equivalent to the 
Jarlskog parameter $J_{\rm CP}$ \cite{CJ}, 
the invariant measure of CP violation in the Standard Model, 
$J_{\rm CP}=\lambda(1-\lambda^2/2){\rm Im}\lambda_t$.

\begin{table}
\caption[]{Illustrative example of the determination of CKM
parameters from $K\to\pi\nu\bar\nu$.
\label{tabkb2}}
\vspace{0.4cm}
\begin{center}
\begin{tabular}{|c|c|c|}\hline
&$\sigma(|V_{cb}|)=\pm 0.002$ & $\sigma(|V_{cb}|)=\pm 0.001$
\\ 
\hline
$\sigma(|V_{td}|) $& $\pm 10\% $ & $ \pm 9\% $
 \\ 
\hline 
$\sigma(\bar\varrho) $ & $\pm 0.16$ &$\pm 0.12$
  \\
\hline
$\sigma(\bar\eta)$ & $\pm 0.04$&$\pm 0.03$
 \\
\hline
$\sigma(\sin 2\beta)$ & $\pm 0.05$&$\pm 0.05$
 \\
\hline
$\sigma({\rm Im}\lambda_t)$&$\pm 5\%$ &$\pm 5\%$ 
 \\
\hline
\end{tabular}
\end{center}
\end{table}

The accuracy to which $\sin 2\beta$ can be obtained from
$K\to\pi\nu\bar\nu$ is, in the  example discussed above, 
comparable to the one expected
in determining $\sin 2\beta$ from CP asymmetries in $B$ decays prior to
LHC experiments.  In this case $\sin 2\beta$ is determined best by
measuring CP violation in $B_d\to J/\psi K_{\rm S}$.
Using the formula  for the corresponding time-integrated 
CP asymmetry one finds an
interesting connection between rare $K$ decays and $B$ physics \cite{BB4}
\begin{equation}\label{kbcon}
{2 r_s(B_1,B_2)\over 1+r^2_s(B_1,B_2)}=
-a_{\mbox{{\scriptsize CP}}}(B_d\to J/\psi K_{\mbox{{\scriptsize S}}})
{1+x^2_d\over x_d}
\end{equation}
which must be satisfied in the Standard Model. 
Here $x_d$ is a $B_d^0-\bar B_d^0$ parameter.
We stress that except
for $P_0(X)$  all quantities in
(\ref{kbcon}) can be directly measured in experiment and that this
relationship is essentially independent of $m_t$ and $V_{cb}$.
Due to very small theoretical uncertainties in (\ref{kbcon}), this
relation is particularly suited for tests of CP violation in the
Standard Model and offers a powerful tool to probe the physics
beyond it.

\section{ Express Review of Rare K and B Decays}
         \label{sec:RareKB}
\setcounter{equation}{0}
\subsection{The Decays $B\to X_{s,d}\nu\bar\nu$}
            \label{sec:HeffRareKB:klpinn2}
The decays $B\to X_{s,d}\nu\bar\nu$ are the theoretically
cleanest decays in the field of rare $B$-decays.
They are dominated by the same $Z^0$-penguin and box diagrams
involving top quark exchanges which we encountered already
in the case of $\kpn$ and $\klpn$ except for the appropriate
change of the external quark flavours. Since the change of external
quark flavours has no impact on the $m_t$ dependence,
the latter is fully described by the function $X(x_t)$ in
(\ref{xx9}) which includes
the NLO corrections. The charm contribution 
is fully neglegible
here and the resulting effective Hamiltonian is very similar to
the one for $\klpn$ given in (\ref{hxnu}). 
For the decay $B\to X_s\nu\bar\nu$ it reads
\begin{equation}\label{bxnu}
{\cal H}_{\rm eff} = {G_{\rm F}\over \sqrt 2} {\alpha \over
2\pi \sin^2 \Theta_{\rm W}} V^\ast_{tb} V_{ts}
X (x_t) (\bar bs)_{V-A} (\bar\nu\nu)_{V-A} + h.c.   
\end{equation}
with $s$ replaced by $d$ in the
case of $B\to X_d\nu\bar\nu$.
 
The theoretical uncertainties related to the renormalization
scale dependence are as in $\klpn$ and 
can be essentially neglected. The same applies to long distance
contributions considered in \cite{BUC97}.
The calculation of the branching fractions for $B\to X_{s,d}\nu\bar\nu$ 
can be done in the spectator model corrected for short distance QCD effects.
Normalizing to $Br(B\to X_c e\bar\nu)$ and summing over three neutrino 
flavours one finds

\begin{equation}\label{bbxnn}
\frac{Br(B\to X_s\nu\bar\nu)}{Br(B\to X_c e\bar\nu)}=
\frac{3 \alpha^2}{4\pi^2\sin^4\Theta_{\rm W}}
\frac{|V_{ts}|^2}{|V_{cb}|^2}\frac{X^2(x_t)}{f(z)}
\frac{\kappa(0)}{\kappa(z)}\,.
\end{equation}
Here $f(z)$ is the phase-space factor for $B\to X_c
e\bar\nu$ with $z=\mc^2/\mb^2$  and $\kappa(z)=0.88$ 
\cite{CM78,KIMM} is the
corresponding QCD correction. The
factor $\kappa(0)=0.83$ represents the QCD correction to the matrix element
of the $b\to s\nu\bar\nu$ transition due to virtual and bremsstrahlung
contributions.
In the case of $B\to X_d\nu\bar\nu$ one has to replace $V_{ts}$ by
$V_{td}$ which results in a decrease of the branching ratio by
roughly an order of magnitude.

Setting $Br(B\to X_ce\bar\nu)=10.4\%$, $f(z)=0.54$,
$\kappa(z)=0.88$ and using the values in (\ref{alsinbr})
 we have
\begin{equation}
Br(B \to X_s \nu\bar\nu) = 3.7 \cdot 10^{-5} \,
\frac{|V_{ts}|^2}{|V_{cb}|^2} \,
\left[ \frac{\mtb(\mt)}{170\gev} \right]^{2.30} \, .
\label{eq:bxsnnnum}
\end{equation}
Taking next, 
$f(z)=0.54\pm 0.04$ and 
$Br(B\to X_ce\bar\nu)=(10.4\pm 0.4)\%$
and scanning the input parameters of table \ref{tab:inputparams}
we find
\begin{equation}\label{klpnr3}
Br(B \to X_s \nu\bar\nu)=
(3.5 \pm 0.7)\cdot 10^{-5}
\end{equation}
to be compared with the experimental upper bound:
\begin{equation}\label{124}
Br(B\to X_s \nu\bar\nu) < 7.7\cdot 10^{-4} 
\quad
(90\%\,\,\mbox{C.L.})
\end{equation}
obtained for the first time by ALEPH \cite{Aleph96}.
This is only a factor of 20 above the Standard Model expectation.
Even if the actual measurement of this decay is extremly difficult,
all efforts should be made to measure it. One should also 
make attempts to measure $Br(B\to X_d \nu\bar\nu)$. Indeed 

\begin{equation}\label{bratio}
\frac{Br(B\to X_d\nu\bar\nu)}{Br(B\to X_s\nu\bar\nu)}=
\frac{|V_{td}|^2}{|V_{ts}|^2}
\end{equation} 
offers the
cleanest direct determination of $\vtd/\vts$ as all uncertainties related
to $\mt$, $f(z)$ and $Br(B\to X_ce\bar\nu)$ cancel out.
\subsection{The Decays $B_{s,d}\to l^+l^-$}
The decays $B_{s,d}\to l^+l^-$ are after $B\to X_{s,d}\nu\bar\nu$ 
the theoretically cleanest decays in the field of rare $B$-decays.
They are dominated by the $Z^0$-penguin and box diagrams
involving top quark exchanges which we encountered already
in the case of $B\to X_{s,d}\nu\bar\nu$   except that due to
charged leptons in the final state the charge flow in the
internal lepton line present in the box diagram is reversed.
This results in a different $\mt$ dependence summarized
by the function  $Y(x_t)$, the NLO generalization \cite{BB2,MU98,BB98}
of the function $Y_0(x_t)$ given in (\ref{XA0}).
The charm contributions are fully negligible
here and the resulting effective Hamiltonian is given 
for $B_s\to l^+l^-$ as follows:
\begin{equation}\label{hyll}
{\cal H}_{\rm eff} = -{G_{\rm F}\over \sqrt 2} {\alpha \over
2\pi \sin^2 \Theta_{\rm W}} V^\ast_{tb} V_{ts}
Y (x_t) (\bar bs)_{V-A} (\bar ll)_{V-A} + h.c.   \end{equation}
with $s$ replaced by $d$ in the
case of $B_d\to l^+l^-$.

The function $Y(x)$ is given by
\begin{equation}\label{yyx}
Y(x_t) = Y_0(x_t) + \aspi Y_1(x_t) \equiv \eta_Y Y_0(x_t),
\qquad \eta_Y=1.012
\end{equation}
where $Y_1(x_t)$ can be found in \cite{BB2,MU98,BB98}.
The leftover $\mu_t$-dependence in $Y(x_t)$ is tiny and amounts to
an uncertainty of $\pm 1\%$ at the level of the branching ratio.
With $\mt\equiv \mtb(\mt)$ the QCD factor $\eta_Y$
depends only very weakly on $m_t$. 
The dependence on
$\Lambda_{\overline{MS}}$ can be neglected. 

The branching ratio for $B_s\to l^+l^-$ is given by \cite{BB2}
\begin{equation}\label{bbll}
Br(B_s\to l^+l^-)=\tau(B_s)\frac{G^2_{\rm F}}{\pi}
\left(\frac{\alpha}{4\pi\sin^2\Theta_{\rm W}}\right)^2 F^2_{B_s}m^2_l m_{B_s}
\sqrt{1-4\frac{m^2_l}{m^2_{B_s}}} |V^\ast_{tb}V_{ts}|^2 Y^2(x_t)
\end{equation}
where $B_s$ denotes the flavour eigenstate $(\bar bs)$ and $F_{B_s}$ is
the corresponding decay constant. Using
(\ref{alsinbr}) and (\ref{yyx})  we find in the
case of $B_s\to\mu^+\mu^-$
\begin{equation}\label{bbmmnum}
Br(B_s\to\mu^+\mu^-)=3.5\cdot 10^{-9}\left[\frac{\tau(B_s)}{1.6
\mbox{ps}}\right]
\left[\frac{F_{B_s}}{210\mev}\right]^2 
\left[\frac{|V_{ts}|}{0.040}\right]^2 
\left[\frac{\mtb(\mt)}{170\gev}\right]^{3.12}.
\end{equation}

The main uncertainty in this branching ratio results from
the uncertainty in $F_{B_s}$.
Scanning the input parameters of table \ref{tab:inputparams}
together with $\tau(B_s)=1.6$ ps and $F_{B_s}=(210\pm 30)\mev$ 
we find
\begin{equation}\label{klpnr1}
Br(B_s\to\mu^+\mu^-)=
(3.2 \pm 1.5)\cdot 10^{-9}~.
\end{equation}

For $B_d\to\mu^+\mu^-$ a similar formula holds with obvious
replacements of labels $(s\to d)$. Provided the decay constants
$F_{B_s}$ and $F_{B_d}$ will have been calculated reliably by
non-perturbative methods or measured in leading leptonic decays one
day, the rare processes $B_{s}\to\mu^+\mu^-$ and $B_{d}\to\mu^+\mu^-$
should offer clean determinations of $|V_{ts}|$ and $|V_{td}|$. 
In particular the ratio
\begin{equation}
\frac{Br(B_d\to\mu^+\mu^-)}{Br(B_s\to\mu^+\mu^-)}
=\frac{\tau(B_d)}{\tau(B_s)}
\frac{m_{B_d}}{m_{B_s}}
\frac{F^2_{B_d}}{F^2_{B_s}}
\frac{|V_{td}|^2}{|V_{ts}|^2}
\end{equation}
having smaller theoretical uncertainties than the separate
branching ratios should offer a useful measurement of
$\vtd/\vts$. Since $Br(B_d\to\mu^+\mu^-)= {\cal O}(10^{-10})$
this is, however, a very difficult task. For $B_s \to \tau^+\tau^-$
and $B_s\to e^+e^-$ one expects branching ratios ${\cal O}(10^{-6})$
and ${\cal O}(10^{-13})$, respectively, with the corresponding branching 
ratios for $B_d$-decays by one order of magnitude smaller.

The bounds on $B_{s,d}\to l\bar l$ are still
 many orders of magnitude away from Standard Model expectations.
The best bounds come from CDF \cite{CDFMU}. One has:
\begin{equation}\label{MUBOUND}
Br(B_s\to\mu^+\mu^-)\le 
2.6\cdot 10^{-6}~~~~~(95\% C.L.)
\end{equation}
and $Br(B_d\to\mu^+\mu^-)\le 8.6\cdot 10^{-7}$.
CDF should reach in Run II the
sensitivity of $1\cdot 10^{-8}$ and $4\cdot 10^{-8}$ for
$B_d\to \mu\bar\mu$ and $B_s\to \mu\bar\mu$, respectively.
It is hoped that these decays will be observed at
LHC-B. The experimental status of $B\to\tau^+\tau^-$ and its
usefulness in tests of the physics beyond the Standard Model
is discussed in \cite{GLN96}.

\subsection{$B \to X_s \gamma$ and $B\to X_s l^+l^-$}
In view of space limitations I will be very brief on
these two decays.

A lot of efforts have been put into predicting the branching
ratio for the inclusive radiative decay $B \to X_s \gamma$ including NLO
QCD corrections and higher order electroweak corrections. The
relevant references can be found in \cite{AJBLH,ZIMG},
where also theoretical details are given. The final result of these efforts
can be summarized by
\be\label{bsg}
Br(B \to X_s \gamma)_{\rm th}
=( 3.30 \pm 0.15 ({\rm scale})\pm 0.26 ({\rm par}))
                     \cdot 10^{-4}
\ee
where the first error represents residual scale dependences and
the second error is due to uncertainties in input parameters.
The main achievement is the reduction of the scale dependence
through NLO calculations, in particular those given in 
\cite{GREUB} and \cite{CZMM}. In
the leading order the corresponding error would be roughly
$\pm 0.6$.

The theoretical result in (\ref{bsg})
 should be compared with experimental data:
\begin{equation}\label{bsgexp}
 Br(B \to X_s \gamma)_{\rm exp}=\left\{ \begin{array}{ll}
(3.15 \pm 0.35 \pm 0.41)\cdot 10^{-4}~, &~~{\rm CLEO} \\
(3.11 \pm 0.80 \pm 0.72)\cdot 10^{-4}~, & ~~{\rm ALEPH} ,
\end{array} \right.
\end{equation}
which implies the combined branching ratio:
\begin{equation}\label{bsgex}
 Br(B \to X_s \gamma)_{\rm exp}=
(3.14 \pm 0.48)\cdot 10^{-4}~.
\end{equation}
Clearly, the Standard Model result agrees well with the data. In
order to see whether any new physics can be seen in this decay,
the theoretical and in particular experimental errors should
be reduced. This is certainly a very difficult task.

The rare decays $B\to X_{s,d} l^+l^-$ have been the subject of
many theoretical studies.
It is clear that once these decays
have been observed, they will offer useful tests of the
Standard Model and of its extentions. Most recent reviews
can be found in \cite{babar,GUDALI}.

\subsection{$K_L \to \pi^0 e^+e^-$}
There are three contributions to this decay: CP conserving,
indirectly CP violating and directly CP violating.
Unfortunately out of these three contributions only the
directly CP violating can be calculated reliably.
Including NLO corrections \cite{BLMM} and scanning the input parameters 
of table \ref{tab:inputparams}
we find
\be
  Br(\kpe)_{\rm dir}=(4.6 \pm 1.8) \cdot 10^{-12}\,, 
  \label{eq:brkpep}
\ee
where the errors come dominantly from the uncertainties in the CKM
parameters. 
The
remaining two contributions to this decay
are plagued by theoretical
uncertainties \cite{KL}. 
They are expected to be $\ord (10^{-12})$ but generally
smaller than $Br(\kpe)_{\rm dir}$. This implies that within the
Standard Model $Br(\kpe)$ is expected to be at most $10^{-11}$.

Experimentally we have 
 the bound \cite{ke} 
\begin{equation}
  \label{eq:brklexp}
  Br(\kpe) < 4.3 \cdot 10^{-9}.
\end{equation}
and considerable improvements are expected in the coming
years.

\subsection{$\kmm$}
The $\kmm$ branching ratio can be decomposed generally as follows:
\begin{equation}
  \label{eq:deckmm}
  BR(\kmm)=\vert \RE A \vert^2 + \vert \IM A \vert^2\,,
\end{equation}
where $\RE A$ denotes the dispersive contribution and $\IM A$ the
absorptive one. The latter contribution can be determined in a model
independent way from the $K_L \to \gamma \gamma$ branching ratio. The
resulting $\vert \IM A \vert^2$ is very close to the experimental
branching ratio $Br(\kmm)=(7.2 \pm 0.5) \cdot 10^{-9}$ 
\cite{mpmm} so that $\vert
\RE A \vert^2$ is substantially smaller and extracted to be \cite{mpmm}
\begin{equation}
  \label{eq:reaexp}
  \vert \RE A_{\rm exp} \vert^2 < 5.6 \cdot 10^{-10} \qquad {\rm 
  (90\% \, \,C.L.).}
\end{equation}
Now $\RE A$ can be decomposed as 
\begin{equation}
  \label{eq:decomp}
  \RE A = \RE A_{\rm LD} + \RE A_{\rm SD}\,,  
\end{equation}
with 
\begin{equation}
  \label{eq:reasdbr}
  \vert \RE A_{\rm SD} \vert^2 \equiv Br(\kmm)_{\rm SD}
\end{equation}
representing the short-distance contribution 
which can be calculated reliably.
An
improved estimate of the long-distance contribution $\RE A_{\rm LD}$
has been recently presented in
\cite{dambrosio}
\begin{equation}
  \label{eq:reald}
  \vert \RE A_{LD} \vert < 2.9 \cdot 10^{-5} \qquad {\rm (90\% \,\,C.L.).} 
\end{equation}
Together with \r{eq:reaexp} this gives 
\begin{equation}
  \label{eq:sdlimitbr}
  Br(\kmm)_{\rm SD} < 2.8 \cdot 10^{-9}.
\end{equation}
This result is very close to the one presented 
by Gomez Dumm and Pich \cite{pich}. 
More pesimistic view on the extraction of the short
distance part from $Br(\kmm)$  can be found in \cite{GVAL}.

The bound in (\ref{eq:sdlimitbr})
should be compared with the short distance contribution within
the Standard Model for which we find
\begin{equation}
\label{kmusm}
 Br(\kmm)_{\rm SD}=(8.7\pm 3.6)\cdot 10^{-10}.
\end{equation}
This implies that there is a considerable room for new physics
contributions. We will return to this point in section 9.
Reviews of rare K decays are listed in \cite{CPRARE}.
\section{Express Review of CP Violation in B Decays}
\setcounter{equation}{0}
\subsection{CP-Asymmetries in B-Decays: General Picture}
CP violation in B-decays is certainly one of the most important 
targets of B-factories and of dedicated B-experiments at hadron 
facilities. It is well known that CP violating effects are expected
to occur in a large number of channels at a level attainable at 
forthcoming experiments. Moreover there exist channels which
offer the determination of CKM phases essentially without any hadronic
uncertainties. Since extensive reviews on CP violation in B decays can 
be found in the literature \cite{NQ,RF97,BF97} and I am running out
of space, 
let me concentrate only on a few points beginning with a quick
review of classic methods for the determination of the angles
$\alpha$, $\beta$ and $\gamma$ in the unitarity triangle.

The classic determination of $\alpha$ by means of the
time dependent CP  asymmetry in the decay
$B_d^0 \rightarrow \pi^+ \pi^-$ 
is affected by the "QCD penguin pollution" which has to be
taken care of in order to extract $\alpha$. 
The recent CLEO results for penguin dominated decays indicate that
this pollution could be substantial as stressed in particular
in \cite{ITAL}.
The most popular strategy to deal with this "penguin problem''
is the isospin analysis of Gronau and London \cite{CPASYM}. It
requires however the measurement of $Br(B^0\to \pi^0\pi^0)$ which is
expected to be below $10^{-6}$: a very difficult experimental task.
For this reason several, rather involved, strategies \cite{SNYD} 
have been proposed which
avoid the use of $B_d \to \pi^0\pi^0$ in conjunction with
$a_{CP}(\pi^+\pi^-,t)$. They are reviewed in \cite{BF97}. 
 It is to be seen which of these methods
will eventually allow us to measure $\alpha$ with a respectable precision.
It is however clear that the determination of this angle is a real
challenge for both theorists and experimentalists.

The CP-asymmetry in the decay $B_d \rightarrow \psi K_S$ allows
 in the Standard Model
a direct measurement of the angle $\beta$ in the unitarity triangle
without any theoretical uncertainties \cite {BSANDA}.
Of considerable interest \cite{RF97,PHI} is also the pure penguin decay
$B_d \rightarrow \phi K_S$, which is expected to be sensitive
to physics beyond the Standard Model. Comparision of $\beta$
extracted from $B_d \rightarrow \phi K_S$ with the one from
$B_d \rightarrow \psi K_S$ should be important in this
respect. An analogue of $B_d \rightarrow \psi K_S$ in $B_s$-decays
is $B_s \rightarrow \psi \phi$. The CP asymmetry measures here
$\eta$ \cite{B95} in the Wolfenstein parametrization. It is very
small, however, and this fact makes it a good place to look for the 
physics beyond the Standard Model. In particular the CP violation
in $B^0_s-\bar B^0_s$ mixing from new sources beyond the Standard
Model should be probed in this decay.

The two theoretically cleanest methods for the determination of $\gamma$
are: i) the full time dependent analysis of 
$B_s\to D^+_s K^{-}$ and $\bar B_s\to D^-_s K^{+}$  \cite{adk}
and ii) the well known triangle construction due to Gronau and Wyler 
\cite{Wyler}
which uses six decay rates $B^{\pm}\to D^0_{CP} K^{\pm}$,
$B^+ \to D^0 K^+,~ \bar D^0 K^+$ and  $B^- \to D^0 K^-,~ \bar D^0 K^-$.
Both methods are  unaffected by penguin contributions. 
The first method is experimentally very
challenging because of the
expected large $B^0_s-\bar B^0_s$ mixing. The second method is problematic
because of the small
branching ratios of the colour supressed channel $B^{+}\to D^0 K^{+}$
and its charge conjugate,
giving a rather squashed triangle and thereby
making
the extraction of $\gamma$ very difficult. Variants of the latter method
which could be more promising have been proposed in \cite{DUN2,V97}.
It appears that these methods will give useful results at later stages
of CP-B investigations. In particular the first method will be feasible
only at LHC-B. Other recent strategies for $\gamma$ will be mentioned 
below.

\subsection{$B^0$-Decays to CP Eigenstates}
Let us demonstrate some of the statements made above explicitly.

A time dependent asymmetry in the decay $B^0\to f$ with $f$ being
a CP eigenstate is given by
\begin{equation}\label{e8}
a_{CP}(t,f)=
{\cal A}^{dir}_{CP}(B\to f)\cos(\Delta M
t)+{\cal A}^{mix-ind}_{CP}(B\to f)\sin(\Delta M t)
\end{equation}
where we have separated the {\it direct} CP-violating contributions 
from those describing {\it mixing-induced} CP violation:
\begin{equation}\label{e9}
{\cal A}^{dir}_{CP}(B\to f)\equiv\frac{1-\left\vert\xi_f\right\vert^2}
{1+\left\vert\xi_f\right\vert^2},
\qquad
{\cal A}^{mix-ind}_{CP}(B\to f)\equiv\frac{2\mbox{Im}\xi_f}{1+
\left\vert\xi_f\right\vert^2}.
\end{equation}
In (\ref{e8}), $\Delta M$ denotes the mass splitting of the 
physical $B^0$--$\bar B^0$--mixing eigenstates. 
The quantity $\xi_f$ containing essentially all the information
needed to evaluate the asymmetries (\ref{e9}) 
is given by
\begin{equation}\label{e11}
\xi_f=\exp(i2\phi_M)\frac{A(\bar B\to f)}{A(B \to f)}
\end{equation}
with $\phi_M$ denoting the weak phase in the $B-\bar B$ mixing
and $A(B \to f)$ the decay amplitude. 

Generally several decay mechanisms with different weak and
strong phases can contribute to $A(B \to f)$. These are
tree diagram (current-current) contributions, QCD penguin
contributions and electroweak penguin contributions. If they
contribute with similar strength to a given decay amplitude
the resulting CP asymmetries suffer from hadronic uncertainies
related to matrix elements of the relevant operators $Q_i$.

An interesting case arises when a single mechanism dominates the 
decay amplitude or the contributing mechanisms have the same weak 
phases. Then
\begin{equation}\label{e111}
\xi_f=\exp(i2\phi_M) \exp(-i 2 \phi_D),
\qquad
\mid \xi_f \mid^2=1
\end{equation}
where $\phi_D$ is the weak phase in the decay amplitude.
In this particular case the hadronic matrix elements drop out,
 the direct CP violating contribution
vanishes and the mixing-induced CP asymmetry is given entirely
in terms of the weak phases $\phi_M$ and $\phi_D$. In particular
the time integrated asymmetry is given by
\begin{equation}
a_{CP}(f)=\pm \sin(2\phi_D-2\phi_M)
\frac{x_{d,s}}{1+x_{d,s}^2}
\end{equation}
where $\pm$ refers to $f$ being a $CP=\pm$ eigenstate and $x_{d,s}$ are
the $B_{d,s}^0-\bar B_{d,s}^0$ mixing parameters.

If a single tree diagram dominates, the factor $\sin(2\phi_D-2\phi_M)$
can be calculated by using
\begin{equation}\label{dtree}
\phi_D =\left\{ \begin{array}{ll}
\gamma & b\to u \\
0 & b\to c \end{array} \right.
\qquad
\phi_M =\left\{ \begin{array}{ll}
-\beta & B^0_d \\
0 & B^0_s \end{array} \right.
\end{equation}
where we have indicated the basic transition of the b-quark into
a lighter quark.
On the other hand if the penguin diagram with internal
top exchange dominates one has
\begin{equation}\label{pen}
\phi_D =\left\{ \begin{array}{ll}
-\beta & b\to d \\
0 & b\to s \end{array} \right.
\qquad
\phi_M =\left\{ \begin{array}{ll}
-\beta & B^0_d \\
0 & B^0_s \end{array} \right.
\end{equation}
These rules have been obtained using Wolfenstein parametrization
in the leading order.
Let us practice these formulae.
Assuming that 
$B_d\to \psi K_S$ and $B_d\to \pi^+\pi^-$ are dominated by
tree diagrams with $b\to c$ and $b\to u$ transitions respectively
we readly find
\begin{equation}\label{113c}
 a_{CP}(\psi K_S)=-\sin(2\beta) \frac{x_d}{1+x_d^2},
\ee
\be\label{113d}
   a_{CP}(\pi^+\pi^-)=-\sin(2\alpha) \frac{x_d}{1+x_d^2}.
\end{equation}
Now in the case of $B_d\to \psi K_S$ the penguin diagrams have
to a very good approximation the same phase ($\phi_D=0)$ as
the tree contribution and moreover are Zweig suppressed.
Consequently (\ref{113c}) is very accurate. This is not
the case for $B_d\to \pi^+\pi^-$ where the penguin contribution
could be substantial. Heaving weak phase $\phi_D=-\beta$,
which differs from the tree phase $\phi_D=\gamma$, this penguin
contribution changes effectively (\ref{113d}) to
\be\label{113e}
   a_{CP}(\pi^+\pi^-)=-\sin(2\alpha+\theta_P) \frac{x_d}{1+x_d^2}
\end{equation}
where $\theta_P$ is a function of $\beta$ and hadronic parameters.
The isospin analysis \cite{CPASYM} mentioned before is supposed 
to determine $\theta_P$ so that $\alpha$ can be extracted
from $a_{CP}(\pi^+\pi^-)$.

Similarly the pure penguin dominated decay $B_d\to \phi K_S$
is governed by the $b\to s$ penguin  with internal top exchange
which implies that in this decay the angle $\beta$ is measured.
The accuracy of this measurement is a bit lower than using 
$B_d\to \psi K_S$ as penguins with internal u and c exchanges
may introduce a small pollution.

Finally we can consider the asymmetry in $B_s\to\psi\phi$, 
an analog of $B_d \to \psi K_s$. In the leading order of the
Wolfenstein parametrization the asymmetry $a_{CP}(\psi\phi)$ vanishes.
Including higher order terms in $\lambda$ one finds \cite{B95}
\begin{equation}\label{DU}
a_{CP}(\psi\phi)=2\lambda^2\eta\frac{x_s}{1+x_s^2}
\end{equation}
where $\lambda$ and $\eta$ are the Wolfenstein parameters.

\subsection{Recent Developments}
All this has been known already for some time and is well documented
in the literature.
The most recent developments are related to the extraction of
the angle $\gamma$ from the decays $B\to PP$ (P=pseudoscalar) 
 and their charge conjugates  
\cite{FM}--\cite{GPAR1}. Some of these modes have been observed 
by the CLEO collaboration \cite{cleo}. In the future they should allow 
us to obtain direct 
information on $\gamma$ at $B$-factories (BaBar, 
BELLE, CLEO III) (for interesting feasibility studies, see 
\cite{GRRO,wuegai,babar}). At present, there are only experimental results 
available for the combined branching ratios of these modes, i.e.\ averaged 
over decay and its charge conjugate, suffering from large hadronic
uncertainties. 

There has been large activity in this field during the last two years.
The main issues here are the final state interactions, SU(3) symmetry
breaking effects and the importance of electroweak penguin
contributions. Several interesting ideas have been put forward
to extract the angle $\gamma$ in spite of large hadronic
uncertainties in $B\to \pi K$ decays \cite{FM,GRRO}. 
Also other $B\to PP$
decays have been investigated. As this field became rather
technical, I decided not to include it in these lectures.
A subset of relevant papers is listed in 
\cite{FM,GRRO,GPAR1,FSI,GPAR2,NRBOUND}, where 
further
references can be found. 
In particular in \cite{GPAR1,GPAR2} general parametrizations for the 
study of the final state interactions, SU(3) symmetry
breaking effects and the importance of electroweak penguin
contributions have been presented. Moreover, upper bounds
on the latter contributions following from SU(3) symmetry
have been derived \cite{NRBOUND}.
Recent reviews can be found in \cite{RFA} and \cite{NEU99}.
 New
strategies for $\gamma$ which include $B_s\to\psi K_S$ and
$B_s \to K^+ K^-$ have been suggested very recently in 
\cite{RF99}.

There is no doubt that these
new ideas will be helpful in the future. 
They are, however, rather demanding for experimentalist as
often several branching ratios have to be studied
simultaneously and each has to be measured precisely
in order to obtain an acceptable measurement of $\gamma$. 
On the other hand various suggested bounds on $\gamma$ may either 
exclude the region around $90^\circ$ \cite{FM} or give an
improved lower bound on it \cite{NRBOUND,NEU99,GAWU} which would
remove a large portion of the allowed range from the analysis
of the unitarity triangle. In this context it has been pointed
out in \cite{HHY} (see also \cite{GAWU})
that generally charmless hadronic B decay
results from CLEO seem to prefer negative values of $\cos\gamma$ which
is not the case in the standard analysis of section 4.

Finally I would like to mention a recent interesting paper of Lenz,
Nierste and Ostermaier \cite{LNO}, 
where inclusive direct CP-asymmetries in 
charmless $B^{\pm}$-decays including QCD effects have been studied.
These asymmetries should offer additional useful means to constrain 
the unitarity
triangle.

\subsection{CP-Asymmetries in $B$-Decays versus $K \to \pi \nu\bar\nu$}
Let us next compare the potentials of the CP asymmetries in
determining the parameters of the Standard Model with those
of the cleanest rare $K$-decays: $K_{\rm L}\to\pi^0\nu\bar\nu$ and
$K^+\to\pi^+\nu\bar\nu$.

Measuring $\sin 2\alpha$ and $\sin 2\beta$ from CP asymmetries in
$B$ decays allows, in principle, to fix the 
parameters $\bar\eta$ and $\bar\varrho$, which can be expressed as
\cite{AJB94}
\begin{equation}\label{ersab}
\bar\eta=\frac{r_-(\sin 2\alpha)+r_+(\sin 2\beta)}{1+
r^2_+(\sin 2\beta)}\,,\qquad
\bar\varrho=1-\bar\eta r_+(\sin 2\beta)\,,
\end{equation}
where $r_\pm(z)=(1\pm\sqrt{1-z^2})/z$.
In general the calculation of $\bar\varrho$ and $\bar\eta$ from
$\sin 2\alpha$ and $\sin 2\beta$ involves discrete ambiguities.
As described in \cite{AJB94}
they can be resolved by using further information, e.g.\ bounds on
$|V_{ub}/V_{cb}|$, so that eventually the solution (\ref{ersab})
is singled out.

Let us then consider two scenarios of the measurements of CP asymmetries 
in $B_d\to\pi^+\pi^-$ and $B_d\to J/\psi K_{\rm S}$, expressed in terms 
of $\sin 2\alpha$ and
$\sin 2\beta$:
\begin{equation}\label{sin2a2bI}
\sin 2\alpha=0.40\pm 0.10\,, \qquad \sin 2\beta=0.70\pm 0.06
\qquad ({\rm scenario\ I})
\end{equation}
\begin{equation}\label{sin2a2bII}
\sin 2\alpha=0.40\pm 0.04\,, \qquad \sin 2\beta=0.70\pm 0.02
\qquad ({\rm scenario\ II})\,.
\end{equation}
Scenario I corresponds to the accuracy being aimed for at $B$-factories
and HERA-B prior to the LHC era. An improved precision can be anticipated from
LHC experiments, which we illustrate with the scenario II.
We assume that the problems with the determination
of $\alpha$ will be solved somehow.

In table \ref{tabkb} this way of the determination of
the Standard Model parameters is compared \cite{BB96} 
with the analogous analysis
using $\klpn$ and $\kpn$ which has been presented in section 6. 
As can be seen in table \ref{tabkb}, the CKM determination
using $K\to\pi\nu\bar\nu$ is competitive with the one based
on CP violation in $B$ decays in scenario I, except for $\bar\varrho$ which
is less constrained by the rare kaon processes.
On the other hand as advertised previously ${\rm Im}\lambda_t$ 
is better determined
in $K\to\pi\nu\bar\nu$ even if scenario II is considered.
The virtue of the comparision of the determinations
of various parameters using CP-B asymmetries with the determinations
in very clean decays $K\to\pi\nu\bar\nu$ is that any substantial deviations
from these two determinations would signal new physics beyond the
Standard Model.
 Formula (\ref{kbcon}) is an example of such a comparison.
There are other strategies for determination of the unitarity
triangle using combinations of CP asymmetries and rare
decays. They are reviewed in \cite{AJBLH}.

\begin{table}
\caption[]{Illustrative example of the determination of CKM
parameters from $K\to\pi\nu\bar\nu$ and B-decays.
We use $\sigma(\vcb)=\pm0.002(0.001)$.
\label{tabkb}}
\vspace{0.4cm}
\begin{center}
\begin{tabular}{|c|c|c|c|}\hline
&$K\to\pi\nu\bar\nu$ 
& {\rm Scenario I} & {\rm Scenario II}
\\ 
\hline
$\sigma(|V_{td}|) $& $\pm 10\% (9\% )$
& $\pm 5.5\% (3.5\%)$ & $\pm 5.0\% (2.5\%)$\\ 
\hline 
$\sigma(\bar\varrho) $ & $\pm 0.16 (0.12)$
& $\pm 0.03$  & $\pm 0.01$\\
\hline
$\sigma(\bar\eta)$ & $\pm 0.04(0.03)$
&$\pm 0.04 $ & $\pm 0.01 $\\
\hline
$\sigma(\sin 2\beta)$ & $\pm 0.05$
& $\pm 0.06 $ & $\pm 0.02$\\
\hline
$\sigma({\rm Im}\lambda_t)$&$\pm 5\%$ 
& $\pm 14\%(11\%)$ & $\pm 10\%(6\%)$\\
\hline
\end{tabular}
\end{center}
\end{table}

\section{A Brief Look Beyond the Standard Model}
\setcounter{equation}{0}
\subsection{General Remarks}
We begin the discussion of the Physics beyond the Standard
Model with a few general remarks. As the new particles
in the extensions of the Standard Model
are generally substantally heavier than $W^\pm$, the
impact of new physics on charged current tree level decays should
be marginal. On the other hand these new contributions
could have in principle an important impact on
loop induced decays. From these two observations we
conclude:

\bi
\item
New physics should have only marginal impact on the determination
of $|V_{us}|$, $\vcb$ and $|V_{ub}|$.
\item
There is no impact on the calculations of the low energy non-perturbative
parameters $B_i$ except that new physics can bring new local
operators implying new parameters $B_i$.
\item
New physics could have substantial impact on rare and CP violating
decays and consequently on the determination of the unitarity triangle.
\ei

\subsection{Classification of New Physics}
Let us then group the extensions of the Standard Model in three
classes.

{\bf Class A}

\bi
\item
 There are no new complex phases and quark mixing is described
 by the CKM matrix.
\item
 There are new contributions to rare and CP violating decays
 through diagrams involving new internal particles.
\ei
 These new contributions will have impact on the determination
 of $\alpha$, $\beta$, $\gamma$, $\vtd$ and $\lambda_t$ and
 will be signaled by
\bi
\item
Inconsistencies in the determination of $(\bar\varrho,\bar\eta)$
through $\varepsilon$, $B^0_{s,d}-\bar B^0_{s,d}$ mixing and
rare decays.
\item
Disagreement of $(\bar\varrho,\bar\eta)$ extracted 
from loop induced decays
with $(\bar\varrho,\bar\eta)$ extracted using
CP asymmetries.
\ei
Examples are two Higgs doublet model II and the constrained MSSM. 

{\bf Class B}

\bi
\item
 Quark mixing is described by the CKM matrix.
\item
 There are new phases in the new contributions to rare and 
 CP violating decays.
 \ei
 This kind of new physics will also be signaled by inconsistencies
 in the $(\bar\varrho,\bar\eta)$ plane. However, new complication
 arises. Because of new phases CP violating asymmetries measure
 generally different quantities than $\alpha$, $\beta$ and $\gamma$.
 For instance the CP asymmetry in $B\to \psi K_S$ will no longer
 measure $\beta$ but $\beta+\theta_{NP}$ where $\theta_{NP}$
 is a new phase. Strategies for dealling with such situation
 have been developed. See for instance \cite{NIR96,BNEW} 
and references  therein.

 Examples are multi-Higgs models with complex phases in the
 Higgs sector, general SUSY models, models with spontaneous
 CP violation and left-right symmetric  models.

 {\bf Class C}

\bi
\item
The unitarity of the three generation CKM matrix does not
hold. 
\ei
Examples are four generation models and models with tree
level FCNC transitions. If this type of physics is present,
the unitarity triangle will not close or some inconsistencies
in the $(\bar\varrho,\bar\eta)$ plane take place.

Clearly in order to sort out which type of new physics is
responsible for deviations from the Standard Model
expectations one has to study many loop induced decays
and many CP asymmeteries. Some ideas in this direction
can be found in \cite{BNEW,NIR96}. 

\subsection{Upper Bounds on $K\to\pi\nu\bar\nu$ and $K_L\to\pi^0 e^+e^-$
from $\epe$ and $K_L\to\mu^+\mu^-$}
\label{sec:intro}
We have seen in previous sections that
 the rare kaon decays $\klpn$, $\kppn$ and $\kpe$
are governed by $Z$-penguin diagrams. Within the Standard Model
the branching ratios for these decays 
have been found to be
\begin{eqnarray}
  \label{eq:brklpn}
  Br(\klpn)&=&(2.8 \pm 1.1) \cdot 10^{-11}\,, \\
  \label{eq:brkpnn}
  Br(\kppn)&=&(7.9 \pm 3.1) \cdot 10^{-11}\,, \\
  Br(\kpe)_{\rm dir}&=&(4.6 \pm 1.8) \cdot 10^{-12}\,, 
  \label{eq:brkpe}
\end{eqnarray}
where the errors come dominantly from the uncertainties in the CKM
parameters. 
The branching ratio in (\ref{eq:brkpe}) represents
the so-called direct CP-violating contribution to $\kpe$. The
remaining two contributions to this decay, the CP-conserving one and
the indirect CP-violating one are plagued by theoretical
uncertainties \cite{KL}. 
They are expected to be $\ord (10^{-12})$ but generally
smaller than $Br(\kpe)_{\rm dir}$. This implies that within the
Standard Model $Br(\kpe)$ is expected to be at most $10^{-11}$.

In this context a very interesting claim has been made by
Colangelo and Isidori \cite{ISI}, who analyzing rare kaon decays in
supersymmetric theories pointed out a possible large enhancement of
the effective $\bar s d Z$ vertex leading 
to an enhancement of $Br(\kppn)$ by one order of
magnitude and of $Br(\klpn)$ and $Br(\kpe)$ by two orders of magnitude
relative to the Standard Model expectations.
Not surprisingly these results brought a lot of
excitement among experimentalists. 

Whether substantial enhancements of the branching ratios in question
are indeed possible in supersymmetric theories is being investigated
at present. On the other hand it can be shown \cite{BSII} that
in models
in which the dominant new effect is an enhanced $\bar s d Z$ vertex,
enhancements 
of $Br(\klpn)$ and $Br(\kpe)$ as large as claimed in \cite{ISI} 
are already excluded by the existing
data on $\epe$ in spite of large theoretical
uncertainties. Similarly the large enhancement of $Br(\kppn)$ can be
excluded by the data on $\epe$ and in particular by the present
information on the short distance contribution to $\kmm$. The latter
can be bounded by analysing the data on $Br(\kmm)$ in conjunction with
improved estimates of long distance dispersive contributions 
\cite{dambrosio,pich}. 
In \cite{ISI} only constraints from
$\kmm$, the $K_L$--$K_S$ mass difference $\Delta
M_K$ and $\eps$ have been taken into account. 
As $\epe$  depends sensitively on the size of $Z$-penguin
contributions and generally on the size of the effective $\bar s d Z$
vertex it is clear that the inclusion of the constraints from
$\epe$ should have an important impact on the bounds for the
rare decays in question.
I will only describe the basic idea of \cite{BSII} and give
numerical results. The relevant expressions can be found in
this paper. Here we go.

In the Standard Model $Z$-penguins are represented by the function
$C_0$ which enters the functions $X_0$, $Y_0$ and $Z_0$.
In order to study the effect of an enhanced $\bar s d Z$ vertex
one simply makes the following replacement in the formulae
for $\epe$, $\kmm$ and rare decays in question:
\begin{equation}
  \label{eq:Wsm}
   \lambda_t C_0(x_t)~ \Longrightarrow~  Z_{ds} 
\end{equation}
where $Z_{ds}$ denotes an effective $\bar s d Z$ vertex.
The remaining contributions to $\epe$, $\kmm$ and rare K decays
are evaluated in the Standard model as we assume that they
are only marginally affected by new physics. We will, however,
consider three scenarios for $\lambda_t$, which enters these
remaining contributions.

Indeed there is the possibility that the value of $\lambda_t$
 is modified
by new contributions to $\eps$ and $B_{d,s}^0-\bar B_{d,s}^0$
mixings. We consider therefore three scenarios:
\bi
\item
{\bf Scenario A}: $\lambda_t$ is taken from the standard analysis of
the unitarity triangle
\item
{\bf Scenario B}: $\IM\lambda_t=0$ and $\RE\lambda_t$ is varied
            in the full range consistent with the unitarity
            of the CKM matrix. In this scenario CP violation
            comes entirely from new physics contributions.
\item
{\bf Scenario C}: $\lambda_t$ is varied
            in the full range consistent with the unitarity
            of the CKM matrix. This means in particular that
            $\IM\lambda_t$ can be negative.
\ei

\begin{table}[htbp]
  \caption{Upper bounds for the rare decays $\klpn$, $\kpe$ and
    $\kppn$, obtained in various scenarios by imposing 
     $\epe\ge 2.5 \cdot 10^{-3}$, in the case $\IM Z_{ds}>0$. 
  \label{tab:rarepo1}}
\begin{center}
  \begin{tabular}{|c|c|c|c|c|}
  \hline 
  Scenario &  A & B  & C & SM  \\ \hline 
  $Br(\klpn)[10^{-10}]$ & $0.5$ &$-$ & $0.7$ & 0.4 \\ \hline 
  $Br(\kpe)[10^{-11}]$  & $0.8$ &$-$ & $1.0$ & 0.7  \\ \hline 
  $Br(\kppn)[10^{-10}]$ & $1.8$ &$-$ & $2.2$ & 1.1  \\ \hline 
  \end{tabular}
\end{center}
\end{table}

\begin{table}[htbp]
  \caption{Upper bounds for the rare decays $\klpn$, $\kpe$ and
    $\kppn$, obtained in various scenarios by imposing 
     $\epe\ge 1.5 \cdot 10^{-3}$, in the case $\IM Z_{ds}>0$. 
  \label{tab:rarepo2}}
\begin{center}
  \begin{tabular}{|c|c|c|c|c|}
  \hline 
  Scenario &  A & B  & C & SM  \\ \hline 
  $Br(\klpn)[10^{-10}]$ & $1.1$ &$-$ & $1.2$ & 0.4 \\ \hline 
  $Br(\kpe)[10^{-11}]$  & $1.5$ &$-$ & $1.8$ & 0.7  \\ \hline 
  $Br(\kppn)[10^{-10}]$ & $1.9$ &$-$ & $2.3$ & 1.1  \\ \hline 
  \end{tabular}
\end{center}
\end{table}

Now $Z_{ds}$ is a complex number. $\IM Z_{ds}$ can be best
bounded by $\epe$. This implies bounds for $Br(\klpn)$ and $Br(\kpe)$
which are sensitive functions of $\IM Z_{ds}$. 
$\RE Z_{ds}$ can be bounded by the present
information on the short distance contribution to $\kmm$. This
bound implies a bound on $Br(\kppn)$. Since $Br(\kppn)$
depends on both $\RE Z_{ds}$ and $\IM Z_{ds}$ also the bound
on $\IM Z_{ds}$ from $\epe$ matters in cases where $\IM Z_{ds}$
is very enhanced over the Standard Model value.

The branching ratios $Br(\klpn)$ and $Br(\kpe)$ are dominated
by  $(\IM Z_{sd})^2$.
Yet, the outcome of this analysis depends sensitively on the
sign of $\IM Z_{sd}$. Indeed, $\IM Z_{sd}>0$
results in the suppression of $\epe$ and as in the
Standard Model the value for $\epe$ is generally below the data
substantial enhancements of $\IM Z_{sd}$ with $\IM Z_{sd}>0$
are not possible. The situation changes if new physics
reverses the sign of $\IM Z_{sd}$ so that it becomes negative.
Then the upper bound on $\IM Z_{sd}$ is governed by the
upper bound on $\epe$ and with suitable choice of hadronic
parameters and $\IM\lambda_t$ (in particular in scenario C)
large enhancements of $-\IM Z_{sd}$ and of rare decay 
branching ratios are possible. 
The largest branching ratios are found when
the neutral meson mixing is
dominated by new physics contributions 
which force $\IM\lambda_t$ to be as negative as possible within
the unitarity of the CKM matrix.
This possibility
is quite remote. However, if
this situation could be realized in some exotic model, then 
the branching ratios in question  could be very  high.

In table~\ref{tab:rarepo1}  we show the upper bounds on 
rare decays for
$\IM Z_{sd}>0$ for three scenarios in question
and $\epe\ge 2.5 \cdot 10^{-3}$. In table~\ref{tab:rarepo2}
 the corresponding
bounds for $\epe\ge 1.5 \cdot 10^{-3}$ are given. To this
end all parameters relevant for $\epe$ have been scanned
in the ranges used in section 5. 
In tables \ref{tab:rarepo3} and \ref{tab:rarepo4}
the case $\IM Z_{sd}<0$ for $\epe\le 2.0 \cdot 10^{-3}$ and
$\epe\le 3.0 \cdot 10^{-3}$ is considered respectively.
In the last column we always give the upper bounds obtained
in the Standard Model. Evidently for positive $\IM Z_{sd}$ 
the enhancement of branching ratios are moderate but
they can be very large when $\IM Z_{sd}<0$.

\begin{table}[htbp]
\caption{Upper bounds for the rare decays $\klpn$, $\kpe$ and
    $\kppn$, obtained in various scenarios by imposing 
     $\epe\le 2.0 \cdot 10^{-3}$, in the case $\IM Z_{ds}<0$. 
  \label{tab:rarepo3}}
\begin{center}
  \begin{tabular}{|c|c|c|c|c|}
  \hline 
  Scenario &  A & B  & C & SM  \\ \hline 
  $BR(\klpn)[10^{-10}]$ & $1.3$ &$2.9$ & $11.2$ & 0.4 \\ \hline 
  $BR(\kpe)[10^{-11}]$  & $2.9$ &$5.1$ & $18.2$ & 0.7  \\ \hline 
  $BR(\kppn)[10^{-10}]$ & $2.0$ &$2.7$ & $4.6$ & 1.1  \\ \hline 
  \end{tabular}
  \end{center}
\end{table}

\begin{table}[htbp]
  \caption{Upper bounds for the rare decays $\klpn$, $\kpe$ and
    $\kppn$, obtained in various scenarios by imposing 
     $\epe\le 3.0 \cdot 10^{-3}$, in the case $\IM Z_{ds}<0$. 
  \label{tab:rarepo4}}
\begin{center}
  \begin{tabular}{|c|c|c|c|c|}
  \hline 
  Scenario &  A & B  & C & SM  \\ \hline 
  $BR(\klpn)[10^{-10}]$ & $3.9$ &$6.5$ & $17.6$ & 0.4 \\ \hline 
  $BR(\kpe)[10^{-11}]$  & $7.9$ &$11.5$ & $28.0$ & 0.7  \\ \hline 
  $BR(\kppn)[10^{-10}]$ & $2.6$ &$3.5$ & $6.1$ & 1.1  \\ \hline 
  \end{tabular}
\end{center}
\end{table}

Other recent 
extensive analyses of supersymmetry effects in $K\to\pi\nu\bar\nu$
 have been presented in \cite{NIR96,GN1,BRS}
where further references can be found. 
Model independent studies of these decays
can be found in \cite{NIR96,BRS}. The corresponding analyses in
various no--supersymmetric extensions of the Standard Model are 
listed in \cite{KLBSM}.
In particular, enhancement of $Br(K_L\to\pi^0\nu\bar\nu)$
by 1--2 orders of magnitude above the Standard Model
expectations is according to \cite{HHW98} 
still possible in four-generation models.
\section{Summary and Outlook}
\setcounter{equation}{0}
I hope that I have convinced the students that the field of CP violation
and rare decays plays an important 
role in the deeper understanding of the Standard Model 
and particle physics in general.
Indeed the field of weak decays and of CP violation is one of the least
understood sectors of the Standard Model.
Even if the Standard Model is still consistent with the existing data for
weak decay processes, the near future could change 
this picture
dramatically through the advances in experiment and theory.
In particular the experimental work
done in the next ten
years at BNL, CERN, CORNELL, DA$\Phi$NE, DESY, 
FNAL, KEK, SLAC and eventually LHC will certainly 
have considerable impact on this field.

Let us then
make a list of things we could expect in the next ten years.
This list is certainly very biased by my own interests but could
be useful anyway. Here we go:

\begin{itemize}
\item
The error on the CKM elements $\vcb$ and $\vub$ could be decreased 
below 0.002 and 0.01, respectively. This progress should come mainly from
Cornell, $B$-factories and new theoretical efforts. It would have
considerable impact on the unitarity triangle and would improve
theoretical predictions for rare and CP-violating decays sensitive
to these elements.
\item
The error on $\mt$ should be decreased down to $\pm 3\gev$
at Tevatron in the Main Injector era and to $\pm 1\gev$ at LHC.
\item
The measurement of non-vanishing ratio of $\epe$ by NA31
and KTeV, excluding confidently the superweak models, 
has been an important achievement.
The improved measurements of $\epe$ with the accuraccy of
 $\pm (1-2) \cdot 10^{-4}$ 
from NA48, KTeV and KLOE should give some insight into the 
physics of 
direct CP violation inspite of large theoretical uncertainties. 
In this respect measurements of CP-violating asymmetries in charged $B$
decays will also play an outstanding role. These experiments can be
performed e.g.\ at CLEO since no time-dependences
are needed. The situation concerning hadronic uncertainties is quite similar
to $\epe$. 
Therefore one should hope 
that some definite progress in calculating relevant hadronic matrix elements 
will also be made. 
\item
More events for $K^+\to\pi^+\nu\bar\nu$ could in principle
be reported from BNL already this year. In view of the theoretical 
cleanliness of this decay an observation of events at the $2\cdot 10^{-10}$
level would signal physics beyond the Standard Model.
A detailed study of this very
important decay requires, however, new experimental ideas and
new efforts. The new efforts \cite{AGS2,Cooper} in this direction allow 
to hope that
a measurement of $Br(\kpn)$ with an accuracy of $\pm 10 \%$ should
be possible before 2005. This would have a very important impact
on the unitarity triangle and would constitute an important test of
the Standard Model.
\item
The future improved inclusive $B \to X_{s,d} \gamma$ measurements
confronted with improved Standard Model predictions could
give the first signals of new physics. It appears that the errors
on the input parameters could be lowered further and the
theoretical error on $Br(B\to X_s\gamma)$ could be decreased
confidently down to $\pm 8 \%$ in the next years. The same
accuracy in the experimental branching ratio will hopefully
come  from Cornell and later from KEK and SLAC. 
This may, however, be insufficient to
disentangle new physics contributions although such an accuracy
should put important constraints on the physics beyond the Standard
Model. It would also be desirable to look for $B \to X_d \gamma$,
but this is clearly a much harder task.
\item
Similar comments apply to transitions $B \to X_s l^+l^-$ 
which appear to be even  more sensitive to new physics contributions
than $ B \to X_{s,d} \gamma$. An observation of
$B \to X_s \mu\bar\mu$ is expected from D0 and $B$-physics dedicated
experiments at the beginning of the next 
decade. The distributions of various kind when measured should
be very useful in the tests of the Standard Model and its extensions.
\item
The theoretical status of $K_{\rm L}\to \pi^0 e^+ e^-$ and of 
$K_{\rm L}\to \mu\bar\mu$, 
should be improved to confront future
data. Experiments at DA$\Phi$NE should be very helpful in this
respect. The first events of $K_{\rm L}\to \pi^0 e^+ e^-$ should
come in the first years of the next decade from KAMI at FNAL.
The experimental status of $K_{\rm L}\to \mu\bar\mu$, with the 
experimental error of $\pm 7\%$ to be decreased soon down to $\pm 1\%$,
is truly impressive.
\item
The newly approved experiment at BNL to
measure $Br(\klpn)$ at the $\pm 10\%$ level before 2005 may make a decisive
impact on the field of CP violation. 
In particular $\klpn$ seems to allow the
cleanest determination of $\imlt$. Taken together with $\kpn$
a very clean determination of $\sin 2 \beta$ can be obtained.
\item
The measurement of the $B^0_s-\bar B^0_s$ mixing and in particular of
$B \to X_{s,d}\nu\bar\nu$ and 
$B_{s,d}\to \mu\bar\mu$ will take most probably longer time but
as stressed in these lectures all efforts should be made to measure
these transitions. Considerable progress on $B^0_s-\bar B^0_s$ mixing
should be expected from HERA-B, SLAC and TEVATRON in the first years
of the next decade. LHC-B should measure it to a high precision.
With the improved calculations of $\xi$ in (\ref{107b}) this will have
important impact on the determination of $\vtd$ and on the
unitarity triangle. 
\item
Clearly future precise studies of CP violation at SLAC-B, KEK-B, 
HERA-B, CORNELL, FNAL and  LHC-B providing first
direct measurements of $\alpha$, $\beta$ and $\gamma$ may totally
revolutionize our field. In particular the first signals
of new physics could be found in the $(\bar\varrho,\bar\eta)$ plane.
During the recent years several, in some cases quite sophisticated and
involved, strategies have been developed to extract these angles with
small or even no hadronic uncertainties. Certainly the future will bring
additional methods to determine $\alpha$, $\beta$ and $\gamma$. 
Obviously it is very desirable to have as many such strategies as possible
available in order to overconstrain the unitarity triangle and to resolve 
certain discrete ambiguities which are a characteristic feature of these 
methods.
\item
The forbidden or strongly suppressed transitions such as
$D^0-\bar D^0$ mixing and $K_{\rm L}\to \mu e$ are also very
important in this respect. Considerable progress in this area
should come from the experiments at BNL, FNAL and KEK.
\item
On the theoretical side,
one should hope that the non-perturbative
methods will be considerably improved so that various $B_i$ parameters
will be calculated with sufficient precision. It is very important
that simultaneously with advances in lattice QCD, further efforts
are being made in finding efficient analytical tools for calculating
QCD effects in the long distance regime. This is, in particular very
important in the field of non-leptonic decays, where one should
not expect too much from our lattice friends in the coming ten years
unless somebody will get a brilliant idea which will revolutionize
lattice calculations. The accumulation of data for non-leptonic $B$
and $D$
decays at Cornell, SLAC, KEK and FNAL should teach us more 
about the role of non-factorizable contributions and in particular
about the final state interactions. 
In this context, in the case of K-decays, important
lessons will come from DA$\Phi$NE which is an excellent machine
for testing chiral perturbation theory and other non-perturbative
methods. 
\end{itemize}

In any case the field of weak decays and in particular of the FCNC 
transitions and of CP violation have a great future and
one should expect that they could dominate particle physics in the first 
part of the next decade. 
Clearly the next ten years should be very exciting in this field.

{\bf Acknowledgements}

I would like to thank Bruce Campbell, Faqir Khanna and Manuella
Vincter for inviting me to
such a wonderful Winter Institute and a great hospitality.
I would also like to thank M. Gorbahn and L. Silvestrini for
comments on the manuscript and the authors of \cite{EP99}
for a most enjoyable collaboration.

\vfill\eject

\end{document}